%% file: main.tex
\PassOptionsToPackage{table}{xcolor}
\documentclass[acmsmall,anonymous=false,review=false,nonacm]{acmart}

\setcopyright{acmlicensed}
\copyrightyear{2025}
\acmYear{2025}
\acmDOI{XXXXXXX.XXXXXXX}

\acmJournal{JACM}
\acmVolume{1}
\acmNumber{1}
\acmArticle{123}
\acmMonth{1}

\usepackage{verbatim}
\usepackage{subcaption}

\usepackage{makecell}
\usepackage{multirow}
\usepackage{enumitem}

\usepackage{siunitx}
\graphicspath{{figures/}}

\newcommand*\rot{\rotatebox{90}}

\begin{document}

\title{Martians Among Us: Observing Private or Reserved IPs on the Public Internet}

\author{Radu Anghel}
\orcid{0000-0002-0556-1742}
\affiliation{%
  \institution{Delft University of Technology}
  \city{Delft}
  \country{Netherlands}
}

\author{Qasim Lone}
\orcid{0009-0007-1742-0844}
\affiliation{%
  \institution{RIPE NCC}
  \city{Amsterdam}
  \country{Netherlands}
}

\author{Matthew Luckie}
\orcid{0000-0002-3872-4624}
\affiliation{%
  \institution{CAIDA}
  \city{}
  \country{United States}
}

\author{Carlos Gañán}
\orcid{0000-0002-4699-3007}
\author{Yury Zhauniarovich}
\orcid{0000-0001-9116-0728}
\affiliation{%
  \institution{Delft University of Technology}
  \city{Delft}
  \country{Netherlands}
}

\input{sections_old/01abstract}

\maketitle

\input{sections_old/02intro}

\input{sections/background.tex}

\input{sections/methodology.tex}
\input{sections/findings.tex}
\input{sections_old/06discussion}
\input{sections_old/07relatedwork}

\input{sections_old/08conclusions}

\clearpage
\bibliographystyle{ACM-Reference-Format}
\bibliography{refs, rfc}

\appendix
\input{sections/apdx_dataset.tex}
\input{sections/apdx_graphs.tex}

\end{document}

%% file: sections_old/01abstract.tex
\begin{abstract}
Spoofed traffic has been identified as one of the main issues of concern for network hygiene nowadays, as it facilitates Distributed Denial-of-Service (DDoS) attacks by hiding their origin and complicating forensic investigations. Some indicators of poor network hygiene are packets with Bogon or Martian source addresses representing either misconfigurations or spoofed packets. Despite the development of Source Address Validation (SAV) techniques and guidelines such as BCP 38 and BCP 84, Bogons are often overlooked in the filtering practices of network operators. This study uses traceroute measurements from the CAIDA Ark dataset, enriched with historical BGP routing information from RIPE RIS and RouteViews, to investigate the prevalence of Bogon addresses over seven years (2017–2023). Our analysis reveals widespread non-compliance with best practices, with Bogon traffic detected across thousands of ASes. Notably, 82.69\%–97.83\% of CAIDA Ark vantage points observe paths containing Bogon IPs, primarily RFC1918 addresses. Additionally, 19.70\% of all analyzed traceroutes include RFC1918 addresses, while smaller proportions involve RFC6598 (1.50\%) and RFC3927 (0.10\%) addresses. We identify more than 13,000 unique ASes transiting Bogon traffic, with only 11.64\% appearing in more than half of the measurements. Cross-referencing with the Spoofer project and MANRS initiatives shows a concerning gap: 62.67\% of ASes that do not filter packets with Bogon sources are marked as non-spoofable, suggesting incomplete SAV implementation. Our contributions include an assessment of network hygiene using the transiting of Bogon packets as a metric, an analysis of the main types of Bogon addresses found in traceroutes, and several proposed recommendations to address the observed gaps, enforcing the need for stronger compliance with best practices to improve global network security.
\end{abstract}

%% file: sections_old/02intro.tex
\section{Introduction}

For years, the persistent lack of network hygiene has been evidenced by the presence of spoofed Internet traffic. This type of traffic serves as a conduit for Distributed Denial-of-Service (DDoS) attacks, which generate significant volumes of data either directly or through amplification~\cite{reflectors,rossow2014amplification}. The obfuscation of attack origins in spoofed traffic complicates forensic investigations, often rendering them infeasible. Addressing this critical issue has been a central focus within the network security community~\cite{hopspoof,yaar2006stackpi,zargar2013survey}, leading to the development of technical solutions collectively known as ``Source Address Validation'' (SAV)~\cite{luckie2019network,beverly2009understanding}. These solutions are designed to authenticate source addresses, thereby mitigating the risks posed by IP spoofing.

Bogons, defined as packets with source addresses either unallocated by IANA or Regional Internet Registries (RIRs), or reserved for private or special use~\cite{rfc3871}, are one manifestation of spoofed traffic. Similarly, Martian packets are described in RFC3871~\cite{rfc3871} as packets with source addresses that cannot have their return traffic routed back to the sender due to the current forwarding tables, refining the earlier definition in RFC1208~\cite{rfc1208}. The IANA IPv4 Special Purpose Address Registry provides further classification of reserved address blocks, detailing their purpose and whether they should be globally reachable.

Since Bogons and Martians typically originate from spoofed traffic~\cite{rfc3871}, they pose significant risks to networks that fail to filter them. Packets originating from addresses that are not globally reachable, and which cannot have valid return routes in the Global Routing Table, align with the definitions of Bogons and Martians. Consequently, filtering such traffic is critical to enhancing network security.

Despite growing awareness of cybersecurity principles like security by design and secure defaults~\cite{secbydefaultcisagov}, these concepts are not yet widely implemented in practice. As a result, network devices often require manual configuration adjustments to improve their security. Guidance documents such as BCP 38~\cite{rfc2827} and BCP 84~\cite{rfc3704,rfc8704} provide recommendations for implementing SAV on both ingress and egress traffic. Researchers have developed several approaches to assess SAV deployment, including active measurements originating from the tested network~\cite{caida_spoofer}, traceroute loop analysis~\cite{lone2017loops}, and sending spoofed packets to DNS resolvers~\cite{korczynski2020don}. Meanwhile, initiatives like MANRS~\cite{MANRS} seek to incentivize network operators to adopt these practices. Yet, Bogon packets are frequently overlooked and dismissed as a minor issue, rather than being treated as a substantial security problem.

This paper challenges the assumption that Bogon packets are inconsequential. By underscoring their risks and measuring their prevalence, it aims to reframe their role in achieving better network hygiene and security.
First, we describe how we looked for the presence of a selection of Bogon addresses in traceroute measurements obtained from CAIDA Ark. The data from CAIDA Ark is enriched with originating AS information from the RIPE RIS and RouteViews projects. Using these datasets, spanning 7 years from January 2017 to December 2023, we identify network hygiene practices focusing on Bogon address usage and filtering practices and try to find correlations with SAV implementation.

Our analysis reveals the presence of Bogon addresses and a lack of filtering beyond the AS borders within thousands of ASes. Between 82.69\% and 97.83\% of CAIDA Ark vantage points detect at least one path containing Bogon IP addresses, with the most prevalent being RFC1918~\cite{rfc1918} space. We find that 19.70\% of the analyzed traceroutes contain RFC1918~\cite{rfc1918} addresses, 1.50\% involve RFC6598~\cite{rfc6598} addresses, and 0.10\% include addresses from RFC3927~\cite{rfc3927}.
Furthermore, our findings indicate that 2.97\% of all issued traceroutes from the CAIDA Ark dataset were directed to the destinations that are not visible in RIS or RouteViews RIBs, prompting further investigation into the accuracy of routed prefixes used by CAIDA Ark.

Our analysis identified \num{14463} unique ASes transiting packets with Bogon addresses as the source, and while some ASes traverse multiple types of Bogons, ASes transiting RFC1918~\cite{rfc1918} addresses dominate. Interestingly, ASes transiting Bogons do not appear consistently across many measurements. The total number of ASes that appeared transiting Bogons across all \num{84} measurements is \num{269}. Only 12.55\% (\num{1743})of ASes are observed in more than half of the measurements. We believe such behavior is either due to the Ark dataset measurement inconsistencies caused by the randomization of vantage points and destinations, or ASes fixing their misconfigurations.

We compared our results with two other sources tackling SAV: the Spoofer project data~\cite{caida_spoofer} and the Mutually Agreed Norms for Routing Security (MANRS)~\cite{MANRS}.
Cross-checking our findings with the Spoofer dataset (considering measurements conducted within 6 months before the corresponding traceroute with an AS transiting packets with Bogon sources, as explained in\autoref{sec:methodology}), we found that 62.67\% of unique ASes transiting Bogons were reported as non-spoofable.
Similarly, cross-referencing with MANRS revealed that 63.66\% of unique ASes adhere to anti-spoofing filtering. This suggests that although some ASes implement SAV checks, they might overlook filtering Bogons.

The main contributions of this paper are:
\begin{itemize}
    \item We present the first comprehensive assessment of network hygiene  --moving beyond conventional concerns strictly related to IP spoofing-- with a focus on packets with Bogon source addresses. Our findings reveal that Private-Use addresses are more prevalent than other types of Bogons appearing on the Internet.

\item We identify widespread non-compliance with best practices across various ASes, mostly in ISP networks. More than 57\% of these ASes are registered in a handful of countries, i.e., the USA, Brazil, and Russia.

\item We provide a detailed breakdown of ASes transiting multiple types of Bogons, offering insights into their distribution and patterns. Over the study period, \num{14463} unique ASNs were found transiting packets sourced by Bogon addresses.

\item We characterize the intersection between spoofable and non-spoofable ASes, as found by the Spoofer project, revealing an overlap with ASes forwarding packets with RFC1918-type Bogons as source address. We provide recommendations to address the gap in the deployment of SAV measures.
\end{itemize}

%% file: sections/background.tex
\section{Background}
\label{sec:background}

\subsection{Bogon Addresses}
\label{subsec:bogon_addresses}
Bogon refer to packets with source or destination addresses that are not supposed to be routed on the public Internet, such as packets coming from private addresses, addresses that are reserved or not yet assigned by a Regional Internet Registry. Some of these addresses are described in the IANA IPv4 Special-Purpose Address Registry~\cite{ianav4special}.

From the IANA IPv4 Special-Purpose Address Registry~\cite{ianav4special}, we considered the following subnets:

\begin{description}[leftmargin=1.5em,labelindent=1.0em]
    \item[240.0.0.0/4] According to RFC1112~\cite{rfc1112}, these addresses, also known as ``Class E'' IPs, are reserved for future use and could not be assigned to network interfaces or routed. However, recently, it has been discovered that some large companies, e.g., Amazon AWS, started using them internally~\cite{qasim240atlas}.     
    \item [127.0.0.0/8] According to RFC1122~\cite{rfc1122}, these addresses are used as Loopback IPs that always point to the local host.
    \item[10.0.0.0/8, 172.16.0.0/12, 192.168.0.0/16] These blocks represent traditional Private-Use addresses. According to RFC1918~\cite{rfc1918}, they are utilized in most home and service provider networks behind various types of firewalls and NAT.  
    \item[169.254.0.0/16] According to RFC3927~\cite{rfc3927}, these Link-Local addresses are used to permit IP connectivity between hosts in the same physical network when no static or dynamic IP configuration is provided to them.
    \item [192.0.2.0/24, 198.51.100.0/24, 203.0.113.0/24] These addresses, specified in RFC5737~\cite{rfc5737}, are reserved for use in documentation, examples, specifications or other types of documents.
    \item[100.64.0.0/10] According to RFC6598~\cite{rfc6598}, this block of IP addresses is anticipated to be used as Shared Address Space for Carrier-Grade NAT (CGN) devices. Their purpose is similar to RFC1918 Private-Use addresses, but they are intended to be used on Service Provider networks.
    \item[192.0.0.0/24] This block of addresses, called Protocol Assignments and defined in RFC6890~\cite{rfc6890}, is reserved for use by various protocols, such as NAT64/DNS64 Discovery (RFC8880~\cite{rfc8880}/RFC7050~\cite{rfc7050}), Port Control Protocol Anycast (RFC7723~\cite{rfc7723}).
    \item[192.88.99.0/24] According to RFC7526~\cite{rfc7526}, this block of 6to4 Relay Anycast addresses was initially designed to aid transition to IPv6, now deprecated due to high failure rates.
\end{description}

Other special purpose addresses such as 192.175.48.0/24, Direct Delegation AS112 Service~\cite{rfc7534}, were excluded as they are expected to be visible on the Internet.

\subsection{Datasets}
\label{subsec:datasets}
In this work, we rely on several datasets to study Bogon addresses. In particular, we leverage datasets from CAIDA, including the IPv4 Routed /24 Topology~\cite{caidav4routedtopodataset} and the Spoofer~\cite{caida_spoofer} data collected by the Ark platform~\cite{ark}, and historical routing information provided by RIPE Routing Information Service (RIS)~\cite{ripeRoutingInformation} and the University of Oregon Route Views Project~\cite{RouteViews}. These datasets collectively provide valuable insights into networks engaged in Bogon filtering practices and the deployment of Source Address Validation (SAV) mechanisms.

\subsubsection{CAIDA IPv4 Routed /24 Topology Dataset}
\label{subsubsec:ipv4_routed_topology_dataset}
For our research, we utilize snapshots of the CAIDA IPv4 Routed /24 Topology Dataset~\cite{caidav4routedtopodataset}, which is collected through the Ark measurement infrastructure~\cite{ark}. This infrastructure is comprised of monitors deployed globally that send scamper~\cite{luckie2010scamper} probes to randomly selected IP addresses from each routed IPv4 /24 prefix. This dataset provides traceroute-like information for the selected destinations and is collected in daily cycles. One cycle contains traceroutes to all routed /24s. We refer to this dataset as \emph{Ark Dataset}.

\subsubsection{RIPE Routing Information Service (RIS)}
\label{subsubsec:ripe_ris_dataset}
In order to identify ASes that transit packets with bogon source addresses we use BGP routing information snapshots. Such information changes over time due to changes in ownership of resources; thus, for better accuracy, the BGP data should be closer to the collection time of other data involved (traceroutes).

The RIS project~\cite{ripeRoutingInformation} is maintained by the RIPE NCC and plays an important role in the collection and retention of historical BGP routing data obtained from various ASes across the globe. RIS operates multiple route collectors, 26 at the time of writing, named \texttt{RRCxx} and located at various Internet Exchange points around the globe, strategically distributed across Europe, North America, Asia, South America, and Africa. Three of these route collectors operate as ``multi-hop'', meaning that they collect routes not only from networks present at that particular IXP but also from the rest of the world. Due to this type of setup, the RIS project can offer global views from the multi-hop RRCs as well as geographically unique information from local networks directly connected at IXPs with an RRC present.

The BGP data is collected as Multi-threaded Routing Toolkit raw \texttt{MRT} files (RFC6396~\cite{rfc6396}) and offered as either full dumps or updates. The dumps represent snapshots of BGP routing information, while updates contain incremental updates to the snapshots. Further, we refer to this dataset as \emph{RIPE RIS Dataset}.

\subsubsection{University of Oregon Route Views}
\label{subsubsec:oregon_route_views_dataset}
Like the RIPE RIS~\cite{ripeRoutingInformation} dataset, University of Oregon's Route Views project~\cite{RouteViews} also collects global routing information. However, one of the notable differences between them is that the original purpose of the Route Views project was to provide a tool for network operators to obtain near real-time BGP information about global routing as seen by the contributing networks. Moreover, it also allowed network operators to perform traceroutes from the route collectors. The current setup of the University of Oregon's Route Views  project is comprised of 42 such collectors distributed worldwide. The BGP data provided by Route Views is also in the \texttt{MRT} format with both snapshots (dumps) and incremental updates. In what follows, we refer to this dataset as \emph{UOregon RVIEWS Dataset}.

\subsubsection{CAIDA Spoofer}
\label{subsubsec:spoofer_dataset}
The second CAIDA dataset used within this work is provided by the Spoofer project~\cite{caida_spoofer}. This project aims to evaluate and report on the implementation of best practices for source address validation (SAV) to mitigate spoofing attacks. To achieve this goal, the project uses a combination of client and server software to send and receive IP packets with a spoofed source address. If a router does not drop inbound packets with the source address internal to client's network or outbound packets with the source address external to client's network, then the corresponding network is vulnerable to the spoofing attack. The results of these tests with the anonymized IP addresses are publicly available. The complete results with real IP addresses are shared privately with network operators that require it to identify equipment where SAV measures are not deployed.

\subsubsection{MANRS for Network Operators}
\label{subsubsec:manrs}
The \emph{Mutually Agreed Norms for Routing Security (MANRS)}~\cite{MANRS} initiative, led by Internet Society, brings together network operators, Internet Exchange Points (IXPs), equipment vendors, CDN and Cloud Providers. It aims at fostering collaboration among participants to enhance the resilience and security of the routing infrastructure. By joining the initiative, they show their commitment to achieving this goal. The initiative tracks participants' commitment by performing a set of regular measurements, showing their readiness to adhere to \emph{Compulsory} and \emph{Recommended} actions. These actions stem from recognized industry best practices, chosen by weighing incremental costs incurred by an individual participant and potential common benefits, and are not exhaustive. The results of the measurements are publicly available online. 

One group of MANRS participants is \emph{Network Operators}~\cite{MANRSNetworkOperators} and ``Prevent traffic with spoofed source IP addresses'', or \texttt{anti\_spoofing}, is among the recommended (non-compulsory for implementation) actions for it. This action suggests network operators check their networks on whether they can send packets with spoofed IP addresses. To check the conformance to this action, MANRS also uses the CAIDA Spoofer~\cite{caida_spoofer} data.

%% file: sections/methodology.tex
\section{Methodology}
\label{sec:methodology}
To identify ASes that allow Bogons to traverse their networks, we rely on the Ark dataset, which contains path information collected with a traceroute tool from the monitor to the destination -- a randomly selected IP address from each routed IPv4 /24 prefix. To collect this information, the tool sends several packets (typically, \emph{ICMP Echo Requests}) to the same destination. The first packet is assigned a Time-to-Live (TTL) value of 1, and each subsequent packet increments the TTL by 1. The TTL field prevents packets from circulating indefinitely: each router along the path decreases the TTL by 1. When a router receives a packet with a TTL of 1, it discards the packet and sends back an \emph{ICMP Time Exceeded} error message, with the router's IP address as the source. By interpreting these responses, the traceroute tool reconstructs the route the packet takes to reach the destination.

\begin{figure}
    \centering
    \includegraphics[width=\linewidth]{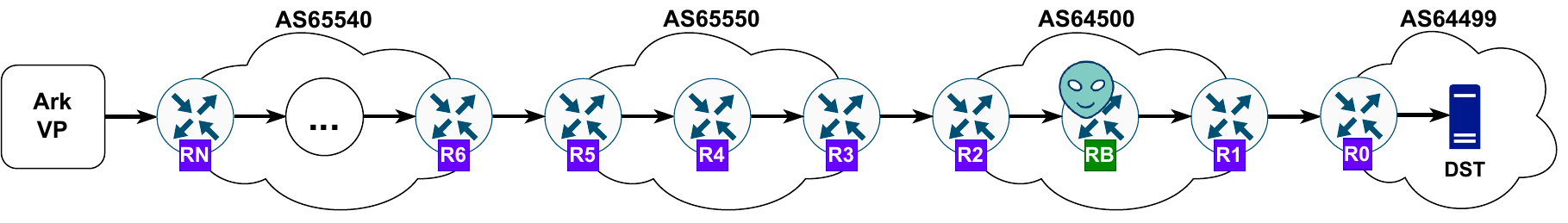}
    \caption{Approach idea}
    \label{fig:idea}
\end{figure}

Consider a case of path collection illustrated in \autoref{fig:idea}, where the monitor, Ark Vantage Point (\emph{Ark VP}), traces the route to the destination (\emph{DST}). The collected path -- RN -> ... -> R6 -> R5 -> R4 -> R3 -> R2 -> \textbf{RB} -> R1 -> R0 -- includes the IP addresses of the routers encountered along the way. Among these, \textbf{RB} has a bogus IP address (Bogon). This means that the \emph{ICMP Time Exceeded} error message is sent from this router with its bogus IP address in the source field. Such IP addresses are non-routable and should be filtered out. If they appear in the collected path, it indicates that the routers between the Bogon and Ark VP (and the ASes correspondingly) fail to filter these packets.

\begin{figure}
    \centering
    \includegraphics[width=\linewidth]{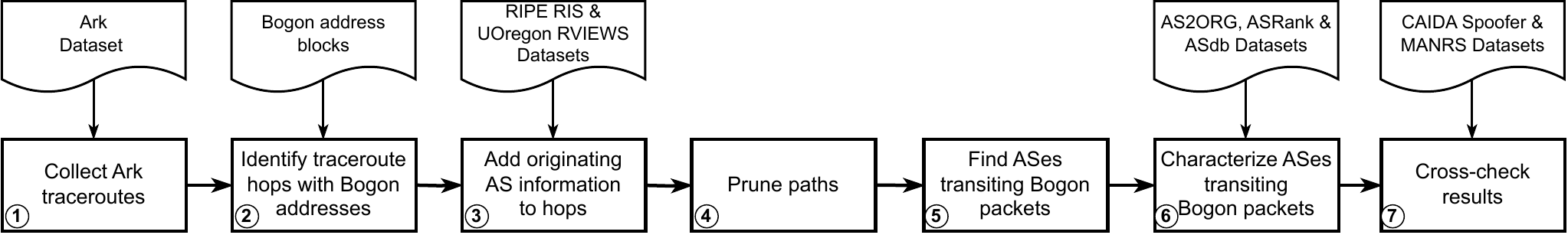}
    \caption{Bogon addresses identification overview}
    \label{fig:methodology}
\end{figure}

\autoref{fig:methodology} outlines the methodology used in this study to identify and characterize ASes transiting Bogons. The rectangles with numbered circles denote specific methodology steps, while the document icons represent the data employed to execute each respective step.

\vspace{3pt}
\noindent \textbf{\textit{Step 1:} Collect Ark traceroutes.} At this step, we extract traceroutes from the Ark dataset~\cite{caidav4routedtopodataset} for the same day of each month for the period of 84 months, from January 2017 up to December 2023. We chose the $18^{th}$ day of each month because we found that day to contain traceroute cycles in all analyzed months. \footnote{We checked all days from $1^{st}$ to $12^{th}$, but there were either incomplete cycles (very few traceroutes) or the data for the whole day was missing. Then, we randomly checked some other days and discovered that for $18^{th}$ day, we have the data for all months.} As described in \autoref{subsubsec:ipv4_routed_topology_dataset}, it may happen that during the same day several cycles of measurements are executed. In these cases, we extract the traceroutes from the cycle with the lowest number.

\vspace{3pt}
\noindent \textbf{\textit{Step 2:} Identify traceroute hops with bogon addresses.} At this step, we check if an IPv4 address of a traceroute hop belongs to one of the selected IANA IPv4 Special-Purpose Addresses~\cite{ianav4special}. We use a subset of address blocks from this registry as listed in \autoref{tab:special_purpose_addresses}, which also determines the reference RFC and a brief description of the corresponding block used further within this work. We do not search for addresses coming from prefixes such as 0.0.0.0/8~\cite{rfc791}\cite{rfc1122}, or 192.31.196.0/24 (AS112-v4)~\cite{rfc7535}, and 192.175.48.0/24 (Direct Delegation AS112 Service)~\cite{rfc7534} due to the low probability of these addresses being visible as intermediary hops in traceroutes.

\begin{table}
\centering
\caption{Bogon address blocks}
\label{tab:special_purpose_addresses}
\resizebox{.6\columnwidth}{!}{%
\begin{tabular}{llr}
\toprule
\textbf{RFC} & \textbf{Description} & \textbf{CIDR} \\
\midrule
1112 & Former Class E & 240.0.0.0/4 \\
1122 & Loopback & 127.0.0.0/8 \\
1918 & Private-Use & 10.0.0.0/8, 172.16.0.0/12, 192.168.0.0/16 \\
3927 & Link-Local & 169.254.0.0/16 \\
5737 & Documentation & 192.0.2.0/24, 198.51.100.0/24, 203.0.113.0/24 \\
6598 & Shared Address Space & 100.64.0.0/10 \\
6890 & Protocol Assignments & 192.0.0.0/24 \\
7526 & 6to4 Relay Anycast & 192.88.99.0/24 \\
\bottomrule
\end{tabular}}
\end{table}

\vspace{3pt}
\noindent \textbf{\textit{Step 3:} Add originating AS information to traceroute hops.} During this step, we map each hop's IPv4 address in a traceroute to the corresponding origin ASN. To perform this task, we use two datasets containing historical routing information, namely RIPE RIS (\autoref{subsubsec:ripe_ris_dataset}) and UOregon RVIEWS (\autoref{subsubsec:oregon_route_views_dataset}). For this, we download the \texttt{MRT} dump files collected at \textit{00:00} on the same day as the Ark traceroute cycle.  From the RIPE RIS dataset, we used RIBs gathered by the \texttt{RRC00} collector. This collector stands out because it is one of the multi-hop RIS route collectors that consolidates information from a wide array of peers found in various locations globally. From the UOregon RVIEWS dataset, we use the dumps from the \texttt{route-views2} collector.
In some cases, RIPE RIS and UOregon RVIEWS datasets also have a mapping between some special-purpose IP addresses and a corresponding AS. These are prefixes leaked to route collectors, and we ignore the origin AS found for them in these datasets.
Based on this information, we build an \emph{AS path} for the data plane of each traceroute -- a sequence of ASNs, each of which corresponds to a hop in the traceroute.

\vspace{3pt}
\noindent \textbf{\textit{Step 4:} Clean the AS path.} We remove consecutive duplicate ASNs (multiple hops in the same ASN), hops with \emph{unknown} origin that are not bogon addresses (public addresses not found in the Global Routing Table, usually IXP prefixes).

\vspace{3pt}
\noindent \textbf{\textit{Step 5:} Categorization.} After obtaining a clean AS path, we look for bogon addresses that are found at more than one AS-hop away from the source of the traceroute, ignoring the bogons inside the network originating the traceroute. We do this since it is expected to have private addresses inside your own network, however they should not be visible when `crossing the border' to a different AS. We specifically look for the following cases:

\begin{itemize}
  \item[\textbf{BA:}] \textbf{All ASes on the path found before a Bogon address.} Since the packets with Bogon addresses as source reach the origin of the traceroute (Ark VP), this case shows that all ASes on that path forward these packets, against best practices. It is, however, possible that due to the asymmetry of routing on the Internet, the actual reverse path these packets take to the origin of the traceroute could be different from the forward path. Thus, there is less certainty about the filtering practices of these ASes, except for the one closest to the origin of the traceroute. In \autoref{fig:idea}, AS64500, AS65550, and AS65540 belong to this category.
    
  \item[\textbf{BB:}] \textbf{The AS found right before a Bogon address.} The AS originating the address found on the traceroute right before a Bogon address could mean that Bogon addresses are in use inside that AS, or a neighboring AS is using them. In both cases though the AS found immediately before the bogon address is transiting those packets with a bogon address as a source and forwarding them beyond their AS towards the Ark VP running the traceroute. In \autoref{fig:idea}, AS64500 belongs to this category.

  \item[\textbf{BC:}] \textbf{An AS sandwich with a Bogon inside.} If the same AS is found both before and after Bogon addresses, this signals that, with a high probability, the Bogon addresses are in use inside this AS. Due to BGP loop prevention mechanisms, it is highly unlikely that packets would leave the AS preceding the Bogon address, traverse to a different AS, and then return to the original AS for the traceroute hop following the Bogon. That AS also does not filter the corresponding packets as they are forwarded towards the Ark VP. For instance, in \autoref{fig:idea}, AS64500 belongs to this category. 
\end{itemize}

Note that according to this definition during a measurement $BC \subseteq BB \subseteq BA$.

\vspace{3pt}
\noindent \textbf{\textit{Step 6:} Characterize ASes transiting bogons.} During this step we characterize the ASes identified in previous steps as allowing packets with bogon source or destination to pass through their network by using publicly available datasets. The datasets used are ASdb~\cite{ziv2021asdb}, AS2ORG~\cite{as2org} and ASRank~\cite{ASRank}. The ASdb dataset~\cite{ziv2021asdb} categorizes organizations associated with an AS according to the North American Industry Classification System (NAICSlite). For our characterization we use the `Category 1 - Layer 1' and `Category 1 - Layer 2' fields of the ASdb dataset.\footnote{\url{https://asdb.stanford.edu/}}  We utilize the AS2ORG dataset~\cite{as2org} to provide a mapping between the AS and the Regional Internet Registry (RIR) where the AS is registered, and identify the country of the organization. Finally, we use the ASRank~\cite{ASRank} dataset to obtain the geographical coordinates of the registration place of the AS in order to build the map graph.

\vspace{3pt}
\noindent \textbf{\textit{Step 7:} Cross-check the results.}
Here we compare the results of our analysis with results in two relevant datasets: CAIDA Spoofer~\cite{caida_spoofer} and MANRS for Network Operators~\cite{MANRSNetworkOperators}.

Transiting packets with bogon addresses as source or destination is a clear sign of a misconfiguration, but it does not necessarily mean that IP spoofing is also possible on that network. For this, the Spoofer dataset acts as ground truth due to the active measurements performed from within the tested AS. We consider an AS as spoofable if we can find at least one match in the Spoofer dataset during a period of up to 6 months before\footnote{We check if a routable spoofed packet is received for the corresponding AS.} the corresponding Ark traceroute measurement. We use such a large timespan of Spoofer data because older versions of the Spoofer client required checks to be manually started, i.e. the end-user has to run the software and push a button to start the test, the result of this approach being irregular measurements. Recent versions of the Spoofer client perform the tests automatically and run the measurements regularly. In order to match with the period of Ark traceroutes, we downloaded through the Spoofer API results for tests run between June 2016 and December 2023.
Although MANRS \texttt{anti-spoofing} action checks rely on data from the Spoofer project, the added information is that the network operator actively pledged to undertake anti-spoofing measures. For MANRS, we downloaded the latest list of conformance for all network operators participating in MANRS by using the MANRS API,\footnote{Measurement date: ``2024-05-01''}  we then cross checked that data with the set of ASes we identified as transiting packets with bogon source or destination.\footnote{\url{https://manrs.stoplight.io/docs/manrs-public-api/97379961794c7-list-net-ops-conformance}}

Note that the Spoofer measurements are done on a voluntary basis and thus are not performed from all ASes, and participants can opt out of sharing the results publicly, thus, the Spoofer dataset may be incomplete.

\subsection{Ethics}
\label{subsec:ethics}
Our study presents no ethical concerns for several reasons. Firstly, we do not collect any additional data; instead, we re-analyze existing data from multiple established sources, including traceroutes from the CAIDA IPv4 Routed /24 Topology Dataset and historical routing information from RIPE RIS and UOregon RVIEWS, as detailed in \autoref{subsec:datasets}. Other databases used to enrich the information about ASes transiting packets with Bogon sources are also publicly available. Consequently, our research entails no supplementary risk in data collection. Our analysis focuses on network topology and filtering, thereby posing no direct risk to individuals. 

%% file: sections/findings.tex
\section{Prevalence of Bogons and characteristics of the networks transiting them}
\label{sec:findings}
To identify the existence and prevalence of Bogon addresses on the public Internet, we conducted a longitudinal study of traceroutes gathered by the CAIDA Ark project over a period of seven years. We downloaded the traceroute data for the $18^{th}$ day of each month within this period and analyzed them following the methodology described in \autoref{sec:methodology}. \autoref{tab:stats_by_month} in \autoref{apdx:measurement_results} reports on the data used in this study and summarizes our findings. In this section we present the main results of our analysis: (i) prevalence of Bogon addresses in traceroutes, (ii) ASes transiting packets from Bogon addresses, (iii) compare ASes with Bogon vs. Spoofable, and (iv) characteristics of ASes transiting Bogons.

\subsection{Traceroutes with Bogon Addresses}
\label{subsec:traceroutes_analysis}
\begin{figure}
\centering
\includegraphics[width=.7\linewidth]{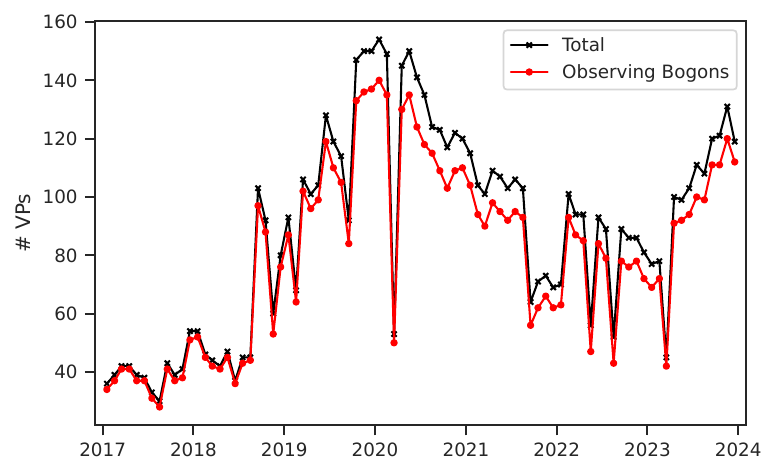}
\caption{Number of Vantage Points}
\label{fig:number_of_vps}
\end{figure}

\autoref{fig:number_of_vps} shows the number of Ark Vantage Points (VPs) used to collect traceroutes. The black line with cross markers shows the total number of Vantage Points used to perform the measurements (the ``\# VPs'' column in \autoref{tab:stats_by_month} reports the same numbers). As we can see, the number of VPs was pretty low (under 40 VPs) till August 2018, then started to grow, reaching $154$ VPs in January 2020, and afterward, decreasing till March 2023. In general, the number of VPs is quite unstable ($88.1\pm35.2$), with deep drops happening from time to time (e.g., in March 2020).

The red line with dot markers shows the number of VPs observing at least one Bogon address (the ``\# VPs observing Bogons'' column in \autoref{tab:stats_by_month}). Between 82.69\% and 97.83\% of CAIDA Ark VPs observed at least one traceroute path containing a Bogon address. 

\begin{figure}[!t]
\centering
\includegraphics[width=.7\linewidth]{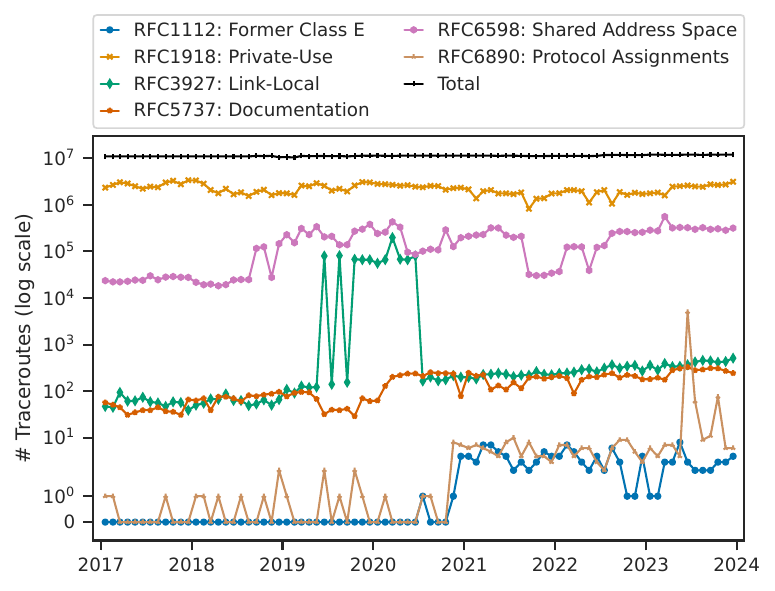}
\caption{Number of Traceroutes with Bogon addresses}
\label{fig:number_of_traceroutes_with_bogons}
\end{figure}

The black line with vertical line markers in \autoref{fig:number_of_traceroutes_with_bogons} shows the total number of analyzed traceroutes per month. We analyzed \num{944180867} traceroutes in total. The average number of traceroutes per measurement is \num{11240248}, with a maximum of \num{11938094} recorded in September 2023 and a minimum of \num{10330500} observed in February 2019.   

\begin{figure}
\centering
\includegraphics[width=.7\linewidth]{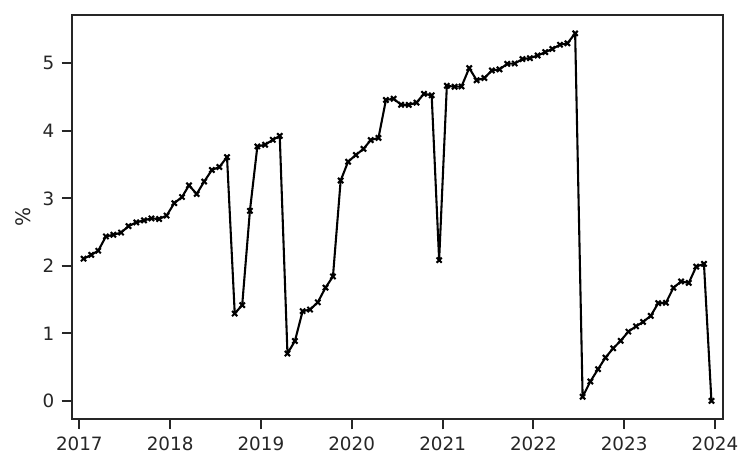}
\caption{\% of Traceroutes with Unrouted Destinations}
\label{fig:perc_of_traceroutes_with_unrouted_dest}
\end{figure}

Out of \num{944180867} traceroutes issued by the CAIDA Ark platform, \num{28000671} are to destinations not found in the RIPE RIS or UOregon RVIEWS Global Routing Tables. This means that 2.97\% of all issued traceroutes in the Ark dataset during the analyzed period were directed towards unrouted destinations. \autoref{fig:perc_of_traceroutes_with_unrouted_dest} shows the percentage of such traceroutes per measurement. We reached out to CAIDA regarding this issue, and they confirmed that the prefixes used to generate the /24 topology dataset may be outdated. Currently, CAIDA is actively working on utilizing the prefix list obtained from a snapshot of the Global Routing Table collected on the day of measurement.

The other lines in \autoref{fig:number_of_traceroutes_with_bogons} correspond to the number of traceroutes containing Bogon addresses of particular types defined in \autoref{tab:special_purpose_addresses}. As we can see, the most common traceroutes with Bogon IPs contain Private-Use addresses (RFC1918). They are found in \num{185958544} (19.70\%) of all traceroutes, with an average of \num{2213792} per measurement. Most often, these Private-Use addresses are located close to the source of the traceroute and do not cross the AS border. Therefore, we believe they are used in the infrastructure of the ASes hosting the Ark nodes.

Protocol Assignments (RFC6890) addresses were observed in \num{5389} traceroutes. By analyzing the paths, we believe that these addresses are used in a similar way to the Private-Use (RFC1918) addresses and not for the specific protocols they were reserved for. We make this assumption because we observe IPs such as 192.0.0.201, a part of 192.0.0.0/24 prefix, which is currently not assigned to any specific protocol. Most of the observed paths lead to the ASN of a telecom provider in Brazil, to an ASN belonging to a cloud service provider in the US, and to four ASNs located in Spain.

The Former Class E addresses (RFC1112) appear in \num{91} traceroutes over the observed period. All these traces target only two ASNs, the largest AS in Japan, according to IHR~\cite{ihrjp}, and one large US-based AS. Although it is known that networks such as Amazon AWS use this address space internally~\cite{qasim240atlas}, these were not visible in the Ark dataset, suggesting proper filtering is in place.

Documentation (RFC5737) prefixes are found in \num{8315} traceroutes to 57 unique destination ASNs.

From \num{944180867} analyzed traceroutes, 1.50\% contain Shared Address Space IPs (RFC6598), and 0.10\% include Link-Local addresses (RFC3927). Interestingly, there is a spike in the number of traces with Link-Local (RFC3927) addresses between June 2019 and July 2020. Moreover, at the start of this period, namely in July and September, the number of traceroutes with this type of Bogon addresses returned to their previous values. We assume that these spikes are due to temporary misconfigurations inside these ASN(s), which were fixed later.  

Other types of Bogon addresses are found in less than 0.01\% of the traceroutes. Loopback (RFC1122) and the 6to4 Relay Anycast (RFC7526) addresses do not appear in any traceroutes, therefore, we exclude them from further consideration.

\subsection{ASNs Transiting packets with Bogon sources}
\label{subsec:asns_transiting_bogons}
\begin{figure}[!t]
\centering
\includegraphics[width=.7\linewidth]{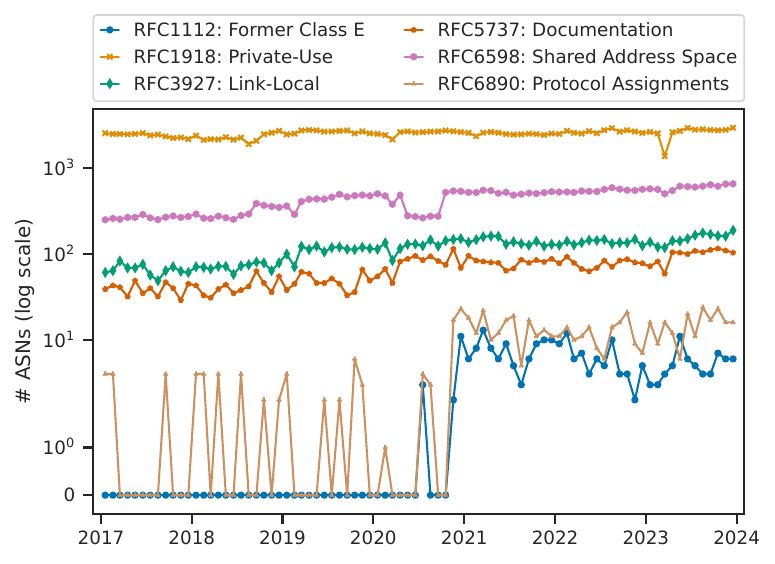}
\caption{\textbf{BA:} Number of ASNs Transiting Bogons per RFC}
\label{fig:number_of_asns_with_bogons}
\end{figure}

In this subsection, we report the results of the analysis of ASNs transiting Bogons as described in \autoref{sec:methodology} (see Step 5). \autoref{fig:number_of_asns_with_bogons} presents the breakdown of ASes transiting Bogons per RFC over the study period (the \textbf{BA} case). Results show that most ASNs transiting packets with Bogon sources do not filter out Private-Use (RFC1918), Shared Address Space (RFC6598), Link-Local (RFC3927) and Documentation (RFC5737) addresses.

For the whole period of the analyzed dataset we have found \num{14463} unique ASes transiting packets with Bogon addresses. \autoref{tab:unique_asns_per_year_left} reports the breakdown of the number of unique ASNs transiting Bogons per year and per RFC (the \textbf{BA} case). Over the 84 months of measurements, we observed a total of \num{13883} ASNs transiting Private-Use (RFC1918), \num{2869} not filtering Shared Address Space (RFC6598), \num{747} with Link-Local addresses (RFC3927) and \num{416} transiting packets coming from Documentation (RFC5737) addresses.

\begin{table}
    \centering
    \caption{\textbf{BA:} Breakdown of the Number of Unique ASNs Transiting Bogons per Year and per RFC}
    \scalebox{.8}{
        \input{tabulars/left/unique_asns_per_year.tex}
    }
    \label{tab:unique_asns_per_year_left}
\end{table}

When using the more conservative approach of only recording the AS immediately before the Bogon address (the \textbf{BB} case), the number of ASes transiting Bogons is slightly lower, as shown in \autoref{tab:unique_asns_per_year_firstleft}, with \num{13084} ASNs forwarding packets for Private-Use (RFC1918) addresses, \num{2188} forwarding Shared Address Space (RFC6598), \num{382} allowing packets from Link-Local addresses (RFC3927), and \num{92} ASes forwarding packets coming from Documentation (RFC5737) addresses.

\begin{table}
    \centering
    \caption{\textbf{BB:} Breakdown of the Number of Unique ASNs Transiting Bogons per Year and per RFC}
    \scalebox{.8}{
        \input{tabulars/firstleft/unique_asns_per_year.tex}
    }
    \label{tab:unique_asns_per_year_firstleft}
\end{table}

Similarly, \autoref{tab:unique_asns_per_year_sandwich} shows the number of ASes transiting packets of Bogon addresses that we can confidently say they also use those addresses in their infrastructure (the \textbf{BC} case). %
For this case, we identify \num{7762} ASes transiting packets with Private-Use (RFC1918) addresses, \num{879} ASes with Shared Address Space (RFC6598), \num{104} ASes with Link-Local addresses (RFC3927), and \num{21} ASes using Documentation (RFC5737) IPs.

\begin{table}
    \centering
    \caption{\textbf{BC:} Breakdown of the Number of Unique ASNs Transiting Bogons per Year and per RFC}
    \scalebox{.8}{
        \input{tabulars/sandwich/unique_asns_per_year.tex}
    }
    \label{tab:unique_asns_per_year_sandwich}
\end{table}

The number of unique ASNs transiting packets with Bogon source addresses has steadily increased over the years, possibly due to the lack of IPv4 addresses since the IPv4 runout. At the same time, the exact number of ASes transiting packets with Bogon addresses varies considerably (see \autoref{tab:stats_by_month} in \autoref{apdx:measurement_results}). For instance, the number of ASNs transiting Private-Use (RFC1918) addresses varies from \num{1382} to \num{2973}. Over the 84 months of measurements, out of the total \num{13883} ASNs transiting the Private-Use (RFC1918) addresses, \num{3213} (23.14\%) ASNs are observed only once. Only \num{9166} (66.02\%) ASNs are present in more than two measurements, and \num{1743} (12.55\%) appear in at least half of the measurements. We find 652 (4.7\%) ASNs appear in over 75 measurements (90\% of the measurements) and 519 (3.74\%) ASNs appear in over 95\% of the measurements. The number of ASNs found in all 84 measurements is only 269, representing 1.94\% of the total observed ASNs. We also observe significant variations for the other two most popular Bogon address types: Shared Address Space (RFC6598) and Link-Local (RFC3927).

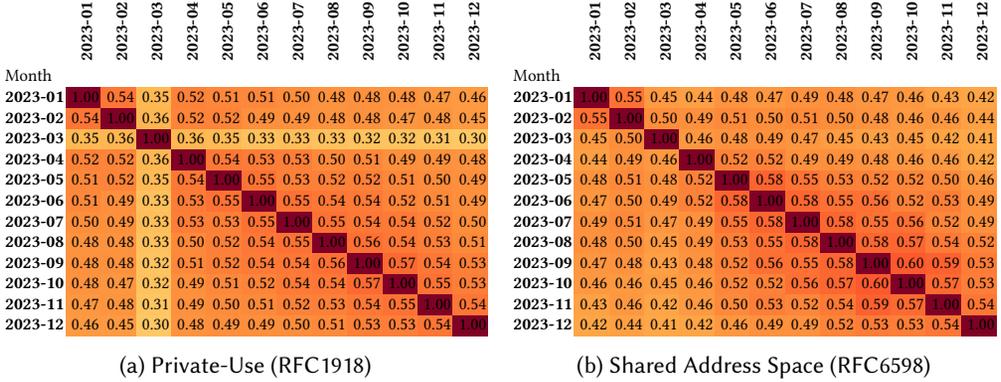
\begin{figure}
\centering
\begin{subfigure}[b]{0.48\linewidth}
    \resizebox{\linewidth}{!}{
        \setlength\tabcolsep{1pt}
        \input{tabulars/left/jaccard_similarity_rfc1918_202301-202312.tex}
    }
    \caption{Private-Use (RFC1918)}
    \label{fig:jaccard_similarity_rfc1918}
\end{subfigure}
\begin{subfigure}[b]{0.48\linewidth}
    \resizebox{\linewidth}{!}{
        \setlength\tabcolsep{1pt}
        \input{tabulars/left/jaccard_similarity_rfc6598_202301-202312.tex}
    }
    \caption{Shared Address Space (RFC6598)}
    \label{fig:jaccard_similarity_rfc6598}
\end{subfigure}
\caption{\textbf{BA}: Jaccard Similarity of ASNs Transiting Bogon packets of Particular Type across Months in 2023}
\label{fig:jaccard_similarity_rfc1918_rfc6598}
\end{figure}

These results indicate that while a large number of unique ASNs are observed, the frequency of appearance varies significantly, with only a small portion of ASNs being observed across multiple measurements. To exemplify this, we built a matrix containing the Jaccard similarity of the sets of ASNs transiting Bogons across different months in 2023. \autoref{eq:jaccard_sim} provides a formula of the Jaccard similarity for two sets. We have chosen the Jaccard similarity because the sets have similar sizes and, as this metric is widely used in scientific literature, it is easy to understand and apply.  
\begin{equation}
    Jaccard(A, B) = \frac{|A \cap B|}{|A \cup B|}
    \label{eq:jaccard_sim}
\end{equation}
\autoref{fig:jaccard_similarity_rfc1918_rfc6598} shows the results for Private-Use (RFC1918) and Shared Address Space (RFC6598) Bogons. (\autoref{fig:jaccard_similarity_rest} presents the Jaccard similarity matrices for other types of Bogons). As we can see, the Jaccard similarity for Private-Use (RFC1918) addresses (\autoref{fig:jaccard_similarity_rfc1918}) across two different months fluctuates around $0.5$, slightly decreasing over time, meaning that only around half of ASNs are the same in the two measurements. The results for March 2023 exhibit noticeable differences compared to other months of the same year. During this month, the number of ASNs transiting Private-Use (RFC1918) addresses dropped almost two times. Compared to Private-Use (RFC1918), the Jaccard similarity for Shared Address Space (RFC6598) exhibits even lower values.

\begin{figure}
\centering
\begin{subfigure}[b]{0.48\linewidth}
    \resizebox{\linewidth}{!}{
        \setlength\tabcolsep{1pt}
        \input{tabulars/firstleft/jaccard_similarity_rfc1918_202301-202312.tex}
    }
    \caption{Private-Use (RFC1918)}
    \label{fig:jaccard_similarity_rfc1918_firstleft}
\end{subfigure}
\begin{subfigure}[b]{0.48\linewidth}
    \resizebox{\linewidth}{!}{
        \setlength\tabcolsep{1pt}
        \input{tabulars/left/jaccard_similarity_rfc6598_202301-202312.tex}
    }
    \caption{Shared Address Space (RFC6598)}
    \label{fig:jaccard_similarity_rfc6598_firstleft}
\end{subfigure}
\caption{\textbf{BB}: Jaccard Similarity of ASNs Transiting Bogon packets of Particular Type across Months in 2023}
\label{fig:jaccard_similarity_rfc1918_rfc6598_firstleft}
\end{figure}
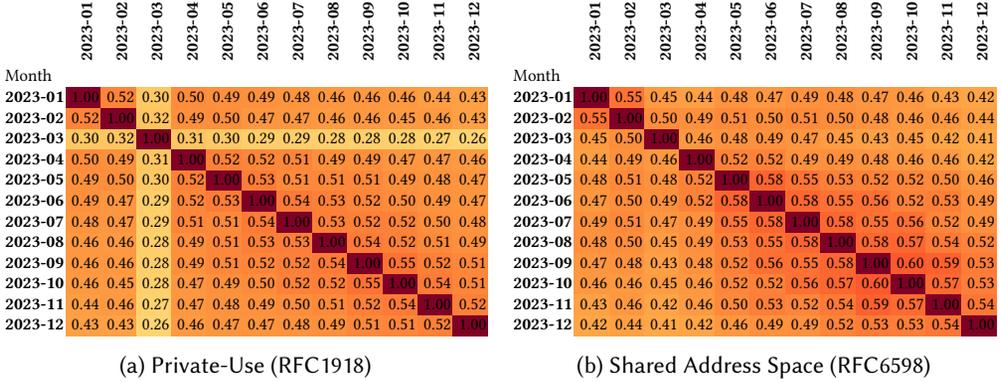

\begin{figure}
\centering
\begin{subfigure}[b]{0.48\linewidth}
    \resizebox{\linewidth}{!}{
        \setlength\tabcolsep{1pt}
        \input{tabulars/sandwich/jaccard_similarity_rfc1918_202301-202312.tex}
    }
    \caption{Private-Use (RFC1918)}
    \label{fig:jaccard_similarity_rfc1918_sandwich}
\end{subfigure}
\begin{subfigure}[b]{0.48\linewidth}
    \resizebox{\linewidth}{!}{
        \setlength\tabcolsep{1pt}
        \input{tabulars/sandwich/jaccard_similarity_rfc6598_202301-202312.tex}
    }
    \caption{Shared Address Space (RFC6598)}
    \label{fig:jaccard_similarity_rfc6598_sandwich}
\end{subfigure}
\caption{\textbf{BC}: Jaccard Similarity of ASNs Transiting Bogon packets of Particular Type across Months in 2023}
\label{fig:jaccard_similarity_rfc1918_rfc6598_sandwich}
\end{figure}
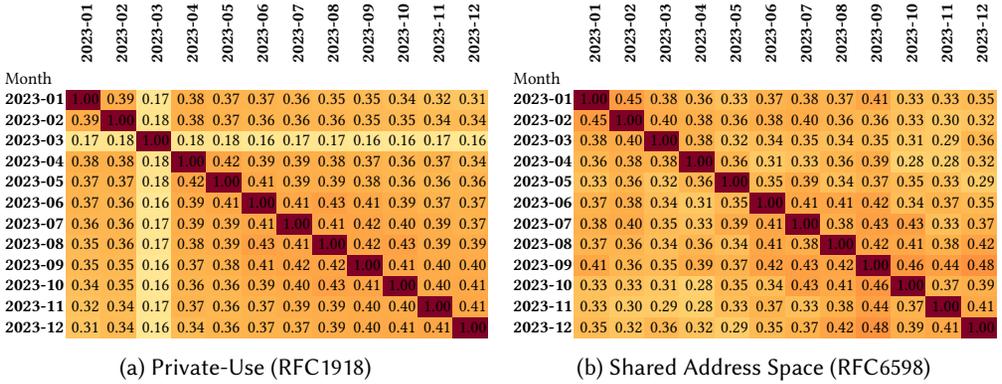

A possible explanation for this is the methodology used by CAIDA Ark to collect the traceroutes, which is optimized for finding alternate paths by randomizing destination IPs inside a /24 and performing the traceroute from different vantage points. Indeed, in order for Bogon addresses to be visible as hops, all ASNs in the path to/from the destination need to transit the Bogon packets without filtering them. The use of another vantage point may result in a different path, including an ASN performing filtering in the path. Thus, the corresponding traceroute will no longer contain Bogon addresses.

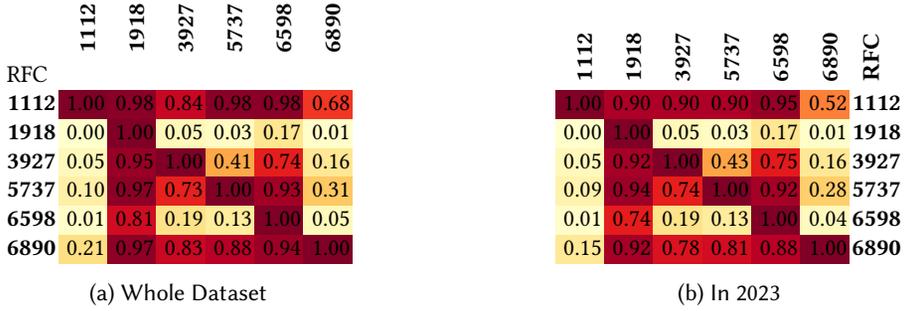
\begin{figure}
    \centering
    \begin{subfigure}[b]{0.47\linewidth}
        \centering
        \scalebox{.9}{
            \setlength\tabcolsep{1pt}
            \input{tabulars/left/rfc_containment_similarity_all.tex}
        }
        \caption{Whole Dataset}
        \label{fig:rfc_containment_similarity_all}
    \end{subfigure}
    \qquad
    \begin{subfigure}[b]{0.47\linewidth}
        \centering
        \scalebox{.9}{
            \setlength\tabcolsep{1pt}
            \input{tabulars/left/rfc_containment_similarity_2023.tex}
        }
        \caption{In 2023}
        \label{fig:rfc_containment_similarity_2023}
    \end{subfigure}
    \caption{\textbf{BA:} Containment Similarity of ASN Sets Transiting Bogons per RFC}
    \label{fig:rfc_containment_similarity}
\end{figure}

We also compared the sets of ASNs transiting different types of Bogons across each other. For this task, we chose the \emph{Containment} similarity, an asymmetrical metric that compares the extent to which one set contains elements of the other set. \autoref{eq:containment_sim} presents the formula for this metric. The rationale behind choosing this metric is as follows. The sets of ASNs transiting Bogon addresses of different types are quite unbalanced. For instance, there are \num{13883} unique ASNs transiting Private-Use (RFC1918) Bogons, and only \num{44} not filtering Former Class E (RFC1112) addresses.
\begin{equation}
    Containment(A, B) = \frac{|A \cap B|}{|A|}
    \label{eq:containment_sim}
\end{equation}
\autoref{fig:rfc_containment_similarity_all} shows the containment similarity matrix of ASNs transiting Bogon packets separated per RFCs for the whole dataset, while \autoref{fig:rfc_containment_similarity_2023} reports the numbers only for measurements done in 2023. As expected, the smaller the size of a corresponding set, the higher the probability it will be contained in a set of the larger size. However, while being the largest, the set covering Private-Use (RFC1918) addresses does not fully cover the other datasets, except the Former Class E (RFC1112). This means that it is worth considering all types of Bogons when evaluating the filtering practices of ASNs.

\subsection{Characterization of ASNs Transiting Bogons}
\label{subsec:characterization_of_asns}
In this section, we characterize the ASNs identified as transiting packets with Bogon address sources (the \textbf{BA} case) according to the methodology explained in \autoref{sec:methodology} (see Step 5). We concentrate specifically on last year's results, as the findings for the preceding period might be outdated. 

First, using the CAIDA AS2ORG~\cite{as2org} dataset, we associated each ASN transiting Bogons with the information about the Regional Internet Registry (RIR), which registered the corresponding entry. As we can see in \autoref{tab:as2org_2023_categ}, most (34.02\%) of the identified ASNs transiting packets with Bogon sources in 2023 are registered with RIPE.

\begin{table}
\centering
\caption{\textbf{BA:} RIR for ASNs Transiting Bogons in 2023}
\scalebox{.8}{
\begin{tabular}{lrr}
    \toprule
    \textbf{RIR} & \textbf{ASNs} & \textbf{\%} \\
    \midrule
    RIPE & 2,355 & 34.02\% \\
    LACNIC & 1,466 & 21.92\% \\
    APNIC & 1,376 & 19.88\% \\
    ARIN & 1,365 & 19.72\% \\
    AFRINIC & 337 & 4.87\% \\
    Not found & 23 & 0.33\% \\
    \midrule
    \textbf{Total ASNs:} & \textbf{6,922} & \textbf{100\%} \\
    \bottomrule
\end{tabular}
}
\label{tab:as2org_2023_categ}
\end{table}

Using the same CAIDA AS2ORG~\cite{as2org} dataset, we also mapped ASNs to the corresponding countries. \autoref{tab:top_10_countries_by_asns} lists the Top 10 countries by the number of unique ASNs found in 2023 as transiting Bogon packets. The USA and Brazil occupy the first two positions with huge margins correspondingly. Interestingly, these findings echo trends observed in the Spoofer dataset, where Autonomous Systems in the USA and Brazil also rank prominently in terms of the percentage of spoofable ASes~\cite{spoofer2023}. Such alignment suggests a potential correlation with suboptimal filtering practices in these regions. Further exploration and research may show the underlying factors contributing to this phenomenon.

\begin{table}[t!]
    \centering
    \caption{\textbf{BA:} Top 10 Countries by Number of ASNs Transiting Bogons}
    \scalebox{.8}{
        \input{tabulars/left/top_10_countries_by_asns.tex}
    }
    \label{tab:top_10_countries_by_asns}
\end{table}

\begin{figure}[h!]
\centering
\begin{subfigure}[b]{0.48\linewidth}
    \includegraphics[width=\linewidth]{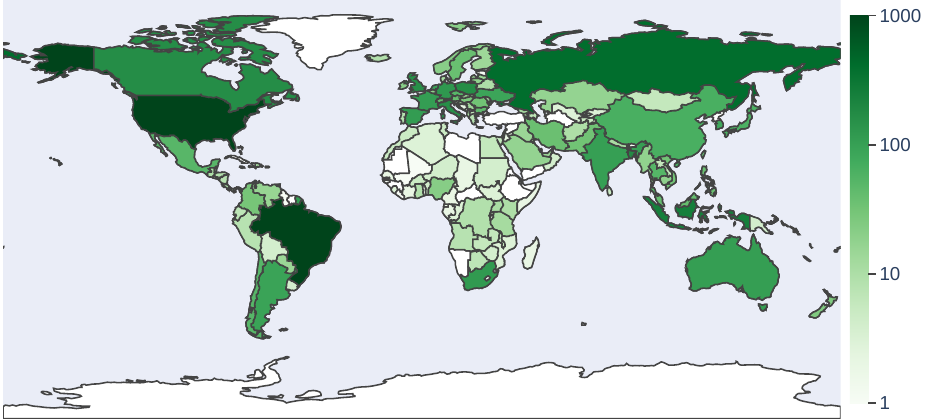}
    \caption{Private-Use (RFC1918)}
    \label{fig:asns_per_country_rfc1918}
\end{subfigure}
\quad
\begin{subfigure}[b]{0.48\linewidth}
    \includegraphics[width=\linewidth]{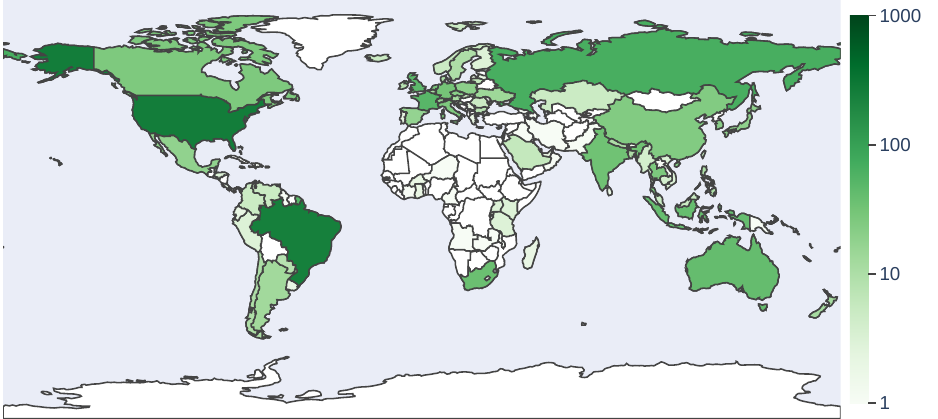}
    \caption{Shared Address Space (RFC6598)}
    \label{fig:asns_per_country_rfc6598}
\end{subfigure}
\caption{\textbf{BA:} Colormap of the Number of ASNs per Country Transiting Bogons of Particular Type}
\label{fig:asns_per_country_rfc1918_rfc6598}
\end{figure}

The other columns in \autoref{tab:top_10_countries_by_asns} break down the unique number of ASNs by the corresponding Bogon RFC types. \autoref{fig:asns_per_country_rfc1918_rfc6598} shows the colormap of the number of ASNs per country transiting (a) Private-Use (RFC1918) and (b) Shared Address Space (RFC6598) addresses (\autoref{fig:asns_per_country_rest} presents them for the rest of Bogon types). 

\begin{figure}[h!]
\centering
\begin{subfigure}[b]{0.48\linewidth}
    \includegraphics[width=\linewidth]{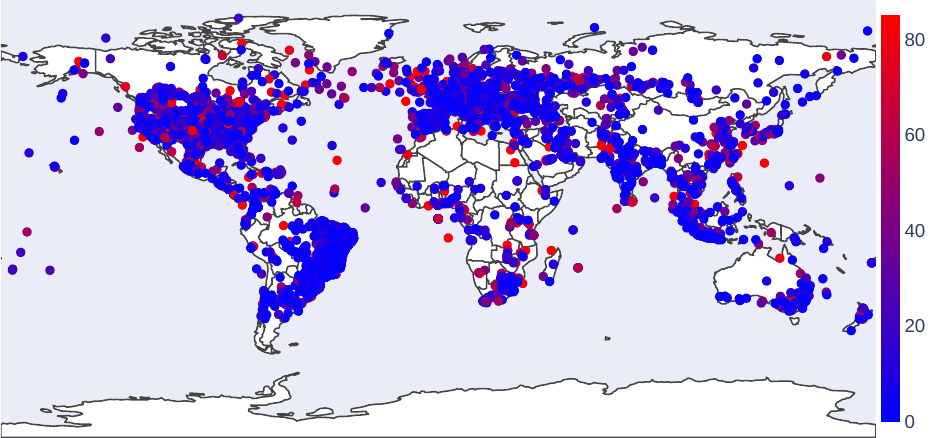}
    \caption{Private-Use (RFC1918)}
    \label{fig:occurences_map_rfc1918}
\end{subfigure}
\quad
\begin{subfigure}[b]{0.48\linewidth}
    \includegraphics[width=\linewidth]{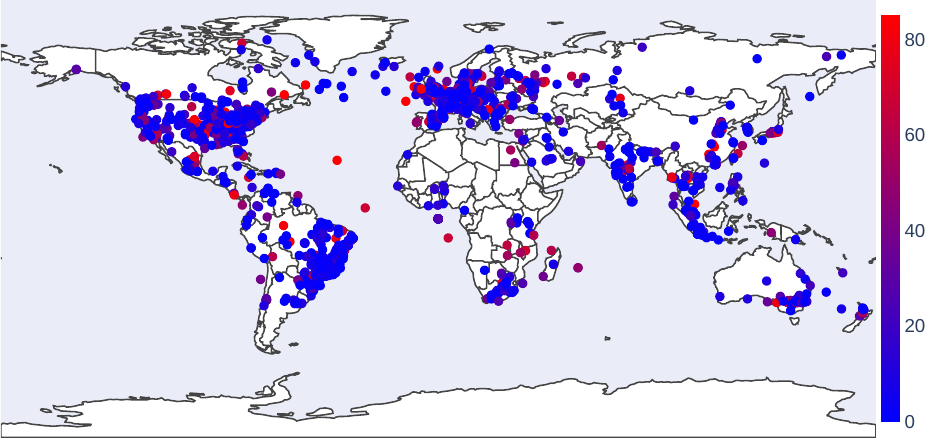}
    \caption{Shared Address Space (RFC6598)}
    \label{fig:occurences_map_rfc6598}
\end{subfigure}
\caption{\textbf{BA:} Map of ASNs Occurrences Transiting Bogons of Particular Type}
\label{fig:occurences_map_rfc1918_rfc6598}
\end{figure}

Using the ASRank~\cite{ASRank} dataset, we associated each ASN identified as transiting packets from Bogon addresses in 2023 with its geographical coordinates representing the registration address of the ASN. \autoref{fig:occurences_map_rfc1918_rfc6598} presents the scatterplots of ASNs transiting (a) Private-Use (RFC1918) and (b) Shared Address Space (RFC6598) Bogons (\autoref{fig:occurences_map_rest} presents the scatterplots for the rest of Bogon types found). The color represents the number of ASN occurrences in our measurements.

We can draw two conclusions from the figure. First, most ASNs appear in the results occasionally -- the majority of dots have a blueish color, representing rare occurrences of the corresponding ASNs across the measurements (see \autoref{subsec:asns_transiting_bogons} for detailed analysis). As we already noted, such behavior is most probably caused by frequent rotation of the vantage points.

Second, most ASes transiting Bogons are located in Europe, South and North America. This outcome comes as no surprise because a large number of organizations are also operating in these regions. We used the ASdb~\cite{ziv2021asdb} January 2024 dataset to characterize the ASNs that were identified as transiting Bogon packets in the 2023 measurements. \autoref{tab:asdb_2023_categ} reports on the results of this analysis.

\begin{table}[h]
\centering
\caption{\textbf{BA:} Categories of ASNs Identied as Transiting Bogons in 2023 according to ASdb}
\scalebox{.8}{
\begin{tabular}{lrr}
    \toprule
    \textbf{Category} & \textbf{ASNs} & \textbf{\%} \\
    \midrule
    Computer and Information Technology - Internet Service Provider (ISP) & 4,701 & 67.91\% \\
    Computer and Information Technology - (no second category found) & 471 & 6.80\% \\
    Computer and Information Technology - Hosting and Cloud Provider & 328 & 4.74\% \\
    Education and Research - Colleges, Universities, and Professional Schools & 161 & 2.33\% \\
    Other & 1,130 & 16.33\% \\
    Not found & 131 & 1.89\% \\
    \midrule
    \textbf{Total ASNs 2023} & \textbf{6,922} & \textbf{100\%} \\
    \bottomrule
\end{tabular}
}
\label{tab:asdb_2023_categ}
\end{table}

We found that the majority of ASNs transiting Bogons (79.45\%) are in the Computer and Information Technology category, with 67.91\% being Internet Service Providers (see \autoref{tab:asdb_2023_categ}). This is not an unusual finding, as it is expected that ISPs provide transit services, similar to NRENs (National Research and Education Networks) providing transit to Educational institutions.

\subsection{Results of Cross-check}
\label{subsec:spoofer_comparison}

The Spoofer dataset contains results of \num{894995} individual measurements performed during our analysis. These measurements are related to \num{8431} unique ASNs. This figure represents roughly 11\% of all active ASNs, indicating low coverage for the dataset. Moreover, it is considerably lower than the number (\num{14463}) of unique ASNs identified in this work as transiting Bogons.

While handling IP spoofing and filtering Bogon addresses entails similar but slightly different technical approaches, the correlation of these network hygiene practices provides interesting results. Out of \num{8431} unique ASNs tested by Spoofer, \num{5284} are non-spoofable (the corresponding routable spoofed packets are \texttt{blocked}), and \num{2617} are spoofable (the matching \texttt{routedspoof} packets are \texttt{received}). Out of the spoofable and not spoofable ASNs, \num{1717} have both spoofable and not spoofable results. For the rest \num{530} ASNs in the Spoofer dataset, the respective \texttt{routedspoof} packets are either \texttt{rewritten}, \texttt{unknown} or \texttt{N/A}, with the latter signifying that the measurement is for IPv6 only.  

As we describe in \autoref{sec:methodology}, for each identified ASN transiting Bogon packets, we look for the data about it in the Spoofer dataset during a period of six months before the corresponding measurement. We report the ASN as \emph{spoofable} if the data is found and the corresponding entry for \texttt{routedspoof} from this ASN is \texttt{received}. Respectively, we consider the ASN as \emph{non-spoofable} if the data is found and the matching \texttt{routedspoof} packet to this ASN is \texttt{blocked}. 

\autoref{tab:spoofer_data} reports the results. In the ``Identified Unique ASNs'', we report the number of unique ASNs identified as transiting Bogon packets of the corresponding ``Bogon Type''. The ``\# in Spoofer'' shows how many of those ASNs are also found in the Spoofer dataset. The next three columns, namely ``Only Spoofable'', ``Only Non-Spoofable'' and ``Both Spoofable Non-Spoofable'', report the numbers and percentages of the ASNs, all measurements for which are either \emph{spoofable}, \emph{non-spoofable} or \emph{both} correspondingly.

For instance, the analysis reveals that out of \num{13883} ASNs identified as transiting Private-Use (RFC1918) addresses, \num{2402} are also found in the Spoofer dataset. Of them, \num{334} (13.91\%) are spoofable in all the corresponding measurements. This means that, with a high confidence level, we can affirm the corresponding ASNs as not implementing current best practices for SAV. Interestingly, \num{1272} (52.96\%) ASNs are found only non-spoofable, while they transit Bogon packets, and \num{690} (28.73\%) ASNs are found as both spoofable and non-spoofable in different measurements. There could be multiple reasons why the corresponding ASNs are identified as non-spoofable. First, this may show that the SAV best practices are implemented only in some parts of that network. Second, the issue might have been fixed sometime after it was detected, thus not appearing in the intersection. Moreover, such a result might also be caused by issues in the Spoofer dataset that we cannot identify.

\begin{table}
    \centering
    \caption{\textbf{BA:} Comparison of ASNs Identified as Transiting Bogons with the CAIDA Spoofer Data}
    \scalebox{.8}{
        \input{tabulars/left/spoofer_data.tex}
    }
    \label{tab:spoofer_data}
\end{table}

We also cross-check our findings with the MANRS for Network Operators dataset~\cite{MANRSNetworkOperators}. This dataset contains \num{940} entries. However, the ``ASNs'' column may contain several ASNs for one network operator. Thus, the dataset describes \num{1261} unique ASNs. It is worth noting that we discovered two records with conflicting information for 6 ASNs. This inconsistency can be attributed to the dataset issues. 

As we explained in \autoref{sec:methodology} (see Step 7), although the MANRS \texttt{anti\_spoofing} data is a derivative from the Spoofer dataset, there is one core difference between them, namely the commitment of the MANRS members to filter out spoofed packets. So, as the MANRS dataset contains the snapshot of the latest measurement results (in our case, for May 2024), we compared them with the list of ASNs transiting Bogon packets in 2023. \autoref{tab:manrs_results} reports on the comparison results.

\begin{table}
    \centering
    \caption{\textbf{BA:} Comparison of ASNs Identified as Transiting Bogons in 2023 with the MANRS Data}
    \scalebox{.8}{
        \input{tabulars/left/manrs_results.tex}
    }
    \label{tab:manrs_results}
\end{table}

As we can see, among \num{6922} unique ASNs identified as transiting Bogons in 2023, the overlap with MANRS participating networks is only \num{359}. The majority of these ASNs, \num{235}, claim to be conformant with the MANRS \texttt{anti\_spoofing} action. It is possible that due to the difference in the datasets' collection time, these ASNs were non-conformant in 2023 but have fixed their networks by the last MANRS measurement. In the future, we plan to collect the data regularly and compare the results collected at a closer timeframe.

To draw some conclusions, we separately report on the results for the participants approved as a MANRS member before January 1, 2023. By the start of 2023, we assume that these participants would have been notified and implemented all best practices, including Bogon filtering. However, \num{209} ASNs marked as conforming to the MANRS \texttt{anti\_spoofing} action were still transiting packets with Bogon sources or destinations in 2023. 

%% file: tabulars/left/unique_asns_per_year.tex
\begin{tabular}{rrrrrrrr}
\toprule
\multirow{2}{*}{\textbf{\makecell[c]{Year}}} & \multicolumn{6}{c}{\textbf{RFC}} & \multirow{2}{*}{\textbf{\makecell[r]{Unique\\per Year}}} \\ \cmidrule(lr){2-7}
& \textbf{\makecell[c]{1112}} & \textbf{\makecell[c]{1918}} & \textbf{\makecell[c]{3927}} & \textbf{\makecell[c]{5737}} & \textbf{\makecell[c]{6598}} & \textbf{\makecell[c]{6890}} & \\
\midrule
\textbf{2017} & 0 & 5,624 & 180 & 96 & 609 & 8 & 5,748 \\
\textbf{2018} & 0 & 5,785 & 225 & 114 & 792 & 12 & 5,950 \\
\textbf{2019} & 0 & 6,507 & 306 & 157 & 1,078 & 12 & 6,754 \\
\textbf{2020} & 13 & 6,489 & 323 & 193 & 1,057 & 37 & 6,735 \\
\textbf{2021} & 30 & 6,129 & 354 & 196 & 1,246 & 60 & 6,485 \\
\textbf{2022} & 23 & 6,493 & 333 & 165 & 1,338 & 61 & 6,865 \\
\textbf{2023} & 21 & 6,514 & 369 & 213 & 1,469 & 74 & 6,922 \\
\midrule
\textbf{\makecell[r]{Unique\\per RFC}} & 44 & 13,883 & 747 & 416 & 2,869 & 144 & 14,463 \\
\bottomrule
\end{tabular}

%% file: tabulars/firstleft/unique_asns_per_year.tex
\begin{tabular}{rrrrrrrr}
\toprule
\multirow{2}{*}{\textbf{\makecell[c]{Year}}} & \multicolumn{6}{c}{\textbf{RFC}} & \multirow{2}{*}{\textbf{\makecell[r]{Unique\\per Year}}} \\ \cmidrule(lr){2-7}
& \textbf{\makecell[c]{1112}} & \textbf{\makecell[c]{1918}} & \textbf{\makecell[c]{3927}} & \textbf{\makecell[c]{5737}} & \textbf{\makecell[c]{6598}} & \textbf{\makecell[c]{6890}} & \\
\midrule
\textbf{2017} & 0 & 5,164 & 74 & 23 & 385 & 1 & 5,295 \\
\textbf{2018} & 0 & 5,293 & 74 & 15 & 498 & 2 & 5,464 \\
\textbf{2019} & 0 & 6,016 & 104 & 23 & 736 & 3 & 6,254 \\
\textbf{2020} & 2 & 6,014 & 140 & 31 & 764 & 3 & 6,268 \\
\textbf{2021} & 2 & 5,642 & 167 & 33 & 935 & 3 & 6,010 \\
\textbf{2022} & 2 & 6,008 & 173 & 33 & 1,027 & 5 & 6,381 \\
\textbf{2023} & 1 & 6,070 & 179 & 43 & 1,121 & 11 & 6,497 \\
\midrule
\textbf{\makecell[r]{Unique\\per RFC}} & 2 & 13,084 & 382 & 92 & 2,188 & 15 & 13,664 \\
\bottomrule
\end{tabular}

%% file: tabulars/sandwich/unique_asns_per_year.tex
\begin{tabular}{rrrrrrrr}
\toprule
\multirow{2}{*}{\textbf{\makecell[c]{Year}}} & \multicolumn{6}{c}{\textbf{RFC}} & \multirow{2}{*}{\textbf{\makecell[r]{Unique\\per Year}}} \\ \cmidrule(lr){2-7}
& \textbf{\makecell[c]{1112}} & \textbf{\makecell[c]{1918}} & \textbf{\makecell[c]{3927}} & \textbf{\makecell[c]{5737}} & \textbf{\makecell[c]{6598}} & \textbf{\makecell[c]{6890}} & \\
\midrule
\textbf{2017} & 0 & 2,807 & 21 & 4 & 133 & 1 & 2,887 \\
\textbf{2018} & 0 & 2,770 & 20 & 5 & 181 & 1 & 2,886 \\
\textbf{2019} & 0 & 3,126 & 27 & 5 & 239 & 0 & 3,264 \\
\textbf{2020} & 1 & 3,072 & 27 & 4 & 256 & 1 & 3,233 \\
\textbf{2021} & 0 & 2,861 & 27 & 6 & 312 & 1 & 3,070 \\
\textbf{2022} & 0 & 3,024 & 32 & 6 & 353 & 2 & 3,249 \\
\textbf{2023} & 0 & 3,051 & 32 & 7 & 382 & 2 & 3,296 \\
\midrule
\textbf{\makecell[r]{Unique\\per RFC}} & 1 & 7,762 & 104 & 21 & 879 & 3 & 8,146 \\
\bottomrule
\end{tabular}

%% file: tabulars/left/jaccard_similarity_rfc1918_202301-202312.tex
\begin{tabular}{lrrrrrrrrrrrr}
 & \makecell[c]{\rot{\textbf{2023-01}}} & \makecell[c]{\rot{\textbf{2023-02}}} & \makecell[c]{\rot{\textbf{2023-03}}} & \makecell[c]{\rot{\textbf{2023-04}}} & \makecell[c]{\rot{\textbf{2023-05}}} & \makecell[c]{\rot{\textbf{2023-06}}} & \makecell[c]{\rot{\textbf{2023-07}}} & \makecell[c]{\rot{\textbf{2023-08}}} & \makecell[c]{\rot{\textbf{2023-09}}} & \makecell[c]{\rot{\textbf{2023-10}}} & \makecell[c]{\rot{\textbf{2023-11}}} & \makecell[c]{\rot{\textbf{2023-12}}} \\
Month &  &  &  &  &  &  &  &  &  &  &  &  \\
\textbf{2023-01} & {\cellcolor[HTML]{800026}} \color[HTML]{000000} 1.00 & {\cellcolor[HTML]{FD7636}} \color[HTML]{000000} 0.54 & {\cellcolor[HTML]{FEB953}} \color[HTML]{000000} 0.35 & {\cellcolor[HTML]{FD8239}} \color[HTML]{000000} 0.52 & {\cellcolor[HTML]{FD883B}} \color[HTML]{000000} 0.51 & {\cellcolor[HTML]{FD883B}} \color[HTML]{000000} 0.51 & {\cellcolor[HTML]{FD8E3C}} \color[HTML]{000000} 0.50 & {\cellcolor[HTML]{FD933F}} \color[HTML]{000000} 0.48 & {\cellcolor[HTML]{FD923E}} \color[HTML]{000000} 0.48 & {\cellcolor[HTML]{FD933F}} \color[HTML]{000000} 0.48 & {\cellcolor[HTML]{FD9740}} \color[HTML]{000000} 0.47 & {\cellcolor[HTML]{FD9A42}} \color[HTML]{000000} 0.46 \\
\textbf{2023-02} & {\cellcolor[HTML]{FD7636}} \color[HTML]{000000} 0.54 & {\cellcolor[HTML]{800026}} \color[HTML]{000000} 1.00 & {\cellcolor[HTML]{FEB651}} \color[HTML]{000000} 0.36 & {\cellcolor[HTML]{FD8439}} \color[HTML]{000000} 0.52 & {\cellcolor[HTML]{FD8439}} \color[HTML]{000000} 0.52 & {\cellcolor[HTML]{FD8F3D}} \color[HTML]{000000} 0.49 & {\cellcolor[HTML]{FD903D}} \color[HTML]{000000} 0.49 & {\cellcolor[HTML]{FD933F}} \color[HTML]{000000} 0.48 & {\cellcolor[HTML]{FD923E}} \color[HTML]{000000} 0.48 & {\cellcolor[HTML]{FD9640}} \color[HTML]{000000} 0.47 & {\cellcolor[HTML]{FD953F}} \color[HTML]{000000} 0.48 & {\cellcolor[HTML]{FD9A42}} \color[HTML]{000000} 0.45 \\
\textbf{2023-03} & {\cellcolor[HTML]{FEB953}} \color[HTML]{000000} 0.35 & {\cellcolor[HTML]{FEB651}} \color[HTML]{000000} 0.36 & {\cellcolor[HTML]{800026}} \color[HTML]{000000} 1.00 & {\cellcolor[HTML]{FEB852}} \color[HTML]{000000} 0.36 & {\cellcolor[HTML]{FEBA55}} \color[HTML]{000000} 0.35 & {\cellcolor[HTML]{FEC15D}} \color[HTML]{000000} 0.33 & {\cellcolor[HTML]{FEBF5A}} \color[HTML]{000000} 0.33 & {\cellcolor[HTML]{FEC15D}} \color[HTML]{000000} 0.33 & {\cellcolor[HTML]{FEC45F}} \color[HTML]{000000} 0.32 & {\cellcolor[HTML]{FEC35E}} \color[HTML]{000000} 0.32 & {\cellcolor[HTML]{FEC561}} \color[HTML]{000000} 0.31 & {\cellcolor[HTML]{FEC965}} \color[HTML]{000000} 0.30 \\
\textbf{2023-04} & {\cellcolor[HTML]{FD8239}} \color[HTML]{000000} 0.52 & {\cellcolor[HTML]{FD8439}} \color[HTML]{000000} 0.52 & {\cellcolor[HTML]{FEB852}} \color[HTML]{000000} 0.36 & {\cellcolor[HTML]{800026}} \color[HTML]{000000} 1.00 & {\cellcolor[HTML]{FD7836}} \color[HTML]{000000} 0.54 & {\cellcolor[HTML]{FD7C37}} \color[HTML]{000000} 0.53 & {\cellcolor[HTML]{FD7E38}} \color[HTML]{000000} 0.53 & {\cellcolor[HTML]{FD8A3B}} \color[HTML]{000000} 0.50 & {\cellcolor[HTML]{FD883B}} \color[HTML]{000000} 0.51 & {\cellcolor[HTML]{FD903D}} \color[HTML]{000000} 0.49 & {\cellcolor[HTML]{FD913E}} \color[HTML]{000000} 0.49 & {\cellcolor[HTML]{FD933F}} \color[HTML]{000000} 0.48 \\
\textbf{2023-05} & {\cellcolor[HTML]{FD883B}} \color[HTML]{000000} 0.51 & {\cellcolor[HTML]{FD8439}} \color[HTML]{000000} 0.52 & {\cellcolor[HTML]{FEBA55}} \color[HTML]{000000} 0.35 & {\cellcolor[HTML]{FD7836}} \color[HTML]{000000} 0.54 & {\cellcolor[HTML]{800026}} \color[HTML]{000000} 1.00 & {\cellcolor[HTML]{FD7234}} \color[HTML]{000000} 0.55 & {\cellcolor[HTML]{FD8038}} \color[HTML]{000000} 0.53 & {\cellcolor[HTML]{FD8038}} \color[HTML]{000000} 0.52 & {\cellcolor[HTML]{FD8038}} \color[HTML]{000000} 0.52 & {\cellcolor[HTML]{FD883B}} \color[HTML]{000000} 0.51 & {\cellcolor[HTML]{FD8C3C}} \color[HTML]{000000} 0.50 & {\cellcolor[HTML]{FD8F3D}} \color[HTML]{000000} 0.49 \\
\textbf{2023-06} & {\cellcolor[HTML]{FD883B}} \color[HTML]{000000} 0.51 & {\cellcolor[HTML]{FD8F3D}} \color[HTML]{000000} 0.49 & {\cellcolor[HTML]{FEC15D}} \color[HTML]{000000} 0.33 & {\cellcolor[HTML]{FD7C37}} \color[HTML]{000000} 0.53 & {\cellcolor[HTML]{FD7234}} \color[HTML]{000000} 0.55 & {\cellcolor[HTML]{800026}} \color[HTML]{000000} 1.00 & {\cellcolor[HTML]{FD7234}} \color[HTML]{000000} 0.55 & {\cellcolor[HTML]{FD7836}} \color[HTML]{000000} 0.54 & {\cellcolor[HTML]{FD7A37}} \color[HTML]{000000} 0.54 & {\cellcolor[HTML]{FD8439}} \color[HTML]{000000} 0.52 & {\cellcolor[HTML]{FD8A3B}} \color[HTML]{000000} 0.51 & {\cellcolor[HTML]{FD903D}} \color[HTML]{000000} 0.49 \\
\textbf{2023-07} & {\cellcolor[HTML]{FD8E3C}} \color[HTML]{000000} 0.50 & {\cellcolor[HTML]{FD903D}} \color[HTML]{000000} 0.49 & {\cellcolor[HTML]{FEBF5A}} \color[HTML]{000000} 0.33 & {\cellcolor[HTML]{FD7E38}} \color[HTML]{000000} 0.53 & {\cellcolor[HTML]{FD8038}} \color[HTML]{000000} 0.53 & {\cellcolor[HTML]{FD7234}} \color[HTML]{000000} 0.55 & {\cellcolor[HTML]{800026}} \color[HTML]{000000} 1.00 & {\cellcolor[HTML]{FD7234}} \color[HTML]{000000} 0.55 & {\cellcolor[HTML]{FD7836}} \color[HTML]{000000} 0.54 & {\cellcolor[HTML]{FD7A37}} \color[HTML]{000000} 0.54 & {\cellcolor[HTML]{FD8239}} \color[HTML]{000000} 0.52 & {\cellcolor[HTML]{FD8E3C}} \color[HTML]{000000} 0.50 \\
\textbf{2023-08} & {\cellcolor[HTML]{FD933F}} \color[HTML]{000000} 0.48 & {\cellcolor[HTML]{FD933F}} \color[HTML]{000000} 0.48 & {\cellcolor[HTML]{FEC15D}} \color[HTML]{000000} 0.33 & {\cellcolor[HTML]{FD8A3B}} \color[HTML]{000000} 0.50 & {\cellcolor[HTML]{FD8038}} \color[HTML]{000000} 0.52 & {\cellcolor[HTML]{FD7836}} \color[HTML]{000000} 0.54 & {\cellcolor[HTML]{FD7234}} \color[HTML]{000000} 0.55 & {\cellcolor[HTML]{800026}} \color[HTML]{000000} 1.00 & {\cellcolor[HTML]{FD7034}} \color[HTML]{000000} 0.56 & {\cellcolor[HTML]{FD7836}} \color[HTML]{000000} 0.54 & {\cellcolor[HTML]{FD8038}} \color[HTML]{000000} 0.53 & {\cellcolor[HTML]{FD8A3B}} \color[HTML]{000000} 0.51 \\
\textbf{2023-09} & {\cellcolor[HTML]{FD923E}} \color[HTML]{000000} 0.48 & {\cellcolor[HTML]{FD923E}} \color[HTML]{000000} 0.48 & {\cellcolor[HTML]{FEC45F}} \color[HTML]{000000} 0.32 & {\cellcolor[HTML]{FD883B}} \color[HTML]{000000} 0.51 & {\cellcolor[HTML]{FD8038}} \color[HTML]{000000} 0.52 & {\cellcolor[HTML]{FD7A37}} \color[HTML]{000000} 0.54 & {\cellcolor[HTML]{FD7836}} \color[HTML]{000000} 0.54 & {\cellcolor[HTML]{FD7034}} \color[HTML]{000000} 0.56 & {\cellcolor[HTML]{800026}} \color[HTML]{000000} 1.00 & {\cellcolor[HTML]{FC6A32}} \color[HTML]{000000} 0.57 & {\cellcolor[HTML]{FD7836}} \color[HTML]{000000} 0.54 & {\cellcolor[HTML]{FD8038}} \color[HTML]{000000} 0.53 \\
\textbf{2023-10} & {\cellcolor[HTML]{FD933F}} \color[HTML]{000000} 0.48 & {\cellcolor[HTML]{FD9640}} \color[HTML]{000000} 0.47 & {\cellcolor[HTML]{FEC35E}} \color[HTML]{000000} 0.32 & {\cellcolor[HTML]{FD903D}} \color[HTML]{000000} 0.49 & {\cellcolor[HTML]{FD883B}} \color[HTML]{000000} 0.51 & {\cellcolor[HTML]{FD8439}} \color[HTML]{000000} 0.52 & {\cellcolor[HTML]{FD7A37}} \color[HTML]{000000} 0.54 & {\cellcolor[HTML]{FD7836}} \color[HTML]{000000} 0.54 & {\cellcolor[HTML]{FC6A32}} \color[HTML]{000000} 0.57 & {\cellcolor[HTML]{800026}} \color[HTML]{000000} 1.00 & {\cellcolor[HTML]{FD7234}} \color[HTML]{000000} 0.55 & {\cellcolor[HTML]{FD7E38}} \color[HTML]{000000} 0.53 \\
\textbf{2023-11} & {\cellcolor[HTML]{FD9740}} \color[HTML]{000000} 0.47 & {\cellcolor[HTML]{FD953F}} \color[HTML]{000000} 0.48 & {\cellcolor[HTML]{FEC561}} \color[HTML]{000000} 0.31 & {\cellcolor[HTML]{FD913E}} \color[HTML]{000000} 0.49 & {\cellcolor[HTML]{FD8C3C}} \color[HTML]{000000} 0.50 & {\cellcolor[HTML]{FD8A3B}} \color[HTML]{000000} 0.51 & {\cellcolor[HTML]{FD8239}} \color[HTML]{000000} 0.52 & {\cellcolor[HTML]{FD8038}} \color[HTML]{000000} 0.53 & {\cellcolor[HTML]{FD7836}} \color[HTML]{000000} 0.54 & {\cellcolor[HTML]{FD7234}} \color[HTML]{000000} 0.55 & {\cellcolor[HTML]{800026}} \color[HTML]{000000} 1.00 & {\cellcolor[HTML]{FD7636}} \color[HTML]{000000} 0.54 \\
\textbf{2023-12} & {\cellcolor[HTML]{FD9A42}} \color[HTML]{000000} 0.46 & {\cellcolor[HTML]{FD9A42}} \color[HTML]{000000} 0.45 & {\cellcolor[HTML]{FEC965}} \color[HTML]{000000} 0.30 & {\cellcolor[HTML]{FD933F}} \color[HTML]{000000} 0.48 & {\cellcolor[HTML]{FD8F3D}} \color[HTML]{000000} 0.49 & {\cellcolor[HTML]{FD903D}} \color[HTML]{000000} 0.49 & {\cellcolor[HTML]{FD8E3C}} \color[HTML]{000000} 0.50 & {\cellcolor[HTML]{FD8A3B}} \color[HTML]{000000} 0.51 & {\cellcolor[HTML]{FD8038}} \color[HTML]{000000} 0.53 & {\cellcolor[HTML]{FD7E38}} \color[HTML]{000000} 0.53 & {\cellcolor[HTML]{FD7636}} \color[HTML]{000000} 0.54 & {\cellcolor[HTML]{800026}} \color[HTML]{000000} 1.00 \\
\end{tabular}

%% file: tabulars/left/jaccard_similarity_rfc6598_202301-202312.tex
\begin{tabular}{lrrrrrrrrrrrr}
 & \makecell[c]{\rot{\textbf{2023-01}}} & \makecell[c]{\rot{\textbf{2023-02}}} & \makecell[c]{\rot{\textbf{2023-03}}} & \makecell[c]{\rot{\textbf{2023-04}}} & \makecell[c]{\rot{\textbf{2023-05}}} & \makecell[c]{\rot{\textbf{2023-06}}} & \makecell[c]{\rot{\textbf{2023-07}}} & \makecell[c]{\rot{\textbf{2023-08}}} & \makecell[c]{\rot{\textbf{2023-09}}} & \makecell[c]{\rot{\textbf{2023-10}}} & \makecell[c]{\rot{\textbf{2023-11}}} & \makecell[c]{\rot{\textbf{2023-12}}} \\
Month &  &  &  &  &  &  &  &  &  &  &  &  \\
\textbf{2023-01} & {\cellcolor[HTML]{800026}} \color[HTML]{000000} 1.00 & {\cellcolor[HTML]{FD7234}} \color[HTML]{000000} 0.55 & {\cellcolor[HTML]{FD9A42}} \color[HTML]{000000} 0.45 & {\cellcolor[HTML]{FD9F44}} \color[HTML]{000000} 0.44 & {\cellcolor[HTML]{FD933F}} \color[HTML]{000000} 0.48 & {\cellcolor[HTML]{FD9740}} \color[HTML]{000000} 0.47 & {\cellcolor[HTML]{FD913E}} \color[HTML]{000000} 0.49 & {\cellcolor[HTML]{FD913E}} \color[HTML]{000000} 0.48 & {\cellcolor[HTML]{FD9640}} \color[HTML]{000000} 0.47 & {\cellcolor[HTML]{FD9841}} \color[HTML]{000000} 0.46 & {\cellcolor[HTML]{FEA245}} \color[HTML]{000000} 0.43 & {\cellcolor[HTML]{FEA647}} \color[HTML]{000000} 0.42 \\
\textbf{2023-02} & {\cellcolor[HTML]{FD7234}} \color[HTML]{000000} 0.55 & {\cellcolor[HTML]{800026}} \color[HTML]{000000} 1.00 & {\cellcolor[HTML]{FD8E3C}} \color[HTML]{000000} 0.50 & {\cellcolor[HTML]{FD903D}} \color[HTML]{000000} 0.49 & {\cellcolor[HTML]{FD8A3B}} \color[HTML]{000000} 0.51 & {\cellcolor[HTML]{FD8E3C}} \color[HTML]{000000} 0.50 & {\cellcolor[HTML]{FD8A3B}} \color[HTML]{000000} 0.51 & {\cellcolor[HTML]{FD8E3C}} \color[HTML]{000000} 0.50 & {\cellcolor[HTML]{FD933F}} \color[HTML]{000000} 0.48 & {\cellcolor[HTML]{FD9941}} \color[HTML]{000000} 0.46 & {\cellcolor[HTML]{FD9841}} \color[HTML]{000000} 0.46 & {\cellcolor[HTML]{FEA044}} \color[HTML]{000000} 0.44 \\
\textbf{2023-03} & {\cellcolor[HTML]{FD9A42}} \color[HTML]{000000} 0.45 & {\cellcolor[HTML]{FD8E3C}} \color[HTML]{000000} 0.50 & {\cellcolor[HTML]{800026}} \color[HTML]{000000} 1.00 & {\cellcolor[HTML]{FD9841}} \color[HTML]{000000} 0.46 & {\cellcolor[HTML]{FD923E}} \color[HTML]{000000} 0.48 & {\cellcolor[HTML]{FD913E}} \color[HTML]{000000} 0.49 & {\cellcolor[HTML]{FD9640}} \color[HTML]{000000} 0.47 & {\cellcolor[HTML]{FD9C42}} \color[HTML]{000000} 0.45 & {\cellcolor[HTML]{FEA145}} \color[HTML]{000000} 0.43 & {\cellcolor[HTML]{FD9C42}} \color[HTML]{000000} 0.45 & {\cellcolor[HTML]{FEA546}} \color[HTML]{000000} 0.42 & {\cellcolor[HTML]{FEA848}} \color[HTML]{000000} 0.41 \\
\textbf{2023-04} & {\cellcolor[HTML]{FD9F44}} \color[HTML]{000000} 0.44 & {\cellcolor[HTML]{FD903D}} \color[HTML]{000000} 0.49 & {\cellcolor[HTML]{FD9841}} \color[HTML]{000000} 0.46 & {\cellcolor[HTML]{800026}} \color[HTML]{000000} 1.00 & {\cellcolor[HTML]{FD8239}} \color[HTML]{000000} 0.52 & {\cellcolor[HTML]{FD8239}} \color[HTML]{000000} 0.52 & {\cellcolor[HTML]{FD913E}} \color[HTML]{000000} 0.49 & {\cellcolor[HTML]{FD903D}} \color[HTML]{000000} 0.49 & {\cellcolor[HTML]{FD933F}} \color[HTML]{000000} 0.48 & {\cellcolor[HTML]{FD9841}} \color[HTML]{000000} 0.46 & {\cellcolor[HTML]{FD9A42}} \color[HTML]{000000} 0.46 & {\cellcolor[HTML]{FEA546}} \color[HTML]{000000} 0.42 \\
\textbf{2023-05} & {\cellcolor[HTML]{FD933F}} \color[HTML]{000000} 0.48 & {\cellcolor[HTML]{FD8A3B}} \color[HTML]{000000} 0.51 & {\cellcolor[HTML]{FD923E}} \color[HTML]{000000} 0.48 & {\cellcolor[HTML]{FD8239}} \color[HTML]{000000} 0.52 & {\cellcolor[HTML]{800026}} \color[HTML]{000000} 1.00 & {\cellcolor[HTML]{FC6430}} \color[HTML]{000000} 0.58 & {\cellcolor[HTML]{FD7435}} \color[HTML]{000000} 0.55 & {\cellcolor[HTML]{FD8038}} \color[HTML]{000000} 0.53 & {\cellcolor[HTML]{FD8239}} \color[HTML]{000000} 0.52 & {\cellcolor[HTML]{FD863A}} \color[HTML]{000000} 0.52 & {\cellcolor[HTML]{FD8A3B}} \color[HTML]{000000} 0.50 & {\cellcolor[HTML]{FD9841}} \color[HTML]{000000} 0.46 \\
\textbf{2023-06} & {\cellcolor[HTML]{FD9740}} \color[HTML]{000000} 0.47 & {\cellcolor[HTML]{FD8E3C}} \color[HTML]{000000} 0.50 & {\cellcolor[HTML]{FD913E}} \color[HTML]{000000} 0.49 & {\cellcolor[HTML]{FD8239}} \color[HTML]{000000} 0.52 & {\cellcolor[HTML]{FC6430}} \color[HTML]{000000} 0.58 & {\cellcolor[HTML]{800026}} \color[HTML]{000000} 1.00 & {\cellcolor[HTML]{FC6631}} \color[HTML]{000000} 0.58 & {\cellcolor[HTML]{FD7234}} \color[HTML]{000000} 0.55 & {\cellcolor[HTML]{FD7034}} \color[HTML]{000000} 0.56 & {\cellcolor[HTML]{FD8239}} \color[HTML]{000000} 0.52 & {\cellcolor[HTML]{FD7E38}} \color[HTML]{000000} 0.53 & {\cellcolor[HTML]{FD913E}} \color[HTML]{000000} 0.49 \\
\textbf{2023-07} & {\cellcolor[HTML]{FD913E}} \color[HTML]{000000} 0.49 & {\cellcolor[HTML]{FD8A3B}} \color[HTML]{000000} 0.51 & {\cellcolor[HTML]{FD9640}} \color[HTML]{000000} 0.47 & {\cellcolor[HTML]{FD913E}} \color[HTML]{000000} 0.49 & {\cellcolor[HTML]{FD7435}} \color[HTML]{000000} 0.55 & {\cellcolor[HTML]{FC6631}} \color[HTML]{000000} 0.58 & {\cellcolor[HTML]{800026}} \color[HTML]{000000} 1.00 & {\cellcolor[HTML]{FC6631}} \color[HTML]{000000} 0.58 & {\cellcolor[HTML]{FD7234}} \color[HTML]{000000} 0.55 & {\cellcolor[HTML]{FD7034}} \color[HTML]{000000} 0.56 & {\cellcolor[HTML]{FD863A}} \color[HTML]{000000} 0.52 & {\cellcolor[HTML]{FD913E}} \color[HTML]{000000} 0.49 \\
\textbf{2023-08} & {\cellcolor[HTML]{FD913E}} \color[HTML]{000000} 0.48 & {\cellcolor[HTML]{FD8E3C}} \color[HTML]{000000} 0.50 & {\cellcolor[HTML]{FD9C42}} \color[HTML]{000000} 0.45 & {\cellcolor[HTML]{FD903D}} \color[HTML]{000000} 0.49 & {\cellcolor[HTML]{FD8038}} \color[HTML]{000000} 0.53 & {\cellcolor[HTML]{FD7234}} \color[HTML]{000000} 0.55 & {\cellcolor[HTML]{FC6631}} \color[HTML]{000000} 0.58 & {\cellcolor[HTML]{800026}} \color[HTML]{000000} 1.00 & {\cellcolor[HTML]{FC6631}} \color[HTML]{000000} 0.58 & {\cellcolor[HTML]{FC6631}} \color[HTML]{000000} 0.57 & {\cellcolor[HTML]{FD7A37}} \color[HTML]{000000} 0.54 & {\cellcolor[HTML]{FD8439}} \color[HTML]{000000} 0.52 \\
\textbf{2023-09} & {\cellcolor[HTML]{FD9640}} \color[HTML]{000000} 0.47 & {\cellcolor[HTML]{FD933F}} \color[HTML]{000000} 0.48 & {\cellcolor[HTML]{FEA145}} \color[HTML]{000000} 0.43 & {\cellcolor[HTML]{FD933F}} \color[HTML]{000000} 0.48 & {\cellcolor[HTML]{FD8239}} \color[HTML]{000000} 0.52 & {\cellcolor[HTML]{FD7034}} \color[HTML]{000000} 0.56 & {\cellcolor[HTML]{FD7234}} \color[HTML]{000000} 0.55 & {\cellcolor[HTML]{FC6631}} \color[HTML]{000000} 0.58 & {\cellcolor[HTML]{800026}} \color[HTML]{000000} 1.00 & {\cellcolor[HTML]{FC592D}} \color[HTML]{000000} 0.60 & {\cellcolor[HTML]{FC612F}} \color[HTML]{000000} 0.59 & {\cellcolor[HTML]{FD7C37}} \color[HTML]{000000} 0.53 \\
\textbf{2023-10} & {\cellcolor[HTML]{FD9841}} \color[HTML]{000000} 0.46 & {\cellcolor[HTML]{FD9941}} \color[HTML]{000000} 0.46 & {\cellcolor[HTML]{FD9C42}} \color[HTML]{000000} 0.45 & {\cellcolor[HTML]{FD9841}} \color[HTML]{000000} 0.46 & {\cellcolor[HTML]{FD863A}} \color[HTML]{000000} 0.52 & {\cellcolor[HTML]{FD8239}} \color[HTML]{000000} 0.52 & {\cellcolor[HTML]{FD7034}} \color[HTML]{000000} 0.56 & {\cellcolor[HTML]{FC6631}} \color[HTML]{000000} 0.57 & {\cellcolor[HTML]{FC592D}} \color[HTML]{000000} 0.60 & {\cellcolor[HTML]{800026}} \color[HTML]{000000} 1.00 & {\cellcolor[HTML]{FC6A32}} \color[HTML]{000000} 0.57 & {\cellcolor[HTML]{FD7E38}} \color[HTML]{000000} 0.53 \\
\textbf{2023-11} & {\cellcolor[HTML]{FEA245}} \color[HTML]{000000} 0.43 & {\cellcolor[HTML]{FD9841}} \color[HTML]{000000} 0.46 & {\cellcolor[HTML]{FEA546}} \color[HTML]{000000} 0.42 & {\cellcolor[HTML]{FD9A42}} \color[HTML]{000000} 0.46 & {\cellcolor[HTML]{FD8A3B}} \color[HTML]{000000} 0.50 & {\cellcolor[HTML]{FD7E38}} \color[HTML]{000000} 0.53 & {\cellcolor[HTML]{FD863A}} \color[HTML]{000000} 0.52 & {\cellcolor[HTML]{FD7A37}} \color[HTML]{000000} 0.54 & {\cellcolor[HTML]{FC612F}} \color[HTML]{000000} 0.59 & {\cellcolor[HTML]{FC6A32}} \color[HTML]{000000} 0.57 & {\cellcolor[HTML]{800026}} \color[HTML]{000000} 1.00 & {\cellcolor[HTML]{FD7636}} \color[HTML]{000000} 0.54 \\
\textbf{2023-12} & {\cellcolor[HTML]{FEA647}} \color[HTML]{000000} 0.42 & {\cellcolor[HTML]{FEA044}} \color[HTML]{000000} 0.44 & {\cellcolor[HTML]{FEA848}} \color[HTML]{000000} 0.41 & {\cellcolor[HTML]{FEA546}} \color[HTML]{000000} 0.42 & {\cellcolor[HTML]{FD9841}} \color[HTML]{000000} 0.46 & {\cellcolor[HTML]{FD913E}} \color[HTML]{000000} 0.49 & {\cellcolor[HTML]{FD913E}} \color[HTML]{000000} 0.49 & {\cellcolor[HTML]{FD8439}} \color[HTML]{000000} 0.52 & {\cellcolor[HTML]{FD7C37}} \color[HTML]{000000} 0.53 & {\cellcolor[HTML]{FD7E38}} \color[HTML]{000000} 0.53 & {\cellcolor[HTML]{FD7636}} \color[HTML]{000000} 0.54 & {\cellcolor[HTML]{800026}} \color[HTML]{000000} 1.00 \\
\end{tabular}

%% file: tabulars/firstleft/jaccard_similarity_rfc1918_202301-202312.tex
\begin{tabular}{lrrrrrrrrrrrr}
 & \makecell[c]{\rot{\textbf{2023-01}}} & \makecell[c]{\rot{\textbf{2023-02}}} & \makecell[c]{\rot{\textbf{2023-03}}} & \makecell[c]{\rot{\textbf{2023-04}}} & \makecell[c]{\rot{\textbf{2023-05}}} & \makecell[c]{\rot{\textbf{2023-06}}} & \makecell[c]{\rot{\textbf{2023-07}}} & \makecell[c]{\rot{\textbf{2023-08}}} & \makecell[c]{\rot{\textbf{2023-09}}} & \makecell[c]{\rot{\textbf{2023-10}}} & \makecell[c]{\rot{\textbf{2023-11}}} & \makecell[c]{\rot{\textbf{2023-12}}} \\
Month &  &  &  &  &  &  &  &  &  &  &  &  \\
\textbf{2023-01} & {\cellcolor[HTML]{800026}} \color[HTML]{000000} 1.00 & {\cellcolor[HTML]{FD8439}} \color[HTML]{000000} 0.52 & {\cellcolor[HTML]{FEC863}} \color[HTML]{000000} 0.30 & {\cellcolor[HTML]{FD8C3C}} \color[HTML]{000000} 0.50 & {\cellcolor[HTML]{FD903D}} \color[HTML]{000000} 0.49 & {\cellcolor[HTML]{FD903D}} \color[HTML]{000000} 0.49 & {\cellcolor[HTML]{FD953F}} \color[HTML]{000000} 0.48 & {\cellcolor[HTML]{FD9841}} \color[HTML]{000000} 0.46 & {\cellcolor[HTML]{FD9841}} \color[HTML]{000000} 0.46 & {\cellcolor[HTML]{FD9941}} \color[HTML]{000000} 0.46 & {\cellcolor[HTML]{FD9E43}} \color[HTML]{000000} 0.44 & {\cellcolor[HTML]{FEA044}} \color[HTML]{000000} 0.43 \\
\textbf{2023-02} & {\cellcolor[HTML]{FD8439}} \color[HTML]{000000} 0.52 & {\cellcolor[HTML]{800026}} \color[HTML]{000000} 1.00 & {\cellcolor[HTML]{FEC561}} \color[HTML]{000000} 0.32 & {\cellcolor[HTML]{FD903D}} \color[HTML]{000000} 0.49 & {\cellcolor[HTML]{FD8E3C}} \color[HTML]{000000} 0.50 & {\cellcolor[HTML]{FD9640}} \color[HTML]{000000} 0.47 & {\cellcolor[HTML]{FD9640}} \color[HTML]{000000} 0.47 & {\cellcolor[HTML]{FD9941}} \color[HTML]{000000} 0.46 & {\cellcolor[HTML]{FD9841}} \color[HTML]{000000} 0.46 & {\cellcolor[HTML]{FD9A42}} \color[HTML]{000000} 0.45 & {\cellcolor[HTML]{FD9941}} \color[HTML]{000000} 0.46 & {\cellcolor[HTML]{FEA145}} \color[HTML]{000000} 0.43 \\
\textbf{2023-03} & {\cellcolor[HTML]{FEC863}} \color[HTML]{000000} 0.30 & {\cellcolor[HTML]{FEC561}} \color[HTML]{000000} 0.32 & {\cellcolor[HTML]{800026}} \color[HTML]{000000} 1.00 & {\cellcolor[HTML]{FEC561}} \color[HTML]{000000} 0.31 & {\cellcolor[HTML]{FEC965}} \color[HTML]{000000} 0.30 & {\cellcolor[HTML]{FECE6A}} \color[HTML]{000000} 0.29 & {\cellcolor[HTML]{FECE6A}} \color[HTML]{000000} 0.29 & {\cellcolor[HTML]{FECF6B}} \color[HTML]{000000} 0.28 & {\cellcolor[HTML]{FED06C}} \color[HTML]{000000} 0.28 & {\cellcolor[HTML]{FED06C}} \color[HTML]{000000} 0.28 & {\cellcolor[HTML]{FED36F}} \color[HTML]{000000} 0.27 & {\cellcolor[HTML]{FED572}} \color[HTML]{000000} 0.26 \\
\textbf{2023-04} & {\cellcolor[HTML]{FD8C3C}} \color[HTML]{000000} 0.50 & {\cellcolor[HTML]{FD903D}} \color[HTML]{000000} 0.49 & {\cellcolor[HTML]{FEC561}} \color[HTML]{000000} 0.31 & {\cellcolor[HTML]{800026}} \color[HTML]{000000} 1.00 & {\cellcolor[HTML]{FD8038}} \color[HTML]{000000} 0.52 & {\cellcolor[HTML]{FD863A}} \color[HTML]{000000} 0.52 & {\cellcolor[HTML]{FD863A}} \color[HTML]{000000} 0.51 & {\cellcolor[HTML]{FD903D}} \color[HTML]{000000} 0.49 & {\cellcolor[HTML]{FD903D}} \color[HTML]{000000} 0.49 & {\cellcolor[HTML]{FD9640}} \color[HTML]{000000} 0.47 & {\cellcolor[HTML]{FD9740}} \color[HTML]{000000} 0.47 & {\cellcolor[HTML]{FD9A42}} \color[HTML]{000000} 0.46 \\
\textbf{2023-05} & {\cellcolor[HTML]{FD903D}} \color[HTML]{000000} 0.49 & {\cellcolor[HTML]{FD8E3C}} \color[HTML]{000000} 0.50 & {\cellcolor[HTML]{FEC965}} \color[HTML]{000000} 0.30 & {\cellcolor[HTML]{FD8038}} \color[HTML]{000000} 0.52 & {\cellcolor[HTML]{800026}} \color[HTML]{000000} 1.00 & {\cellcolor[HTML]{FD7C37}} \color[HTML]{000000} 0.53 & {\cellcolor[HTML]{FD8A3B}} \color[HTML]{000000} 0.51 & {\cellcolor[HTML]{FD883B}} \color[HTML]{000000} 0.51 & {\cellcolor[HTML]{FD883B}} \color[HTML]{000000} 0.51 & {\cellcolor[HTML]{FD903D}} \color[HTML]{000000} 0.49 & {\cellcolor[HTML]{FD933F}} \color[HTML]{000000} 0.48 & {\cellcolor[HTML]{FD953F}} \color[HTML]{000000} 0.47 \\
\textbf{2023-06} & {\cellcolor[HTML]{FD903D}} \color[HTML]{000000} 0.49 & {\cellcolor[HTML]{FD9640}} \color[HTML]{000000} 0.47 & {\cellcolor[HTML]{FECE6A}} \color[HTML]{000000} 0.29 & {\cellcolor[HTML]{FD863A}} \color[HTML]{000000} 0.52 & {\cellcolor[HTML]{FD7C37}} \color[HTML]{000000} 0.53 & {\cellcolor[HTML]{800026}} \color[HTML]{000000} 1.00 & {\cellcolor[HTML]{FD7A37}} \color[HTML]{000000} 0.54 & {\cellcolor[HTML]{FD8038}} \color[HTML]{000000} 0.53 & {\cellcolor[HTML]{FD863A}} \color[HTML]{000000} 0.52 & {\cellcolor[HTML]{FD8C3C}} \color[HTML]{000000} 0.50 & {\cellcolor[HTML]{FD913E}} \color[HTML]{000000} 0.49 & {\cellcolor[HTML]{FD953F}} \color[HTML]{000000} 0.47 \\
\textbf{2023-07} & {\cellcolor[HTML]{FD953F}} \color[HTML]{000000} 0.48 & {\cellcolor[HTML]{FD9640}} \color[HTML]{000000} 0.47 & {\cellcolor[HTML]{FECE6A}} \color[HTML]{000000} 0.29 & {\cellcolor[HTML]{FD863A}} \color[HTML]{000000} 0.51 & {\cellcolor[HTML]{FD8A3B}} \color[HTML]{000000} 0.51 & {\cellcolor[HTML]{FD7A37}} \color[HTML]{000000} 0.54 & {\cellcolor[HTML]{800026}} \color[HTML]{000000} 1.00 & {\cellcolor[HTML]{FD7E38}} \color[HTML]{000000} 0.53 & {\cellcolor[HTML]{FD8038}} \color[HTML]{000000} 0.52 & {\cellcolor[HTML]{FD8239}} \color[HTML]{000000} 0.52 & {\cellcolor[HTML]{FD8E3C}} \color[HTML]{000000} 0.50 & {\cellcolor[HTML]{FD933F}} \color[HTML]{000000} 0.48 \\
\textbf{2023-08} & {\cellcolor[HTML]{FD9841}} \color[HTML]{000000} 0.46 & {\cellcolor[HTML]{FD9941}} \color[HTML]{000000} 0.46 & {\cellcolor[HTML]{FECF6B}} \color[HTML]{000000} 0.28 & {\cellcolor[HTML]{FD903D}} \color[HTML]{000000} 0.49 & {\cellcolor[HTML]{FD883B}} \color[HTML]{000000} 0.51 & {\cellcolor[HTML]{FD8038}} \color[HTML]{000000} 0.53 & {\cellcolor[HTML]{FD7E38}} \color[HTML]{000000} 0.53 & {\cellcolor[HTML]{800026}} \color[HTML]{000000} 1.00 & {\cellcolor[HTML]{FD7A37}} \color[HTML]{000000} 0.54 & {\cellcolor[HTML]{FD8038}} \color[HTML]{000000} 0.52 & {\cellcolor[HTML]{FD8A3B}} \color[HTML]{000000} 0.51 & {\cellcolor[HTML]{FD903D}} \color[HTML]{000000} 0.49 \\
\textbf{2023-09} & {\cellcolor[HTML]{FD9841}} \color[HTML]{000000} 0.46 & {\cellcolor[HTML]{FD9841}} \color[HTML]{000000} 0.46 & {\cellcolor[HTML]{FED06C}} \color[HTML]{000000} 0.28 & {\cellcolor[HTML]{FD903D}} \color[HTML]{000000} 0.49 & {\cellcolor[HTML]{FD883B}} \color[HTML]{000000} 0.51 & {\cellcolor[HTML]{FD863A}} \color[HTML]{000000} 0.52 & {\cellcolor[HTML]{FD8038}} \color[HTML]{000000} 0.52 & {\cellcolor[HTML]{FD7A37}} \color[HTML]{000000} 0.54 & {\cellcolor[HTML]{800026}} \color[HTML]{000000} 1.00 & {\cellcolor[HTML]{FD7234}} \color[HTML]{000000} 0.55 & {\cellcolor[HTML]{FD8038}} \color[HTML]{000000} 0.52 & {\cellcolor[HTML]{FD883B}} \color[HTML]{000000} 0.51 \\
\textbf{2023-10} & {\cellcolor[HTML]{FD9941}} \color[HTML]{000000} 0.46 & {\cellcolor[HTML]{FD9A42}} \color[HTML]{000000} 0.45 & {\cellcolor[HTML]{FED06C}} \color[HTML]{000000} 0.28 & {\cellcolor[HTML]{FD9640}} \color[HTML]{000000} 0.47 & {\cellcolor[HTML]{FD903D}} \color[HTML]{000000} 0.49 & {\cellcolor[HTML]{FD8C3C}} \color[HTML]{000000} 0.50 & {\cellcolor[HTML]{FD8239}} \color[HTML]{000000} 0.52 & {\cellcolor[HTML]{FD8038}} \color[HTML]{000000} 0.52 & {\cellcolor[HTML]{FD7234}} \color[HTML]{000000} 0.55 & {\cellcolor[HTML]{800026}} \color[HTML]{000000} 1.00 & {\cellcolor[HTML]{FD7A37}} \color[HTML]{000000} 0.54 & {\cellcolor[HTML]{FD863A}} \color[HTML]{000000} 0.51 \\
\textbf{2023-11} & {\cellcolor[HTML]{FD9E43}} \color[HTML]{000000} 0.44 & {\cellcolor[HTML]{FD9941}} \color[HTML]{000000} 0.46 & {\cellcolor[HTML]{FED36F}} \color[HTML]{000000} 0.27 & {\cellcolor[HTML]{FD9740}} \color[HTML]{000000} 0.47 & {\cellcolor[HTML]{FD933F}} \color[HTML]{000000} 0.48 & {\cellcolor[HTML]{FD913E}} \color[HTML]{000000} 0.49 & {\cellcolor[HTML]{FD8E3C}} \color[HTML]{000000} 0.50 & {\cellcolor[HTML]{FD8A3B}} \color[HTML]{000000} 0.51 & {\cellcolor[HTML]{FD8038}} \color[HTML]{000000} 0.52 & {\cellcolor[HTML]{FD7A37}} \color[HTML]{000000} 0.54 & {\cellcolor[HTML]{800026}} \color[HTML]{000000} 1.00 & {\cellcolor[HTML]{FD8038}} \color[HTML]{000000} 0.52 \\
\textbf{2023-12} & {\cellcolor[HTML]{FEA044}} \color[HTML]{000000} 0.43 & {\cellcolor[HTML]{FEA145}} \color[HTML]{000000} 0.43 & {\cellcolor[HTML]{FED572}} \color[HTML]{000000} 0.26 & {\cellcolor[HTML]{FD9A42}} \color[HTML]{000000} 0.46 & {\cellcolor[HTML]{FD953F}} \color[HTML]{000000} 0.47 & {\cellcolor[HTML]{FD953F}} \color[HTML]{000000} 0.47 & {\cellcolor[HTML]{FD933F}} \color[HTML]{000000} 0.48 & {\cellcolor[HTML]{FD903D}} \color[HTML]{000000} 0.49 & {\cellcolor[HTML]{FD883B}} \color[HTML]{000000} 0.51 & {\cellcolor[HTML]{FD863A}} \color[HTML]{000000} 0.51 & {\cellcolor[HTML]{FD8038}} \color[HTML]{000000} 0.52 & {\cellcolor[HTML]{800026}} \color[HTML]{000000} 1.00 \\
\end{tabular}

%% file: tabulars/sandwich/jaccard_similarity_rfc1918_202301-202312.tex
\begin{tabular}{lrrrrrrrrrrrr}
 & \makecell[c]{\rot{\textbf{2023-01}}} & \makecell[c]{\rot{\textbf{2023-02}}} & \makecell[c]{\rot{\textbf{2023-03}}} & \makecell[c]{\rot{\textbf{2023-04}}} & \makecell[c]{\rot{\textbf{2023-05}}} & \makecell[c]{\rot{\textbf{2023-06}}} & \makecell[c]{\rot{\textbf{2023-07}}} & \makecell[c]{\rot{\textbf{2023-08}}} & \makecell[c]{\rot{\textbf{2023-09}}} & \makecell[c]{\rot{\textbf{2023-10}}} & \makecell[c]{\rot{\textbf{2023-11}}} & \makecell[c]{\rot{\textbf{2023-12}}} \\
Month &  &  &  &  &  &  &  &  &  &  &  &  \\
\textbf{2023-01} & {\cellcolor[HTML]{800026}} \color[HTML]{000000} 1.00 & {\cellcolor[HTML]{FEAD4A}} \color[HTML]{000000} 0.39 & {\cellcolor[HTML]{FFE691}} \color[HTML]{000000} 0.17 & {\cellcolor[HTML]{FEB04B}} \color[HTML]{000000} 0.38 & {\cellcolor[HTML]{FEB54F}} \color[HTML]{000000} 0.37 & {\cellcolor[HTML]{FEB44E}} \color[HTML]{000000} 0.37 & {\cellcolor[HTML]{FEB651}} \color[HTML]{000000} 0.36 & {\cellcolor[HTML]{FEBA55}} \color[HTML]{000000} 0.35 & {\cellcolor[HTML]{FEBB56}} \color[HTML]{000000} 0.35 & {\cellcolor[HTML]{FEBE59}} \color[HTML]{000000} 0.34 & {\cellcolor[HTML]{FEC35E}} \color[HTML]{000000} 0.32 & {\cellcolor[HTML]{FEC662}} \color[HTML]{000000} 0.31 \\
\textbf{2023-02} & {\cellcolor[HTML]{FEAD4A}} \color[HTML]{000000} 0.39 & {\cellcolor[HTML]{800026}} \color[HTML]{000000} 1.00 & {\cellcolor[HTML]{FFE48C}} \color[HTML]{000000} 0.18 & {\cellcolor[HTML]{FEB04B}} \color[HTML]{000000} 0.38 & {\cellcolor[HTML]{FEB34D}} \color[HTML]{000000} 0.37 & {\cellcolor[HTML]{FEB651}} \color[HTML]{000000} 0.36 & {\cellcolor[HTML]{FEB651}} \color[HTML]{000000} 0.36 & {\cellcolor[HTML]{FEB651}} \color[HTML]{000000} 0.36 & {\cellcolor[HTML]{FEB953}} \color[HTML]{000000} 0.35 & {\cellcolor[HTML]{FEBA55}} \color[HTML]{000000} 0.35 & {\cellcolor[HTML]{FEBD57}} \color[HTML]{000000} 0.34 & {\cellcolor[HTML]{FEBB56}} \color[HTML]{000000} 0.34 \\
\textbf{2023-03} & {\cellcolor[HTML]{FFE691}} \color[HTML]{000000} 0.17 & {\cellcolor[HTML]{FFE48C}} \color[HTML]{000000} 0.18 & {\cellcolor[HTML]{800026}} \color[HTML]{000000} 1.00 & {\cellcolor[HTML]{FFE48D}} \color[HTML]{000000} 0.18 & {\cellcolor[HTML]{FFE48D}} \color[HTML]{000000} 0.18 & {\cellcolor[HTML]{FFE793}} \color[HTML]{000000} 0.16 & {\cellcolor[HTML]{FFE691}} \color[HTML]{000000} 0.17 & {\cellcolor[HTML]{FFE590}} \color[HTML]{000000} 0.17 & {\cellcolor[HTML]{FFE794}} \color[HTML]{000000} 0.16 & {\cellcolor[HTML]{FFE793}} \color[HTML]{000000} 0.16 & {\cellcolor[HTML]{FFE691}} \color[HTML]{000000} 0.17 & {\cellcolor[HTML]{FFE794}} \color[HTML]{000000} 0.16 \\
\textbf{2023-04} & {\cellcolor[HTML]{FEB04B}} \color[HTML]{000000} 0.38 & {\cellcolor[HTML]{FEB04B}} \color[HTML]{000000} 0.38 & {\cellcolor[HTML]{FFE48D}} \color[HTML]{000000} 0.18 & {\cellcolor[HTML]{800026}} \color[HTML]{000000} 1.00 & {\cellcolor[HTML]{FEA647}} \color[HTML]{000000} 0.42 & {\cellcolor[HTML]{FEAD4A}} \color[HTML]{000000} 0.39 & {\cellcolor[HTML]{FEAE4A}} \color[HTML]{000000} 0.39 & {\cellcolor[HTML]{FEB04B}} \color[HTML]{000000} 0.38 & {\cellcolor[HTML]{FEB44E}} \color[HTML]{000000} 0.37 & {\cellcolor[HTML]{FEB651}} \color[HTML]{000000} 0.36 & {\cellcolor[HTML]{FEB44E}} \color[HTML]{000000} 0.37 & {\cellcolor[HTML]{FEBB56}} \color[HTML]{000000} 0.34 \\
\textbf{2023-05} & {\cellcolor[HTML]{FEB54F}} \color[HTML]{000000} 0.37 & {\cellcolor[HTML]{FEB34D}} \color[HTML]{000000} 0.37 & {\cellcolor[HTML]{FFE48D}} \color[HTML]{000000} 0.18 & {\cellcolor[HTML]{FEA647}} \color[HTML]{000000} 0.42 & {\cellcolor[HTML]{800026}} \color[HTML]{000000} 1.00 & {\cellcolor[HTML]{FEA848}} \color[HTML]{000000} 0.41 & {\cellcolor[HTML]{FEAE4A}} \color[HTML]{000000} 0.39 & {\cellcolor[HTML]{FEAE4A}} \color[HTML]{000000} 0.39 & {\cellcolor[HTML]{FEB24C}} \color[HTML]{000000} 0.38 & {\cellcolor[HTML]{FEB651}} \color[HTML]{000000} 0.36 & {\cellcolor[HTML]{FEB852}} \color[HTML]{000000} 0.36 & {\cellcolor[HTML]{FEB651}} \color[HTML]{000000} 0.36 \\
\textbf{2023-06} & {\cellcolor[HTML]{FEB44E}} \color[HTML]{000000} 0.37 & {\cellcolor[HTML]{FEB651}} \color[HTML]{000000} 0.36 & {\cellcolor[HTML]{FFE793}} \color[HTML]{000000} 0.16 & {\cellcolor[HTML]{FEAD4A}} \color[HTML]{000000} 0.39 & {\cellcolor[HTML]{FEA848}} \color[HTML]{000000} 0.41 & {\cellcolor[HTML]{800026}} \color[HTML]{000000} 1.00 & {\cellcolor[HTML]{FEA848}} \color[HTML]{000000} 0.41 & {\cellcolor[HTML]{FEA145}} \color[HTML]{000000} 0.43 & {\cellcolor[HTML]{FEA948}} \color[HTML]{000000} 0.41 & {\cellcolor[HTML]{FEAE4A}} \color[HTML]{000000} 0.39 & {\cellcolor[HTML]{FEB44E}} \color[HTML]{000000} 0.37 & {\cellcolor[HTML]{FEB44E}} \color[HTML]{000000} 0.37 \\
\textbf{2023-07} & {\cellcolor[HTML]{FEB651}} \color[HTML]{000000} 0.36 & {\cellcolor[HTML]{FEB651}} \color[HTML]{000000} 0.36 & {\cellcolor[HTML]{FFE691}} \color[HTML]{000000} 0.17 & {\cellcolor[HTML]{FEAE4A}} \color[HTML]{000000} 0.39 & {\cellcolor[HTML]{FEAE4A}} \color[HTML]{000000} 0.39 & {\cellcolor[HTML]{FEA848}} \color[HTML]{000000} 0.41 & {\cellcolor[HTML]{800026}} \color[HTML]{000000} 1.00 & {\cellcolor[HTML]{FEA647}} \color[HTML]{000000} 0.41 & {\cellcolor[HTML]{FEA546}} \color[HTML]{000000} 0.42 & {\cellcolor[HTML]{FEAB49}} \color[HTML]{000000} 0.40 & {\cellcolor[HTML]{FEAE4A}} \color[HTML]{000000} 0.39 & {\cellcolor[HTML]{FEB34D}} \color[HTML]{000000} 0.37 \\
\textbf{2023-08} & {\cellcolor[HTML]{FEBA55}} \color[HTML]{000000} 0.35 & {\cellcolor[HTML]{FEB651}} \color[HTML]{000000} 0.36 & {\cellcolor[HTML]{FFE590}} \color[HTML]{000000} 0.17 & {\cellcolor[HTML]{FEB04B}} \color[HTML]{000000} 0.38 & {\cellcolor[HTML]{FEAE4A}} \color[HTML]{000000} 0.39 & {\cellcolor[HTML]{FEA145}} \color[HTML]{000000} 0.43 & {\cellcolor[HTML]{FEA647}} \color[HTML]{000000} 0.41 & {\cellcolor[HTML]{800026}} \color[HTML]{000000} 1.00 & {\cellcolor[HTML]{FEA546}} \color[HTML]{000000} 0.42 & {\cellcolor[HTML]{FEA245}} \color[HTML]{000000} 0.43 & {\cellcolor[HTML]{FEAD4A}} \color[HTML]{000000} 0.39 & {\cellcolor[HTML]{FEAD4A}} \color[HTML]{000000} 0.39 \\
\textbf{2023-09} & {\cellcolor[HTML]{FEBB56}} \color[HTML]{000000} 0.35 & {\cellcolor[HTML]{FEB953}} \color[HTML]{000000} 0.35 & {\cellcolor[HTML]{FFE794}} \color[HTML]{000000} 0.16 & {\cellcolor[HTML]{FEB44E}} \color[HTML]{000000} 0.37 & {\cellcolor[HTML]{FEB24C}} \color[HTML]{000000} 0.38 & {\cellcolor[HTML]{FEA948}} \color[HTML]{000000} 0.41 & {\cellcolor[HTML]{FEA546}} \color[HTML]{000000} 0.42 & {\cellcolor[HTML]{FEA546}} \color[HTML]{000000} 0.42 & {\cellcolor[HTML]{800026}} \color[HTML]{000000} 1.00 & {\cellcolor[HTML]{FEA647}} \color[HTML]{000000} 0.41 & {\cellcolor[HTML]{FEAB49}} \color[HTML]{000000} 0.40 & {\cellcolor[HTML]{FEAC49}} \color[HTML]{000000} 0.40 \\
\textbf{2023-10} & {\cellcolor[HTML]{FEBE59}} \color[HTML]{000000} 0.34 & {\cellcolor[HTML]{FEBA55}} \color[HTML]{000000} 0.35 & {\cellcolor[HTML]{FFE793}} \color[HTML]{000000} 0.16 & {\cellcolor[HTML]{FEB651}} \color[HTML]{000000} 0.36 & {\cellcolor[HTML]{FEB651}} \color[HTML]{000000} 0.36 & {\cellcolor[HTML]{FEAE4A}} \color[HTML]{000000} 0.39 & {\cellcolor[HTML]{FEAB49}} \color[HTML]{000000} 0.40 & {\cellcolor[HTML]{FEA245}} \color[HTML]{000000} 0.43 & {\cellcolor[HTML]{FEA647}} \color[HTML]{000000} 0.41 & {\cellcolor[HTML]{800026}} \color[HTML]{000000} 1.00 & {\cellcolor[HTML]{FEAB49}} \color[HTML]{000000} 0.40 & {\cellcolor[HTML]{FEA948}} \color[HTML]{000000} 0.41 \\
\textbf{2023-11} & {\cellcolor[HTML]{FEC35E}} \color[HTML]{000000} 0.32 & {\cellcolor[HTML]{FEBD57}} \color[HTML]{000000} 0.34 & {\cellcolor[HTML]{FFE691}} \color[HTML]{000000} 0.17 & {\cellcolor[HTML]{FEB44E}} \color[HTML]{000000} 0.37 & {\cellcolor[HTML]{FEB852}} \color[HTML]{000000} 0.36 & {\cellcolor[HTML]{FEB44E}} \color[HTML]{000000} 0.37 & {\cellcolor[HTML]{FEAE4A}} \color[HTML]{000000} 0.39 & {\cellcolor[HTML]{FEAD4A}} \color[HTML]{000000} 0.39 & {\cellcolor[HTML]{FEAB49}} \color[HTML]{000000} 0.40 & {\cellcolor[HTML]{FEAB49}} \color[HTML]{000000} 0.40 & {\cellcolor[HTML]{800026}} \color[HTML]{000000} 1.00 & {\cellcolor[HTML]{FEA747}} \color[HTML]{000000} 0.41 \\
\textbf{2023-12} & {\cellcolor[HTML]{FEC662}} \color[HTML]{000000} 0.31 & {\cellcolor[HTML]{FEBB56}} \color[HTML]{000000} 0.34 & {\cellcolor[HTML]{FFE794}} \color[HTML]{000000} 0.16 & {\cellcolor[HTML]{FEBB56}} \color[HTML]{000000} 0.34 & {\cellcolor[HTML]{FEB651}} \color[HTML]{000000} 0.36 & {\cellcolor[HTML]{FEB44E}} \color[HTML]{000000} 0.37 & {\cellcolor[HTML]{FEB34D}} \color[HTML]{000000} 0.37 & {\cellcolor[HTML]{FEAD4A}} \color[HTML]{000000} 0.39 & {\cellcolor[HTML]{FEAC49}} \color[HTML]{000000} 0.40 & {\cellcolor[HTML]{FEA948}} \color[HTML]{000000} 0.41 & {\cellcolor[HTML]{FEA747}} \color[HTML]{000000} 0.41 & {\cellcolor[HTML]{800026}} \color[HTML]{000000} 1.00 \\
\end{tabular}

%% file: tabulars/sandwich/jaccard_similarity_rfc6598_202301-202312.tex
\begin{tabular}{lrrrrrrrrrrrr}
 & \makecell[c]{\rot{\textbf{2023-01}}} & \makecell[c]{\rot{\textbf{2023-02}}} & \makecell[c]{\rot{\textbf{2023-03}}} & \makecell[c]{\rot{\textbf{2023-04}}} & \makecell[c]{\rot{\textbf{2023-05}}} & \makecell[c]{\rot{\textbf{2023-06}}} & \makecell[c]{\rot{\textbf{2023-07}}} & \makecell[c]{\rot{\textbf{2023-08}}} & \makecell[c]{\rot{\textbf{2023-09}}} & \makecell[c]{\rot{\textbf{2023-10}}} & \makecell[c]{\rot{\textbf{2023-11}}} & \makecell[c]{\rot{\textbf{2023-12}}} \\
Month &  &  &  &  &  &  &  &  &  &  &  &  \\
\textbf{2023-01} & {\cellcolor[HTML]{800026}} \color[HTML]{000000} 1.00 & {\cellcolor[HTML]{FD9C42}} \color[HTML]{000000} 0.45 & {\cellcolor[HTML]{FEAF4B}} \color[HTML]{000000} 0.38 & {\cellcolor[HTML]{FEB852}} \color[HTML]{000000} 0.36 & {\cellcolor[HTML]{FEC15D}} \color[HTML]{000000} 0.33 & {\cellcolor[HTML]{FEB44E}} \color[HTML]{000000} 0.37 & {\cellcolor[HTML]{FEB04B}} \color[HTML]{000000} 0.38 & {\cellcolor[HTML]{FEB54F}} \color[HTML]{000000} 0.37 & {\cellcolor[HTML]{FEA848}} \color[HTML]{000000} 0.41 & {\cellcolor[HTML]{FEC05B}} \color[HTML]{000000} 0.33 & {\cellcolor[HTML]{FEC05B}} \color[HTML]{000000} 0.33 & {\cellcolor[HTML]{FEBB56}} \color[HTML]{000000} 0.35 \\
\textbf{2023-02} & {\cellcolor[HTML]{FD9C42}} \color[HTML]{000000} 0.45 & {\cellcolor[HTML]{800026}} \color[HTML]{000000} 1.00 & {\cellcolor[HTML]{FEAB49}} \color[HTML]{000000} 0.40 & {\cellcolor[HTML]{FEB04B}} \color[HTML]{000000} 0.38 & {\cellcolor[HTML]{FEB54F}} \color[HTML]{000000} 0.36 & {\cellcolor[HTML]{FEAF4B}} \color[HTML]{000000} 0.38 & {\cellcolor[HTML]{FEAC49}} \color[HTML]{000000} 0.40 & {\cellcolor[HTML]{FEB54F}} \color[HTML]{000000} 0.36 & {\cellcolor[HTML]{FEB852}} \color[HTML]{000000} 0.36 & {\cellcolor[HTML]{FEC15D}} \color[HTML]{000000} 0.33 & {\cellcolor[HTML]{FECA66}} \color[HTML]{000000} 0.30 & {\cellcolor[HTML]{FEC45F}} \color[HTML]{000000} 0.32 \\
\textbf{2023-03} & {\cellcolor[HTML]{FEAF4B}} \color[HTML]{000000} 0.38 & {\cellcolor[HTML]{FEAB49}} \color[HTML]{000000} 0.40 & {\cellcolor[HTML]{800026}} \color[HTML]{000000} 1.00 & {\cellcolor[HTML]{FEB24C}} \color[HTML]{000000} 0.38 & {\cellcolor[HTML]{FEC45F}} \color[HTML]{000000} 0.32 & {\cellcolor[HTML]{FEBE59}} \color[HTML]{000000} 0.34 & {\cellcolor[HTML]{FEBA55}} \color[HTML]{000000} 0.35 & {\cellcolor[HTML]{FEBF5A}} \color[HTML]{000000} 0.34 & {\cellcolor[HTML]{FEBB56}} \color[HTML]{000000} 0.35 & {\cellcolor[HTML]{FEC863}} \color[HTML]{000000} 0.31 & {\cellcolor[HTML]{FECB67}} \color[HTML]{000000} 0.29 & {\cellcolor[HTML]{FEB852}} \color[HTML]{000000} 0.36 \\
\textbf{2023-04} & {\cellcolor[HTML]{FEB852}} \color[HTML]{000000} 0.36 & {\cellcolor[HTML]{FEB04B}} \color[HTML]{000000} 0.38 & {\cellcolor[HTML]{FEB24C}} \color[HTML]{000000} 0.38 & {\cellcolor[HTML]{800026}} \color[HTML]{000000} 1.00 & {\cellcolor[HTML]{FEB651}} \color[HTML]{000000} 0.36 & {\cellcolor[HTML]{FEC863}} \color[HTML]{000000} 0.31 & {\cellcolor[HTML]{FEC15D}} \color[HTML]{000000} 0.33 & {\cellcolor[HTML]{FEB651}} \color[HTML]{000000} 0.36 & {\cellcolor[HTML]{FEAE4A}} \color[HTML]{000000} 0.39 & {\cellcolor[HTML]{FED16E}} \color[HTML]{000000} 0.28 & {\cellcolor[HTML]{FED06C}} \color[HTML]{000000} 0.28 & {\cellcolor[HTML]{FEC35E}} \color[HTML]{000000} 0.32 \\
\textbf{2023-05} & {\cellcolor[HTML]{FEC15D}} \color[HTML]{000000} 0.33 & {\cellcolor[HTML]{FEB54F}} \color[HTML]{000000} 0.36 & {\cellcolor[HTML]{FEC45F}} \color[HTML]{000000} 0.32 & {\cellcolor[HTML]{FEB651}} \color[HTML]{000000} 0.36 & {\cellcolor[HTML]{800026}} \color[HTML]{000000} 1.00 & {\cellcolor[HTML]{FEB953}} \color[HTML]{000000} 0.35 & {\cellcolor[HTML]{FEAD4A}} \color[HTML]{000000} 0.39 & {\cellcolor[HTML]{FEBE59}} \color[HTML]{000000} 0.34 & {\cellcolor[HTML]{FEB44E}} \color[HTML]{000000} 0.37 & {\cellcolor[HTML]{FEBA55}} \color[HTML]{000000} 0.35 & {\cellcolor[HTML]{FEBF5A}} \color[HTML]{000000} 0.33 & {\cellcolor[HTML]{FECC68}} \color[HTML]{000000} 0.29 \\
\textbf{2023-06} & {\cellcolor[HTML]{FEB44E}} \color[HTML]{000000} 0.37 & {\cellcolor[HTML]{FEAF4B}} \color[HTML]{000000} 0.38 & {\cellcolor[HTML]{FEBE59}} \color[HTML]{000000} 0.34 & {\cellcolor[HTML]{FEC863}} \color[HTML]{000000} 0.31 & {\cellcolor[HTML]{FEB953}} \color[HTML]{000000} 0.35 & {\cellcolor[HTML]{800026}} \color[HTML]{000000} 1.00 & {\cellcolor[HTML]{FEA747}} \color[HTML]{000000} 0.41 & {\cellcolor[HTML]{FEA747}} \color[HTML]{000000} 0.41 & {\cellcolor[HTML]{FEA446}} \color[HTML]{000000} 0.42 & {\cellcolor[HTML]{FEBD57}} \color[HTML]{000000} 0.34 & {\cellcolor[HTML]{FEB54F}} \color[HTML]{000000} 0.37 & {\cellcolor[HTML]{FEBA55}} \color[HTML]{000000} 0.35 \\
\textbf{2023-07} & {\cellcolor[HTML]{FEB04B}} \color[HTML]{000000} 0.38 & {\cellcolor[HTML]{FEAC49}} \color[HTML]{000000} 0.40 & {\cellcolor[HTML]{FEBA55}} \color[HTML]{000000} 0.35 & {\cellcolor[HTML]{FEC15D}} \color[HTML]{000000} 0.33 & {\cellcolor[HTML]{FEAD4A}} \color[HTML]{000000} 0.39 & {\cellcolor[HTML]{FEA747}} \color[HTML]{000000} 0.41 & {\cellcolor[HTML]{800026}} \color[HTML]{000000} 1.00 & {\cellcolor[HTML]{FEAF4B}} \color[HTML]{000000} 0.38 & {\cellcolor[HTML]{FEA044}} \color[HTML]{000000} 0.43 & {\cellcolor[HTML]{FEA145}} \color[HTML]{000000} 0.43 & {\cellcolor[HTML]{FEBF5A}} \color[HTML]{000000} 0.33 & {\cellcolor[HTML]{FEB34D}} \color[HTML]{000000} 0.37 \\
\textbf{2023-08} & {\cellcolor[HTML]{FEB54F}} \color[HTML]{000000} 0.37 & {\cellcolor[HTML]{FEB54F}} \color[HTML]{000000} 0.36 & {\cellcolor[HTML]{FEBF5A}} \color[HTML]{000000} 0.34 & {\cellcolor[HTML]{FEB651}} \color[HTML]{000000} 0.36 & {\cellcolor[HTML]{FEBE59}} \color[HTML]{000000} 0.34 & {\cellcolor[HTML]{FEA747}} \color[HTML]{000000} 0.41 & {\cellcolor[HTML]{FEAF4B}} \color[HTML]{000000} 0.38 & {\cellcolor[HTML]{800026}} \color[HTML]{000000} 1.00 & {\cellcolor[HTML]{FEA546}} \color[HTML]{000000} 0.42 & {\cellcolor[HTML]{FEA848}} \color[HTML]{000000} 0.41 & {\cellcolor[HTML]{FEB24C}} \color[HTML]{000000} 0.38 & {\cellcolor[HTML]{FEA446}} \color[HTML]{000000} 0.42 \\
\textbf{2023-09} & {\cellcolor[HTML]{FEA848}} \color[HTML]{000000} 0.41 & {\cellcolor[HTML]{FEB852}} \color[HTML]{000000} 0.36 & {\cellcolor[HTML]{FEBB56}} \color[HTML]{000000} 0.35 & {\cellcolor[HTML]{FEAE4A}} \color[HTML]{000000} 0.39 & {\cellcolor[HTML]{FEB44E}} \color[HTML]{000000} 0.37 & {\cellcolor[HTML]{FEA446}} \color[HTML]{000000} 0.42 & {\cellcolor[HTML]{FEA044}} \color[HTML]{000000} 0.43 & {\cellcolor[HTML]{FEA546}} \color[HTML]{000000} 0.42 & {\cellcolor[HTML]{800026}} \color[HTML]{000000} 1.00 & {\cellcolor[HTML]{FD9841}} \color[HTML]{000000} 0.46 & {\cellcolor[HTML]{FEA044}} \color[HTML]{000000} 0.44 & {\cellcolor[HTML]{FD933F}} \color[HTML]{000000} 0.48 \\
\textbf{2023-10} & {\cellcolor[HTML]{FEC05B}} \color[HTML]{000000} 0.33 & {\cellcolor[HTML]{FEC15D}} \color[HTML]{000000} 0.33 & {\cellcolor[HTML]{FEC863}} \color[HTML]{000000} 0.31 & {\cellcolor[HTML]{FED16E}} \color[HTML]{000000} 0.28 & {\cellcolor[HTML]{FEBA55}} \color[HTML]{000000} 0.35 & {\cellcolor[HTML]{FEBD57}} \color[HTML]{000000} 0.34 & {\cellcolor[HTML]{FEA145}} \color[HTML]{000000} 0.43 & {\cellcolor[HTML]{FEA848}} \color[HTML]{000000} 0.41 & {\cellcolor[HTML]{FD9841}} \color[HTML]{000000} 0.46 & {\cellcolor[HTML]{800026}} \color[HTML]{000000} 1.00 & {\cellcolor[HTML]{FEB44E}} \color[HTML]{000000} 0.37 & {\cellcolor[HTML]{FEAF4B}} \color[HTML]{000000} 0.39 \\
\textbf{2023-11} & {\cellcolor[HTML]{FEC05B}} \color[HTML]{000000} 0.33 & {\cellcolor[HTML]{FECA66}} \color[HTML]{000000} 0.30 & {\cellcolor[HTML]{FECB67}} \color[HTML]{000000} 0.29 & {\cellcolor[HTML]{FED06C}} \color[HTML]{000000} 0.28 & {\cellcolor[HTML]{FEBF5A}} \color[HTML]{000000} 0.33 & {\cellcolor[HTML]{FEB54F}} \color[HTML]{000000} 0.37 & {\cellcolor[HTML]{FEBF5A}} \color[HTML]{000000} 0.33 & {\cellcolor[HTML]{FEB24C}} \color[HTML]{000000} 0.38 & {\cellcolor[HTML]{FEA044}} \color[HTML]{000000} 0.44 & {\cellcolor[HTML]{FEB44E}} \color[HTML]{000000} 0.37 & {\cellcolor[HTML]{800026}} \color[HTML]{000000} 1.00 & {\cellcolor[HTML]{FEA948}} \color[HTML]{000000} 0.41 \\
\textbf{2023-12} & {\cellcolor[HTML]{FEBB56}} \color[HTML]{000000} 0.35 & {\cellcolor[HTML]{FEC45F}} \color[HTML]{000000} 0.32 & {\cellcolor[HTML]{FEB852}} \color[HTML]{000000} 0.36 & {\cellcolor[HTML]{FEC35E}} \color[HTML]{000000} 0.32 & {\cellcolor[HTML]{FECC68}} \color[HTML]{000000} 0.29 & {\cellcolor[HTML]{FEBA55}} \color[HTML]{000000} 0.35 & {\cellcolor[HTML]{FEB34D}} \color[HTML]{000000} 0.37 & {\cellcolor[HTML]{FEA446}} \color[HTML]{000000} 0.42 & {\cellcolor[HTML]{FD933F}} \color[HTML]{000000} 0.48 & {\cellcolor[HTML]{FEAF4B}} \color[HTML]{000000} 0.39 & {\cellcolor[HTML]{FEA948}} \color[HTML]{000000} 0.41 & {\cellcolor[HTML]{800026}} \color[HTML]{000000} 1.00 \\
\end{tabular}

%% file: tabulars/left/rfc_containment_similarity_all.tex
\begin{tabular}{lrrrrrr}
 & \makecell[c]{\rot{\textbf{1112}}} & \makecell[c]{\rot{\textbf{1918}}} & \makecell[c]{\rot{\textbf{3927}}} & \makecell[c]{\rot{\textbf{5737}}} & \makecell[c]{\rot{\textbf{6598}}} & \makecell[c]{\rot{\textbf{6890}}} \\
RFC &  &  &  &  &  &  \\
\textbf{1112} & {\cellcolor[HTML]{800026}} \color[HTML]{000000} 1.00 & {\cellcolor[HTML]{8A0026}} \color[HTML]{000000} 0.98 & {\cellcolor[HTML]{C70723}} \color[HTML]{000000} 0.84 & {\cellcolor[HTML]{8A0026}} \color[HTML]{000000} 0.98 & {\cellcolor[HTML]{8A0026}} \color[HTML]{000000} 0.98 & {\cellcolor[HTML]{F13624}} \color[HTML]{000000} 0.68 \\
\textbf{1918} & {\cellcolor[HTML]{FFFFCC}} \color[HTML]{000000} 0.00 & {\cellcolor[HTML]{800026}} \color[HTML]{000000} 1.00 & {\cellcolor[HTML]{FFF8BA}} \color[HTML]{000000} 0.05 & {\cellcolor[HTML]{FFFBC2}} \color[HTML]{000000} 0.03 & {\cellcolor[HTML]{FFE691}} \color[HTML]{000000} 0.17 & {\cellcolor[HTML]{FFFEC9}} \color[HTML]{000000} 0.01 \\
\textbf{3927} & {\cellcolor[HTML]{FFF8BB}} \color[HTML]{000000} 0.05 & {\cellcolor[HTML]{990026}} \color[HTML]{000000} 0.95 & {\cellcolor[HTML]{800026}} \color[HTML]{000000} 1.00 & {\cellcolor[HTML]{FEA848}} \color[HTML]{000000} 0.41 & {\cellcolor[HTML]{E51E1D}} \color[HTML]{000000} 0.74 & {\cellcolor[HTML]{FFE794}} \color[HTML]{000000} 0.16 \\
\textbf{5737} & {\cellcolor[HTML]{FFF0A8}} \color[HTML]{000000} 0.10 & {\cellcolor[HTML]{8D0026}} \color[HTML]{000000} 0.97 & {\cellcolor[HTML]{E6211E}} \color[HTML]{000000} 0.73 & {\cellcolor[HTML]{800026}} \color[HTML]{000000} 1.00 & {\cellcolor[HTML]{A40026}} \color[HTML]{000000} 0.93 & {\cellcolor[HTML]{FEC863}} \color[HTML]{000000} 0.31 \\
\textbf{6598} & {\cellcolor[HTML]{FFFDC8}} \color[HTML]{000000} 0.01 & {\cellcolor[HTML]{CF0C21}} \color[HTML]{000000} 0.81 & {\cellcolor[HTML]{FEE289}} \color[HTML]{000000} 0.19 & {\cellcolor[HTML]{FFEC9D}} \color[HTML]{000000} 0.13 & {\cellcolor[HTML]{800026}} \color[HTML]{000000} 1.00 & {\cellcolor[HTML]{FFF8BB}} \color[HTML]{000000} 0.05 \\
\textbf{6890} & {\cellcolor[HTML]{FEE084}} \color[HTML]{000000} 0.21 & {\cellcolor[HTML]{8F0026}} \color[HTML]{000000} 0.97 & {\cellcolor[HTML]{C90823}} \color[HTML]{000000} 0.83 & {\cellcolor[HTML]{B90026}} \color[HTML]{000000} 0.88 & {\cellcolor[HTML]{9D0026}} \color[HTML]{000000} 0.94 & {\cellcolor[HTML]{800026}} \color[HTML]{000000} 1.00 \\
\end{tabular}

%% file: tabulars/left/rfc_containment_similarity_2023.tex
\begin{tabular}{rrrrrrl}
\makecell[c]{\rot{\textbf{1112}}} & \makecell[c]{\rot{\textbf{1918}}} & \makecell[c]{\rot{\textbf{3927}}} & \makecell[c]{\rot{\textbf{5737}}} & \makecell[c]{\rot{\textbf{6598}}} & \makecell[c]{\rot{\textbf{6890}}} & \makecell[c]{\rot{\textbf{RFC}}} \\
{\cellcolor[HTML]{800026}} \color[HTML]{000000} 1.00 & {\cellcolor[HTML]{AE0026}} \color[HTML]{000000} 0.90 & {\cellcolor[HTML]{AE0026}} \color[HTML]{000000} 0.90 & {\cellcolor[HTML]{AE0026}} \color[HTML]{000000} 0.90 & {\cellcolor[HTML]{970026}} \color[HTML]{000000} 0.95 & {\cellcolor[HTML]{FD8038}} \color[HTML]{000000} 0.52 & \bfseries 1112 \\
{\cellcolor[HTML]{FFFFCC}} \color[HTML]{000000} 0.00 & {\cellcolor[HTML]{800026}} \color[HTML]{000000} 1.00 & {\cellcolor[HTML]{FFF8BA}} \color[HTML]{000000} 0.05 & {\cellcolor[HTML]{FFFBC2}} \color[HTML]{000000} 0.03 & {\cellcolor[HTML]{FFE793}} \color[HTML]{000000} 0.17 & {\cellcolor[HTML]{FFFEC9}} \color[HTML]{000000} 0.01 & \bfseries 1918 \\
{\cellcolor[HTML]{FFF8BA}} \color[HTML]{000000} 0.05 & {\cellcolor[HTML]{A60026}} \color[HTML]{000000} 0.92 & {\cellcolor[HTML]{800026}} \color[HTML]{000000} 1.00 & {\cellcolor[HTML]{FEA446}} \color[HTML]{000000} 0.43 & {\cellcolor[HTML]{E31A1C}} \color[HTML]{000000} 0.75 & {\cellcolor[HTML]{FFE895}} \color[HTML]{000000} 0.16 & \bfseries 3927 \\
{\cellcolor[HTML]{FFF3AE}} \color[HTML]{000000} 0.09 & {\cellcolor[HTML]{9D0026}} \color[HTML]{000000} 0.94 & {\cellcolor[HTML]{E61F1D}} \color[HTML]{000000} 0.74 & {\cellcolor[HTML]{800026}} \color[HTML]{000000} 1.00 & {\cellcolor[HTML]{A60026}} \color[HTML]{000000} 0.92 & {\cellcolor[HTML]{FECF6B}} \color[HTML]{000000} 0.28 & \bfseries 5737 \\
{\cellcolor[HTML]{FFFDC8}} \color[HTML]{000000} 0.01 & {\cellcolor[HTML]{E41C1D}} \color[HTML]{000000} 0.74 & {\cellcolor[HTML]{FEE38B}} \color[HTML]{000000} 0.19 & {\cellcolor[HTML]{FFEC9D}} \color[HTML]{000000} 0.13 & {\cellcolor[HTML]{800026}} \color[HTML]{000000} 1.00 & {\cellcolor[HTML]{FFF9BD}} \color[HTML]{000000} 0.04 & \bfseries 6598 \\
{\cellcolor[HTML]{FFE998}} \color[HTML]{000000} 0.15 & {\cellcolor[HTML]{A60026}} \color[HTML]{000000} 0.92 & {\cellcolor[HTML]{D9131F}} \color[HTML]{000000} 0.78 & {\cellcolor[HTML]{D00D21}} \color[HTML]{000000} 0.81 & {\cellcolor[HTML]{BB0026}} \color[HTML]{000000} 0.88 & {\cellcolor[HTML]{800026}} \color[HTML]{000000} 1.00 & \bfseries 6890 \\
\end{tabular}

%% file: tabulars/left/top_10_countries_by_asns.tex
\begin{tabular}{crrrrrrr}
\toprule
\multirow{2}{*}{\textbf{\makecell[c]{Country\\Code}}} & \multirow{2}{*}{\textbf{\makecell[c]{\# Unique\\ASNs}}} & \multicolumn{6}{c}{\textbf{\# Unique ASNs per RFCs}} \\
\cmidrule(lr){3-8}
& & \textbf{\makecell[c]{1112}} & \textbf{\makecell[c]{1918}} & \textbf{\makecell[c]{3927}} & \textbf{\makecell[c]{5737}} & \textbf{\makecell[c]{6598}} & \textbf{\makecell[c]{6890}} \\
\midrule
\textbf{US} & \textbf{1192} & 7 & 1123 & 86 & 52 & 264 & 22 \\
\textbf{BR} & \textbf{1096} & 0 & 998 & 14 & 13 & 243 & 1 \\
\textbf{RU} & \textbf{429} & 0 & 414 & 20 & 13 & 67 & 1 \\
\textbf{ID} & \textbf{265} & 0 & 254 & 10 & 1 & 40 & 0 \\
\textbf{BD} & \textbf{176} & 0 & 176 & 0 & 0 & 10 & 0 \\
\textbf{PL} & \textbf{175} & 0 & 170 & 5 & 5 & 20 & 0 \\
\textbf{GB} & \textbf{175} & 0 & 158 & 15 & 7 & 47 & 4 \\
\textbf{CA} & \textbf{172} & 0 & 168 & 4 & 2 & 26 & 1 \\
\textbf{IT} & \textbf{168} & 0 & 160 & 5 & 2 & 26 & 0 \\
\textbf{DE} & \textbf{137} & 2 & 125 & 11 & 4 & 31 & 2 \\
\bottomrule
\end{tabular}

%% file: tabulars/left/spoofer_data.tex
\resizebox{\textwidth}{!}{
\begin{tabular}{lrrrrrrrr}
\toprule
\multirow{2}{*}{\textbf{\makecell[c]{Bogon Type}}} & \multirow{2}{*}{\textbf{\makecell[c]{Identified\\Unique ASNs}}} & \multirow{2}{*}{\textbf{\makecell[c]{\# in\\Spoofer}}} & \multicolumn{2}{c}{\textbf{\makecell[c]{Only\\Spoofable}}} & \multicolumn{2}{c}{\textbf{\makecell[c]{Only\\Non-Spoofable}}} & \multicolumn{2}{c}{\textbf{\makecell[c]{Both Spoofable\\Non-Spoofable}}} \\
\cmidrule(lr){4-5} \cmidrule(lr){6-7} \cmidrule(lr){8-9}
 & & & \textbf{\makecell[c]{\#}} & \textbf{\makecell[c]{\%}} & \textbf{\makecell[c]{\#}} & \textbf{\makecell[c]{\%}} & \textbf{\makecell[c]{\#}} & \textbf{\makecell[c]{\%}} \\
\midrule
\textbf{RFC1112: Former Class E} & 44 & 22 & 1 & 4.55\% & 15 & 68.18\% & 6 & 27.27\% \\
\textbf{RFC1918: Private-Use} & 13,883 & 2,402 & 334 & 13.91\% & 1,272 & 52.96\% & 690 & 28.73\% \\
\textbf{RFC3927: Link-Local} & 747 & 279 & 24 & 8.60\% & 143 & 51.25\% & 106 & 37.99\% \\
\textbf{RFC5737: Documentation} & 416 & 199 & 12 & 6.03\% & 110 & 55.28\% & 75 & 37.69\% \\
\textbf{RFC6598: Shared Address Space} & 2,869 & 740 & 84 & 11.35\% & 376 & 50.81\% & 261 & 35.27\% \\
\textbf{RFC6890: Protocol Assignments} & 144 & 63 & 6 & 9.52\% & 36 & 57.14\% & 21 & 33.33\% \\
\bottomrule
\end{tabular}
}

%% file: tabulars/left/manrs_results.tex
\begin{tabular}{llrrrrrrr}
\toprule
\multirow[c]{2}{*}{\textbf{\makecell[c]{Members}}} & \multirow[c]{2}{*}{\textbf{\makecell[c]{Conformance}}} & \multirow[c]{2}{*}{\textbf{\makecell[c]{\# Unique\\ASNs}}} & \multicolumn{6}{c}{\textbf{\# Unique ASNs per RFC}} \\
\cmidrule(lr){4-9}
 & & & \textbf{\makecell[c]{1112}} & \textbf{\makecell[c]{1918}} & \textbf{\makecell[c]{3927}} & \textbf{\makecell[c]{5737}} & \textbf{\makecell[c]{6598}} & \textbf{\makecell[c]{6890}} \\
\midrule
\multirow[c]{2}{*}{\textbf{All}} & \textbf{conformant} & 235 & 6 & 217 & 56 & 52 & 115 & 19 \\
 & \textbf{non-conformant} & 124 & 3 & 116 & 23 & 17 & 49 & 7 \\
\multirow[c]{2}{*}{\textbf{Before 2023}} & \textbf{conformant} & 209 & 5 & 191 & 51 & 49 & 106 & 18 \\
 & \textbf{non-conformant} & 102 & 3 & 96 & 21 & 15 & 40 & 6 \\
\bottomrule
\end{tabular}

%% file: sections_old/06discussion.tex
\section{Discussion}

Our analysis revealed the lack of filtering packets from Bogon addresses in more than \num{13000} ASes. These findings represent a conservative estimate based on our measurement datasets. As networks grow and the transition to IPv6 does not happen at the same pace, the need to assign IPv4 addresses to equipment increases. With a lack of IPv4 addresses, networks will use more Private-Use or Shared Address Space addresses for their infrastructures. The risks of accepting and forwarding spoofed packets without proper filtering also increase. Thus, network operators should be motivated to implement current best practices.

\subsection{Impact of Incomplete Routing Information}
Internet-wide traceroute measurements, such as those performed by the CAIDA Ark project, are limited in their ability to uncover network paths. It is not feasible to trace all IP addresses from each VP during every measurement cycle. These limitations make it hard to reproduce or pinpoint when specific changes happened~\cite{ark}.

These findings show the need to develop different approaches and tools beyond traceroute measurements for identifying networks not filtering Bogon addresses. Incorporating multiple data sources and methodologies, such as deterministic active probing techniques, could improve the detection of ASNs transiting packets from Bogon addresses. Moreover, collaboration between network operators and researchers is crucial to address the limitations of traceroute measurements and improve the accuracy of Internet topology mapping efforts.

\subsection{Incentives for Bogon Address Filtering}
The technical means for blocking packets with Bogon addresses are similar to those used for blocking spoofed packets~\cite{haeni1997firewall, kashefi2013survey} and also improve the traceability of network traffic. As for spoofed packets, the benefits of filtering them are mainly for the neighboring ASes. At the same time, the difficulties and increased costs generated by the implementation are sustained by the ASes not directly benefiting. While this benefits the Internet ecosystem as a whole, individual ASes might not see reasons for implementing these measures.

To support and standardize efforts in routing security, initiatives such as the Mutually Agreed Norms for Routing Security (MANRS) encourage network operators and equipment vendors to adopt best practices. MANRS recommends that network operators implement Source Address Validation (SAV). In 2023, we observed \num{6922} ASes transiting packets with Bogon addresses, of which \num{359} ASes are participating in MANRS. Many of these members (\num{124} ASes) are known not to implement the anti-spoofing recommended action in their networks. However, \num{235} ASes that report having 100\% anti-spoofing measures in place still transiting packets with Bogon addresses. 

Despite the interest demonstrated by MANRS members through their participation in the initiative, the presence of Bogon addresses visible in traceroutes crossing AS borders shows that some of these networks lack basic security measures. This underscores the substantial work needed to improve routing security and implement effective anti-spoofing measures across all networks on the Internet.

\subsection{Limitations}
Our study is based on an incomplete view of Internet topology discovered from traceroutes. As discussed in \autoref{subsec:datasets}, traceroute data only offers a one-way perspective from the source to the destination, and due to diverse routing policies, it is impossible to determine the reverse path, which is commonly asymmetrical. Ongoing research aims to improve the reach of measurement platforms by diversifying Vantage Point distribution. In this context, our findings should be considered as a snapshot taken at a certain point in time from the available Vantage Points, meaning that we cannot claim to find all networks not filtering Bogon addresses inbound or outbound.

However, the results presented in this study are based on actual observations of the networks and thus reflect a sample of real-world scenarios. Additionally, utilizing traceroute data presents common shortcomings. Some routers may be configured to avoid responding to traceroute probes or rate limit the responses, while others may not respond from the address of the receiving interface.

%% file: sections_old/07relatedwork.tex
\section{Related work}
\label{sec:relatedwork}
Bogon addresses often appear in Internet measurements. Despite their commonality, academic papers tend to perceive these as anomalies or measurement errors~\cite{alaraj2023global}. Also, attributing them to specific network infrastructures without additional information is a difficult task.

\subsection{Bogon Addresses}
Much of the literature on Bogon or Martian addresses is dedicated to outlining strategies for their filtering. For example, filtering routes based on assignments from the Regional Internet Registries in order to reduce the routing table size~\cite{bellovin2001slowing}. A study on Bogon routes~\cite{feamster2005empirical} looks for such routes in the BGP Global Routing Table. In \cite{qasim240atlas}, the usage of former Class E space, 240.0.0.0/4, is identified inside the networks of large companies such as Amazon and Verizon Business. Our research also looks for former Class E address space, but as seen from vantage points outside the AS using them.

\subsection{Topology Discovery with Traceroutes}
Traceroutes are frequently used in Internet measurements in order to uncover network paths otherwise not visible, and large datasets of such traceroutes exist (Ark~\cite{ark}, Atlas~\cite{staff2015ripe}). The CAIDA Ark dataset is found to achieve the most comprehensive coverage of node discovery by~\cite{canbaz2017comparative}. One type of information that can be found using traceroutes is routing loops. \cite{lone2017loops} uses routing loops to infer the ability to send spoofed packets from inside one network based on the loops found between interconnecting ASes. Loops as a security risk for amplification and other types of attacks are found by~\cite{alaraj2023global}, while their research excludes RFC 1918 Private-Use addresses on purpose. In~\cite{salamatian2023squats}, traceroute information is used to identify networks squatting IP addresses assigned to other organizations; this work is similar to the one proposed in our paper; however, because the authors are looking for IP addresses that are assigned to organizations, no implications about SAV can be inferred, as the squatted IPs are expected to be found routed on the Internet.

\subsection{Source Address Validation}
As SAV is a problem affecting the correct operation of the Internet, in current research, various ways of detecting networks not implementing SAV are developed, some are based on active measurements~\cite{lone2017loops}, using spoofed packets~\cite{korczynski2020don} with various protocols/IP header options~\cite{dai2021smap}, some more intrusive using BGP poisoning in combination with passive honeypots~\cite{fonseca2021identifying}. The CAIDA Spoofer project offers active measurements run by volunteers to identify spoofable networks~\cite{beverly2009understanding}; in our research, we also cross-check our findings using this dataset. A study~\cite{lichtblau2017detection} of flow data from a large European IXP finds 72\% of members send packets with Bogon sources, concluding that the majority of members do not filter outbound traffic.

%% file: sections_old/08conclusions.tex
\section{Conclusions}

As shown in \autoref{sec:relatedwork}, Bogon addresses often appear in Internet measurements but are seen and dismissed by researchers as measurement errors or anomalies. Our paper shows that this is not an anomaly but a too common occurrence caused by poor network hygiene that should be considered when doing Internet measurements, and fixed by network operators. Our paper shows the extent of this problem, which has been previously noticed.

Across the 84 measurements we analyzed, a clear upward trend can be noticed in the prevalence of ASes transiting packets with Bogon sources, notably for three main types of Bogon addresses: Private-Use (RFC1918), Link-local (RFC3927), and Shared Address Space (RFC6598). This trend can be explained by the increased use of private addresses within networks, spurred by the exhaustion of IPv4 resources and the lack of filtering at interconnection points between ASes.

By using the available datasets for inferring the ability to spoof from the ASes identified as transiting packets with Bogon sources, we find the number of spoofable is lower than the number of non-spoofable ASes, with overlaps of spoofable/non-spoofable tests from the same AS, suggesting different policies regarding SAV exist per prefix or network segment. However, the available data is limited by the number of ASes tested by the Spoofer project. There is also the possibility that transiting packets of Bogon addresses represents a different type of misconfiguration from SAV. Due to this limited number of tested ASes, we cannot deduce if SAV is implemented for non-private address space inside the ASes found with Bogons.

%% file: sections/apdx_dataset.tex
\section{Measurement results}
\label{apdx:measurement_results}

\begin{table}[h]
    \centering
    \caption{Dataset Analysis Results}
    \resizebox*{!}{0.9\textheight}{
        \renewcommand{\arraystretch}{0.8}
        \setlength\tabcolsep{2pt}
        \input{tabulars/left/stats_by_month.tex}
    }
    \label{tab:stats_by_month}
\end{table}

%% file: tabulars/left/stats_by_month.tex
\begin{tabular}{lrrrrrrrrrrrrrrrr}
\toprule
\multirow{3}{*}{\textbf{\makecell[c]{Month}}} & \multirow{3}{*}{\textbf{\makecell[c]{\# VPs}}} & \multirow{3}{*}{\textbf{\makecell[c]{\# VPs\\observing\\Bogons}}} & \multicolumn{8}{c}{\textbf{\# Traceroutes}} & \multicolumn{6}{c}{\textbf{\# ASNs transiting Bogons}} \\
\cmidrule(lr){4-11}
 &  &  & \multirow{2}{*}{\textbf{\makecell[c]{Total}}} & \multicolumn{6}{c}{\textbf{With Bogon addresses per RFC}} & \multirow{2}{*}{\textbf{\makecell[c]{W/ no\\ASNs}}} & \multicolumn{6}{c}{\textbf{per RFC}} \\
 \cmidrule(lr){5-10} \cmidrule(lr){12-17}
 &  &  &  & \makecell[c]{\textbf{1112}} & \makecell[c]{\textbf{1918}} & \makecell[c]{\textbf{3927}} & \makecell[c]{\textbf{5737}} & \makecell[c]{\textbf{6598}} & \makecell[c]{\textbf{6890}} &  & \makecell[c]{\textbf{1112}} & \makecell[c]{\textbf{1918}} & \makecell[c]{\textbf{3927}} & \makecell[c]{\textbf{5737}} & \makecell[c]{\textbf{6598}} & \makecell[c]{\textbf{6890}}  \\
\midrule
\textbf{2017-01} & 36 & 34 & 10,908,012 & 0 & 2,333,712 & 47 & 57 & 23,535 & 1 & 229,470 & 0 & 2,565 & 61 & 39 & 251 & 4 \\
\textbf{2017-02} & 39 & 37 & 10,908,012 & 0 & 2,665,121 & 45 & 51 & 22,179 & 1 & 235,726 & 0 & 2,508 & 64 & 43 & 261 & 4 \\
\textbf{2017-03} & 42 & 41 & 10,908,012 & 0 & 3,032,516 & 94 & 45 & 22,113 & 0 & 242,343 & 0 & 2,512 & 83 & 41 & 255 & 0 \\
\textbf{2017-04} & 42 & 41 & 10,908,012 & 0 & 2,864,580 & 61 & 31 & 22,769 & 0 & 265,457 & 0 & 2,484 & 69 & 32 & 267 & 0 \\
\textbf{2017-05} & 39 & 37 & 10,903,071 & 0 & 2,496,062 & 62 & 35 & 24,231 & 0 & 267,989 & 0 & 2,517 & 69 & 49 & 268 & 0 \\
\textbf{2017-06} & 38 & 37 & 10,908,012 & 0 & 2,211,989 & 74 & 39 & 23,915 & 0 & 271,619 & 0 & 2,568 & 76 & 35 & 289 & 0 \\
\textbf{2017-07} & 33 & 31 & 10,908,012 & 0 & 2,449,291 & 59 & 39 & 29,861 & 0 & 282,081 & 0 & 2,416 & 57 & 40 & 264 & 0 \\
\textbf{2017-08} & 30 & 28 & 10,908,012 & 0 & 2,369,351 & 56 & 45 & 24,570 & 0 & 288,012 & 0 & 2,472 & 49 & 32 & 252 & 0 \\
\textbf{2017-09} & 43 & 41 & 10,906,323 & 0 & 3,009,100 & 47 & 37 & 28,063 & 1 & 291,330 & 0 & 2,352 & 64 & 47 & 269 & 4 \\
\textbf{2017-10} & 39 & 37 & 10,908,012 & 0 & 3,275,053 & 59 & 36 & 28,838 & 0 & 294,498 & 0 & 2,253 & 71 & 40 & 278 & 0 \\
\textbf{2017-11} & 41 & 38 & 10,906,398 & 0 & 2,753,270 & 57 & 31 & 27,807 & 0 & 293,535 & 0 & 2,287 & 63 & 29 & 267 & 0 \\
\textbf{2017-12} & 54 & 51 & 10,906,262 & 0 & 3,362,171 & 39 & 66 & 27,597 & 0 & 299,026 & 0 & 2,190 & 61 & 45 & 274 & 0 \\
\textbf{2018-01} & 54 & 52 & 10,906,280 & 0 & 3,323,094 & 50 & 63 & 21,706 & 1 & 319,112 & 0 & 2,408 & 71 & 43 & 293 & 4 \\
\textbf{2018-02} & 46 & 45 & 10,904,557 & 0 & 2,829,771 & 55 & 70 & 19,225 & 1 & 328,951 & 0 & 2,138 & 70 & 33 & 262 & 4 \\
\textbf{2018-03} & 44 & 42 & 10,908,012 & 0 & 2,075,940 & 68 & 39 & 19,851 & 0 & 347,956 & 0 & 2,186 & 67 & 31 & 259 & 0 \\
\textbf{2018-04} & 42 & 41 & 10,906,242 & 0 & 1,766,926 & 67 & 76 & 18,263 & 1 & 333,916 & 0 & 2,169 & 72 & 39 & 277 & 4 \\
\textbf{2018-05} & 47 & 45 & 10,906,404 & 0 & 2,209,012 & 87 & 76 & 19,310 & 0 & 353,910 & 0 & 2,299 & 72 & 44 & 265 & 0 \\
\textbf{2018-06} & 37 & 36 & 10,905,421 & 0 & 1,680,677 & 63 & 70 & 24,331 & 0 & 372,640 & 0 & 2,158 & 58 & 35 & 254 & 0 \\
\textbf{2018-07} & 45 & 43 & 10,908,012 & 0 & 1,841,169 & 63 & 60 & 24,849 & 1 & 377,478 & 0 & 2,281 & 73 & 38 & 281 & 4 \\
\textbf{2018-08} & 45 & 44 & 10,906,464 & 0 & 1,538,555 & 49 & 81 & 24,750 & 0 & 393,395 & 0 & 1,914 & 75 & 42 & 293 & 0 \\
\textbf{2018-09} & 103 & 97 & 11,194,865 & 0 & 1,891,850 & 53 & 78 & 115,758 & 0 & 144,702 & 0 & 2,091 & 81 & 63 & 389 & 0 \\
\textbf{2018-10} & 92 & 88 & 11,075,865 & 0 & 2,106,341 & 65 & 84 & 124,939 & 1 & 156,946 & 0 & 2,495 & 79 & 46 & 370 & 2 \\
\textbf{2018-11} & 60 & 53 & 11,204,865 & 0 & 1,608,819 & 50 & 88 & 27,600 & 0 & 315,055 & 0 & 2,595 & 64 & 36 & 359 & 0 \\
\textbf{2018-12} & 80 & 76 & 10,466,865 & 0 & 1,777,668 & 66 & 97 & 145,832 & 2 & 394,152 & 0 & 2,735 & 79 & 55 & 350 & 2 \\
\textbf{2019-01} & 93 & 87 & 10,547,865 & 0 & 1,760,943 & 109 & 77 & 227,752 & 1 & 399,736 & 0 & 2,479 & 100 & 38 & 363 & 4 \\
\textbf{2019-02} & 68 & 64 & 10,330,500 & 0 & 1,621,149 & 91 & 92 & 153,034 & 0 & 399,084 & 0 & 2,532 & 71 & 45 & 290 & 0 \\
\textbf{2019-03} & 106 & 102 & 11,204,730 & 0 & 2,578,371 & 128 & 96 & 309,956 & 0 & 439,286 & 0 & 2,760 & 123 & 62 & 410 & 0 \\
\textbf{2019-04} & 101 & 96 & 10,999,256 & 0 & 2,489,626 & 120 & 94 & 228,744 & 0 & 77,086 & 0 & 2,796 & 113 & 59 & 436 & 0 \\
\textbf{2019-05} & 104 & 99 & 11,107,221 & 0 & 2,907,541 & 122 & 68 & 338,754 & 0 & 98,394 & 0 & 2,766 & 124 & 46 & 439 & 0 \\
\textbf{2019-06} & 128 & 119 & 11,126,128 & 0 & 2,557,159 & 79,620 & 32 & 204,499 & 2 & 147,756 & 0 & 2,676 & 106 & 46 & 437 & 2 \\
\textbf{2019-07} & 119 & 110 & 11,031,628 & 0 & 2,010,701 & 140 & 40 & 210,255 & 0 & 149,073 & 0 & 2,683 & 119 & 52 & 461 & 0 \\
\textbf{2019-08} & 114 & 105 & 11,118,628 & 0 & 2,203,410 & 81,142 & 39 & 138,174 & 1 & 162,203 & 0 & 2,732 & 121 & 45 & 497 & 2 \\
\textbf{2019-09} & 92 & 84 & 10,911,628 & 0 & 1,933,021 & 155 & 42 & 138,969 & 0 & 182,851 & 0 & 2,759 & 114 & 33 & 464 & 0 \\
\textbf{2019-10} & 147 & 133 & 11,119,216 & 0 & 2,588,404 & 67,630 & 29 & 271,362 & 2 & 204,715 & 0 & 2,551 & 113 & 36 & 480 & 6 \\
\textbf{2019-11} & 150 & 136 & 11,308,704 & 0 & 3,066,096 & 66,138 & 70 & 299,519 & 1 & 368,798 & 0 & 2,682 & 120 & 66 & 488 & 3 \\
\textbf{2019-12} & 150 & 137 & 11,269,704 & 0 & 3,015,695 & 66,128 & 61 & 380,526 & 0 & 398,909 & 0 & 2,551 & 116 & 49 & 477 & 0 \\
\textbf{2020-01} & 154 & 140 & 11,331,204 & 0 & 2,795,546 & 55,620 & 63 & 241,246 & 0 & 412,223 & 0 & 2,521 & 114 & 55 & 504 & 0 \\
\textbf{2020-02} & 149 & 135 & 11,259,000 & 0 & 2,760,966 & 66,159 & 129 & 258,399 & 1 & 419,925 & 0 & 2,439 & 135 & 67 & 477 & 1 \\
\textbf{2020-03} & 53 & 50 & 11,118,985 & 0 & 2,676,252 & 198,138 & 202 & 429,860 & 0 & 429,162 & 0 & 2,173 & 84 & 46 & 380 & 0 \\
\textbf{2020-04} & 145 & 130 & 11,332,908 & 0 & 2,568,605 & 67,630 & 219 & 330,081 & 0 & 440,969 & 0 & 2,640 & 116 & 82 & 486 & 0 \\
\textbf{2020-05} & 150 & 135 & 11,331,408 & 0 & 2,635,751 & 66,149 & 238 & 95,954 & 0 & 504,577 & 0 & 2,692 & 130 & 88 & 279 & 0 \\
\textbf{2020-06} & 141 & 124 & 11,334,204 & 0 & 2,446,309 & 79,636 & 239 & 86,206 & 0 & 506,685 & 0 & 2,609 & 131 & 95 & 274 & 0 \\
\textbf{2020-07} & 135 & 118 & 11,325,000 & 1 & 2,375,570 & 165 & 210 & 101,066 & 1 & 496,324 & 3 & 2,637 & 125 & 85 & 264 & 4 \\
\textbf{2020-08} & 124 & 115 & 11,326,500 & 0 & 2,551,634 & 201 & 254 & 111,035 & 1 & 496,130 & 0 & 2,675 & 146 & 94 & 276 & 3 \\
\textbf{2020-09} & 123 & 109 & 11,286,204 & 0 & 2,520,081 & 167 & 243 & 107,487 & 0 & 498,144 & 0 & 2,683 & 124 & 83 & 276 & 0 \\
\textbf{2020-10} & 117 & 103 & 11,326,500 & 0 & 2,106,997 & 176 & 243 & 289,132 & 0 & 514,640 & 0 & 2,757 & 143 & 75 & 524 & 0 \\
\textbf{2020-11} & 122 & 109 & 11,290,908 & 1 & 2,273,141 & 216 & 238 & 126,485 & 8 & 510,543 & 2 & 2,701 & 148 & 115 & 544 & 17 \\
\textbf{2020-12} & 120 & 110 & 11,328,204 & 4 & 2,324,084 & 200 & 78 & 196,482 & 7 & 235,925 & 11 & 2,642 & 151 & 69 & 540 & 23 \\
\textbf{2021-01} & 115 & 104 & 11,329,704 & 4 & 2,140,406 & 198 & 248 & 210,825 & 6 & 528,196 & 6 & 2,587 & 137 & 95 & 524 & 18 \\
\textbf{2021-02} & 104 & 94 & 11,334,914 & 3 & 1,369,941 & 183 & 213 & 224,061 & 7 & 526,893 & 8 & 2,363 & 147 & 84 & 522 & 12 \\
\textbf{2021-03} & 101 & 90 & 11,334,408 & 7 & 1,957,880 & 228 & 226 & 229,972 & 6 & 527,306 & 13 & 2,599 & 159 & 82 & 555 & 22 \\
\textbf{2021-04} & 109 & 98 & 11,329,500 & 7 & 2,064,681 & 231 & 108 & 318,913 & 5 & 557,966 & 8 & 2,648 & 162 & 80 & 549 & 10 \\
\textbf{2021-05} & 107 & 95 & 11,200,500 & 5 & 1,751,453 & 243 & 132 & 317,190 & 4 & 531,416 & 6 & 2,599 & 161 & 79 & 511 & 12 \\
\textbf{2021-06} & 103 & 92 & 11,332,908 & 4 & 1,754,968 & 231 & 108 & 223,049 & 8 & 541,410 & 9 & 2,504 & 130 & 64 & 525 & 17 \\
\textbf{2021-07} & 106 & 95 & 11,314,500 & 2 & 1,694,122 & 205 & 154 & 200,085 & 10 & 553,234 & 5 & 2,471 & 139 & 68 & 487 & 19 \\
\textbf{2021-08} & 103 & 93 & 11,193,204 & 3 & 1,836,343 & 220 & 115 & 210,469 & 4 & 549,035 & 3 & 2,486 & 132 & 86 & 499 & 5 \\
\textbf{2021-09} & 64 & 56 & 11,109,408 & 2 & 822,262 & 220 & 195 & 31,948 & 8 & 554,082 & 6 & 2,531 & 127 & 79 & 514 & 17 \\
\textbf{2021-10} & 71 & 62 & 11,028,204 & 3 & 1,343,907 & 263 & 205 & 30,237 & 4 & 550,561 & 9 & 2,504 & 141 & 85 & 507 & 11 \\
\textbf{2021-11} & 73 & 66 & 11,106,000 & 5 & 1,383,787 & 230 & 185 & 30,622 & 4 & 561,695 & 10 & 2,437 & 124 & 81 & 519 & 13 \\
\textbf{2021-12} & 69 & 62 & 11,172,000 & 4 & 1,739,596 & 224 & 200 & 34,095 & 3 & 566,579 & 10 & 2,539 & 130 & 88 & 534 & 11 \\
\textbf{2022-01} & 70 & 63 & 11,169,000 & 4 & 1,769,730 & 240 & 212 & 36,863 & 7 & 570,746 & 9 & 2,513 & 127 & 78 & 529 & 11 \\
\textbf{2022-02} & 101 & 93 & 11,208,204 & 7 & 2,061,262 & 245 & 190 & 123,164 & 7 & 578,420 & 12 & 2,728 & 141 & 93 & 530 & 14 \\
\textbf{2022-03} & 94 & 87 & 11,205,204 & 5 & 2,068,627 & 256 & 89 & 126,164 & 4 & 583,457 & 6 & 2,594 & 128 & 79 & 523 & 10 \\
\textbf{2022-04} & 94 & 85 & 11,208,204 & 3 & 1,956,568 & 288 & 176 & 124,553 & 6 & 590,618 & 7 & 2,545 & 136 & 67 & 543 & 11 \\
\textbf{2022-05} & 56 & 47 & 11,073,408 & 2 & 1,122,791 & 297 & 206 & 38,957 & 6 & 585,876 & 4 & 2,705 & 145 & 63 & 538 & 14 \\
\textbf{2022-06} & 93 & 84 & 11,209,502 & 4 & 1,866,069 & 259 & 200 & 121,859 & 3 & 609,576 & 6 & 2,567 & 144 & 69 & 536 & 8 \\
\textbf{2022-07} & 89 & 79 & 11,684,594 & 2 & 2,069,687 & 312 & 226 & 132,918 & 2 & 7,011 & 5 & 2,781 & 148 & 84 & 567 & 6 \\
\textbf{2022-08} & 52 & 43 & 11,711,797 & 6 & 1,056,536 & 367 & 242 & 244,498 & 6 & 33,560 & 10 & 2,930 & 132 & 71 & 594 & 14 \\
\textbf{2022-09} & 89 & 78 & 11,791,742 & 3 & 1,876,047 & 311 & 195 & 267,390 & 9 & 55,317 & 4 & 2,675 & 136 & 84 & 570 & 16 \\
\textbf{2022-10} & 86 & 76 & 11,644,094 & 1 & 1,625,380 & 343 & 223 & 266,353 & 9 & 74,671 & 4 & 2,777 & 135 & 87 & 556 & 21 \\
\textbf{2022-11} & 86 & 78 & 11,643,674 & 1 & 1,810,025 & 357 & 211 & 252,845 & 5 & 90,589 & 2 & 2,683 & 149 & 80 & 552 & 9 \\
\textbf{2022-12} & 81 & 72 & 11,621,509 & 4 & 1,704,850 & 270 & 179 & 257,264 & 3 & 103,328 & 5 & 2,583 & 125 & 78 & 568 & 7 \\
\textbf{2023-01} & 77 & 69 & 11,936,594 & 1 & 1,763,610 & 359 & 181 & 284,882 & 6 & 122,340 & 3 & 2,653 & 138 & 72 & 576 & 16 \\
\textbf{2023-02} & 78 & 72 & 11,932,500 & 1 & 1,831,727 & 283 & 195 & 275,044 & 4 & 131,665 & 3 & 2,548 & 121 & 82 & 566 & 9 \\
\textbf{2023-03} & 45 & 42 & 11,794,287 & 3 & 1,587,447 & 391 & 176 & 558,159 & 7 & 137,722 & 4 & 1,382 & 119 & 59 & 505 & 16 \\
\textbf{2023-04} & 100 & 91 & 11,815,297 & 3 & 2,442,528 & 336 & 282 & 318,211 & 7 & 148,763 & 5 & 2,635 & 143 & 105 & 549 & 12 \\
\textbf{2023-05} & 99 & 92 & 11,815,500 & 8 & 2,511,525 & 341 & 308 & 326,387 & 4 & 170,822 & 11 & 2,720 & 142 & 104 & 618 & 6 \\
\textbf{2023-06} & 103 & 94 & 11,930,797 & 3 & 2,585,158 & 368 & 329 & 322,413 & 5,022 & 173,141 & 6 & 2,946 & 151 & 100 & 612 & 20 \\
\textbf{2023-07} & 111 & 100 & 11,929,470 & 2 & 2,475,620 & 423 & 281 & 296,993 & 60 & 199,534 & 5 & 2,816 & 166 & 109 & 604 & 11 \\
\textbf{2023-08} & 108 & 99 & 11,701,297 & 2 & 2,430,242 & 457 & 291 & 323,026 & 9 & 206,814 & 4 & 2,840 & 176 & 105 & 619 & 24 \\
\textbf{2023-09} & 120 & 111 & 11,938,094 & 2 & 2,733,240 & 444 & 312 & 297,203 & 11 & 208,460 & 4 & 2,794 & 170 & 112 & 643 & 17 \\
\textbf{2023-10} & 121 & 111 & 11,848,297 & 3 & 2,656,772 & 417 & 308 & 305,045 & 76 & 235,442 & 7 & 2,767 & 163 & 116 & 617 & 23 \\
\textbf{2023-11} & 131 & 120 & 11,936,797 & 3 & 2,729,388 & 433 & 272 & 283,485 & 6 & 241,965 & 6 & 2,806 & 162 & 110 & 655 & 16 \\
\textbf{2023-12} & 119 & 112 & 11,899,520 & 4 & 3,125,279 & 511 & 244 & 315,932 & 6 & 24 & 6 & 2,973 & 189 & 104 & 658 & 16 \\
\midrule
\textbf{Mean} & \textbf{88.1} & \textbf{80.6} & \textbf{11,240,248.4} & \textbf{1.6} & \textbf{2,213,792.2} & \textbf{10,814.3} & \textbf{144.1} & \textbf{168,189.8} & \textbf{64.2} & \textbf{333,341.3} & \textbf{2.9} & \textbf{2,547.1} & \textbf{115.6} & \textbf{66.5} & \textbf{440.3} & \textbf{7.0} \\
\bottomrule
\end{tabular}

%% file: sections/apdx_graphs.tex
\section{Additional Figures}
\label{apdx:additional_figures}

\begin{figure}[h!]
\centering
\begin{subfigure}[b]{0.48\linewidth}
    \resizebox{\linewidth}{!}{
        \setlength\tabcolsep{1pt}
        \input{tabulars/left/jaccard_similarity_rfc3927_202301-202312.tex}
    }
    \caption{Link-Local (RFC3927)}
    \label{fig:jaccard_similarity_rfc3927}
\end{subfigure}
\begin{subfigure}[b]{0.48\linewidth}
    \resizebox{\linewidth}{!}{
        \setlength\tabcolsep{1pt}
        \input{tabulars/left/jaccard_similarity_rfc3927_202301-202312.tex}
    }
    \caption{Documentation (RFC5737)}
    \label{fig:jaccard_similarity_rfc5737}
\end{subfigure}
\\
\begin{subfigure}[b]{0.48\linewidth}
    \resizebox{\linewidth}{!}{
        \setlength\tabcolsep{1pt}
        \input{tabulars/left/jaccard_similarity_rfc6890_202301-202312.tex}
    }
    \caption{Protocol Assignments (RFC6890)}
    \label{fig:jaccard_similarity_rfc6890}
\end{subfigure}
\begin{subfigure}[b]{0.48\linewidth}
    \resizebox{\linewidth}{!}{
        \setlength\tabcolsep{1pt}
        \input{tabulars/left/jaccard_similarity_rfc1112_202301-202312.tex}
    }
    \caption{Former Class E (RFC1112)}
    \label{fig:jaccard_similarity_rfc1112}
\end{subfigure}
\caption{\textbf{BA:} Jaccard Similarity of ASNs Transiting Bogons of Particular Type across Months in 2023}
\label{fig:jaccard_similarity_rest}
\end{figure}

\begin{figure*}[h!]
\centering
\begin{subfigure}[b]{0.48\linewidth}
    \includegraphics[width=\linewidth]{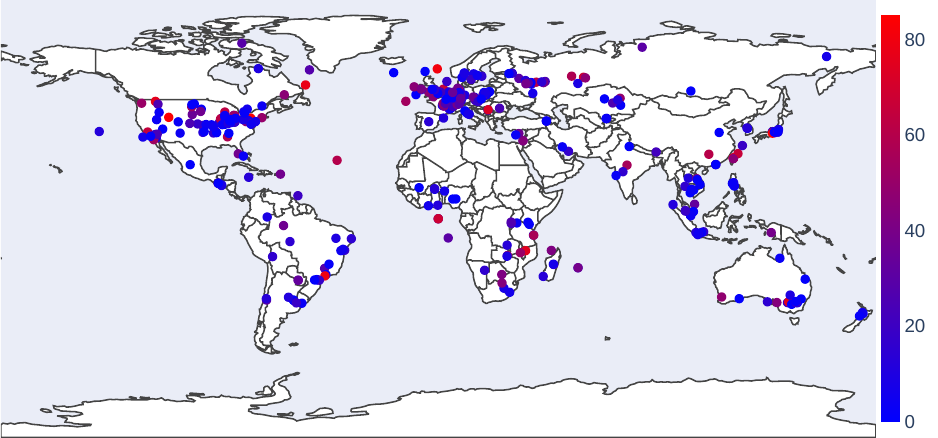}
    \caption{Link-Local (RFC3927)}
    \label{fig:occurences_map_rfc3927}
\end{subfigure}
\quad
\begin{subfigure}[b]{0.48\linewidth}
    \includegraphics[width=\linewidth]{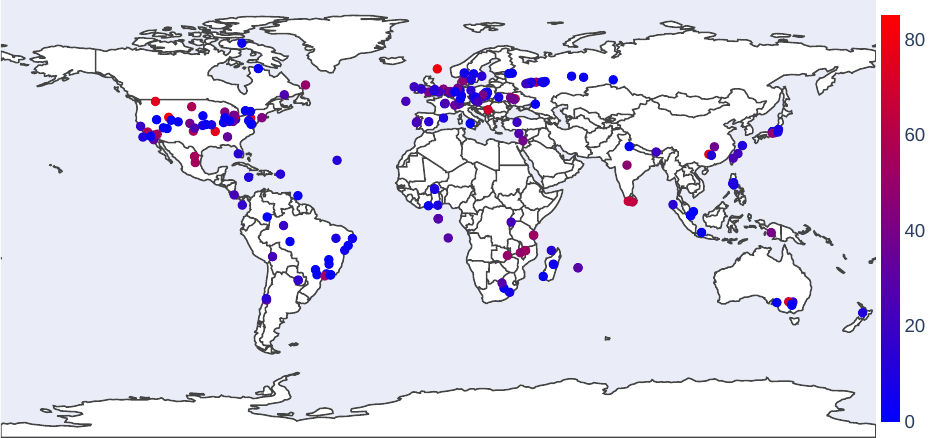}
    \caption{Documentation (RFC5737)}
    \label{fig:occurences_map_rfc5737}
\end{subfigure}
\\
\begin{subfigure}[b]{0.48\linewidth}
    \includegraphics[width=\linewidth]{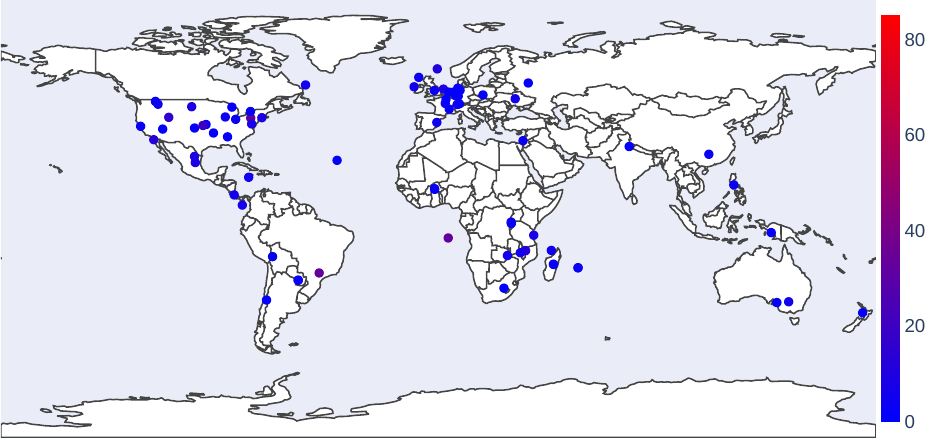}
    \caption{Protocol Assignments (RFC6890)}
    \label{fig:occurences_map_rfc6890}
\end{subfigure}
\quad
\begin{subfigure}[b]{0.48\linewidth}
    \includegraphics[width=\linewidth]{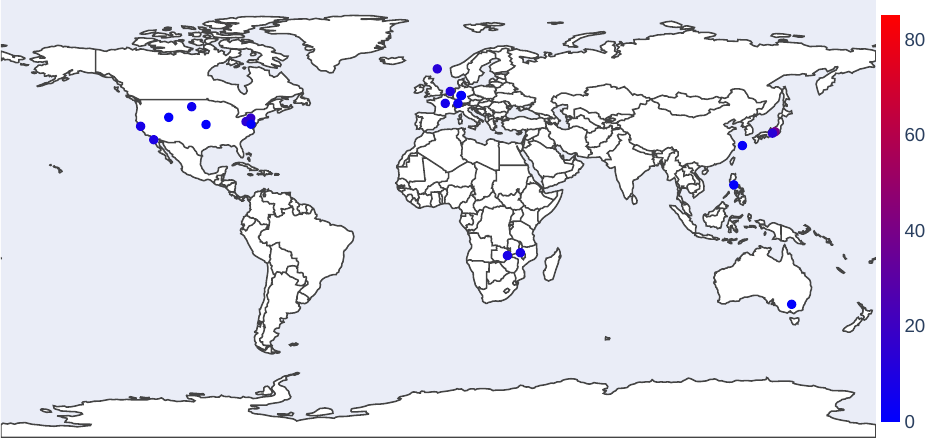}
    \caption{Former Class E (RFC1112)}
    \label{fig:occurences_map_rfc1112}
\end{subfigure}
\caption{\textbf{BA:} Map of ASNs Occurences Transiting Bogons of Particular Type}
\label{fig:occurences_map_rest}
\end{figure*}

\begin{figure*}[h!]
\centering
\begin{subfigure}[b]{0.48\linewidth}
    \includegraphics[width=\linewidth]{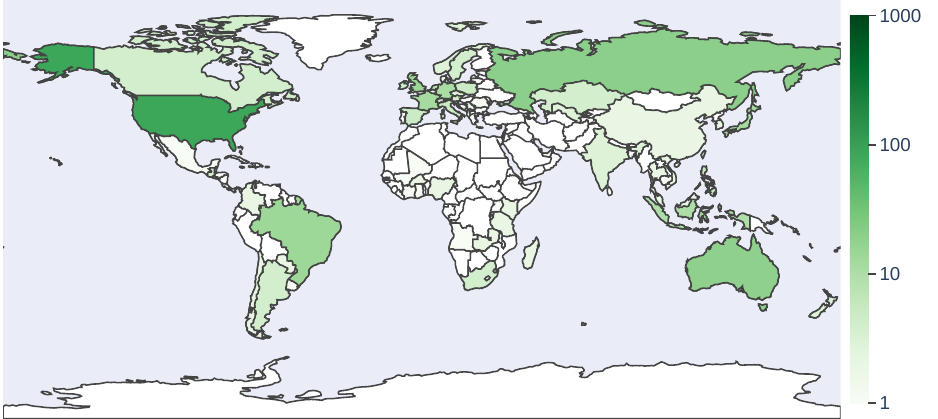}
    \caption{Link-Local (RFC3927)}
    \label{fig:asns_per_country_rfc3927}
\end{subfigure}
\quad
\begin{subfigure}[b]{0.48\linewidth}
    \includegraphics[width=\linewidth]{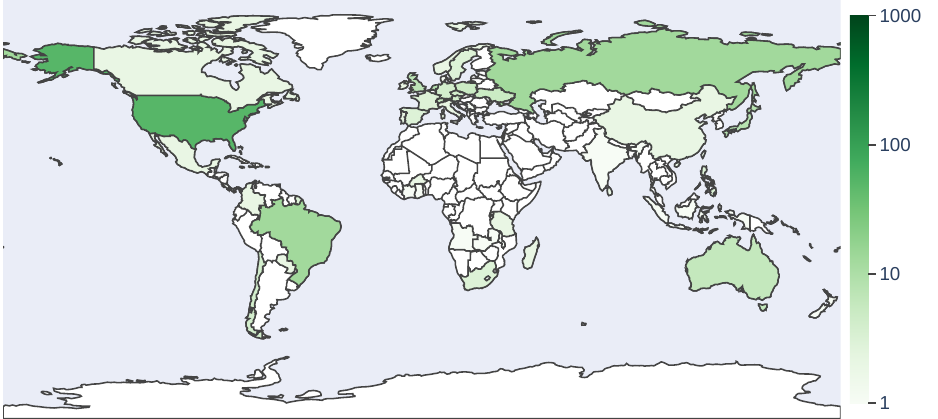}
    \caption{Documentation (RFC5737)}
    \label{fig:asns_per_country_rfc5737}
\end{subfigure}
\\
\begin{subfigure}[b]{0.48\linewidth}
    \includegraphics[width=\linewidth]{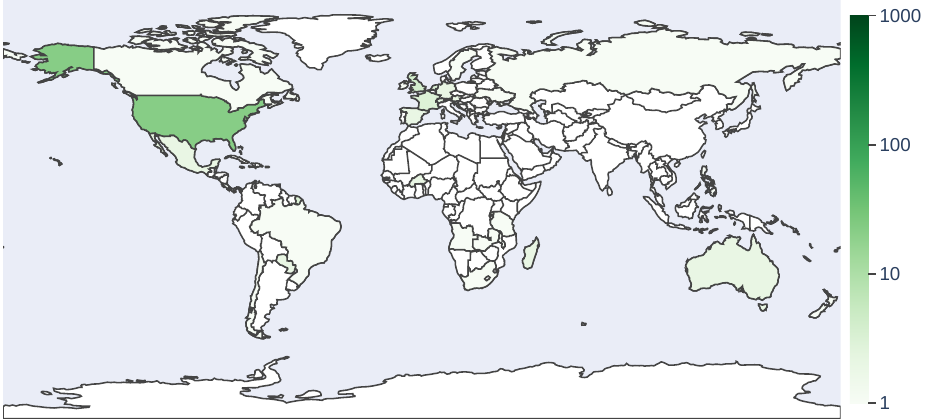}
    \caption{Protocol Assignments (RFC6890)}
    \label{fig:asns_per_country_rfc6890}
\end{subfigure}
\quad
\begin{subfigure}[b]{0.48\linewidth}
    \includegraphics[width=\linewidth]{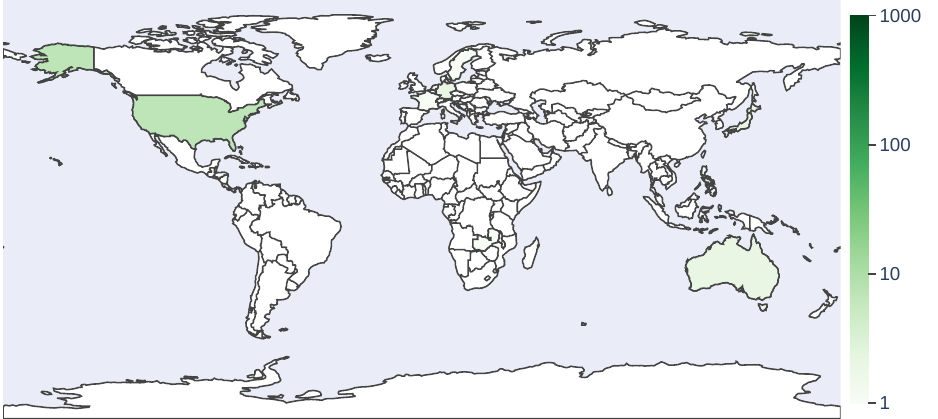}
    \caption{Former Class E (RFC1112)}
    \label{fig:asns_per_country_rfc1112}
\end{subfigure}
\caption{\textbf{BA:} Colormap of the Number of ASNs per Country Transiting Bogons of Particular Type}
\label{fig:asns_per_country_rest}
\end{figure*}

%% file: tabulars/left/jaccard_similarity_rfc3927_202301-202312.tex
\begin{tabular}{lrrrrrrrrrrrr}
\makecell[c]{\textbf{Month}} & \makecell[c]{\rot{\textbf{2023-01}}} & \makecell[c]{\rot{\textbf{2023-02}}} & \makecell[c]{\rot{\textbf{2023-03}}} & \makecell[c]{\rot{\textbf{2023-04}}} & \makecell[c]{\rot{\textbf{2023-05}}} & \makecell[c]{\rot{\textbf{2023-06}}} & \makecell[c]{\rot{\textbf{2023-07}}} & \makecell[c]{\rot{\textbf{2023-08}}} & \makecell[c]{\rot{\textbf{2023-09}}} & \makecell[c]{\rot{\textbf{2023-10}}} & \makecell[c]{\rot{\textbf{2023-11}}} & \makecell[c]{\rot{\textbf{2023-12}}} \\
\textbf{2023-01} & {\cellcolor[HTML]{800026}} \color[HTML]{000000} 1.00 & {\cellcolor[HTML]{FB4B29}} \color[HTML]{000000} 0.63 & {\cellcolor[HTML]{FEA948}} \color[HTML]{000000} 0.40 & {\cellcolor[HTML]{FD8C3C}} \color[HTML]{000000} 0.50 & {\cellcolor[HTML]{FD923E}} \color[HTML]{000000} 0.48 & {\cellcolor[HTML]{FD953F}} \color[HTML]{000000} 0.47 & {\cellcolor[HTML]{FEA044}} \color[HTML]{000000} 0.43 & {\cellcolor[HTML]{FEA044}} \color[HTML]{000000} 0.43 & {\cellcolor[HTML]{FD9C42}} \color[HTML]{000000} 0.45 & {\cellcolor[HTML]{FEA546}} \color[HTML]{000000} 0.42 & {\cellcolor[HTML]{FEA848}} \color[HTML]{000000} 0.41 & {\cellcolor[HTML]{FEAD4A}} \color[HTML]{000000} 0.39 \\
\textbf{2023-02} & {\cellcolor[HTML]{FB4B29}} \color[HTML]{000000} 0.63 & {\cellcolor[HTML]{800026}} \color[HTML]{000000} 1.00 & {\cellcolor[HTML]{FD9D43}} \color[HTML]{000000} 0.45 & {\cellcolor[HTML]{FD9740}} \color[HTML]{000000} 0.47 & {\cellcolor[HTML]{FD9C42}} \color[HTML]{000000} 0.45 & {\cellcolor[HTML]{FD9640}} \color[HTML]{000000} 0.47 & {\cellcolor[HTML]{FD9E43}} \color[HTML]{000000} 0.44 & {\cellcolor[HTML]{FEA546}} \color[HTML]{000000} 0.42 & {\cellcolor[HTML]{FEA747}} \color[HTML]{000000} 0.41 & {\cellcolor[HTML]{FEAF4B}} \color[HTML]{000000} 0.39 & {\cellcolor[HTML]{FEA848}} \color[HTML]{000000} 0.41 & {\cellcolor[HTML]{FEB54F}} \color[HTML]{000000} 0.37 \\
\textbf{2023-03} & {\cellcolor[HTML]{FEA948}} \color[HTML]{000000} 0.40 & {\cellcolor[HTML]{FD9D43}} \color[HTML]{000000} 0.45 & {\cellcolor[HTML]{800026}} \color[HTML]{000000} 1.00 & {\cellcolor[HTML]{FD9841}} \color[HTML]{000000} 0.46 & {\cellcolor[HTML]{FEA546}} \color[HTML]{000000} 0.42 & {\cellcolor[HTML]{FEAF4B}} \color[HTML]{000000} 0.38 & {\cellcolor[HTML]{FEAC49}} \color[HTML]{000000} 0.40 & {\cellcolor[HTML]{FEB34D}} \color[HTML]{000000} 0.37 & {\cellcolor[HTML]{FEB852}} \color[HTML]{000000} 0.36 & {\cellcolor[HTML]{FEBA55}} \color[HTML]{000000} 0.35 & {\cellcolor[HTML]{FEBA55}} \color[HTML]{000000} 0.35 & {\cellcolor[HTML]{FEC35E}} \color[HTML]{000000} 0.32 \\
\textbf{2023-04} & {\cellcolor[HTML]{FD8C3C}} \color[HTML]{000000} 0.50 & {\cellcolor[HTML]{FD9740}} \color[HTML]{000000} 0.47 & {\cellcolor[HTML]{FD9841}} \color[HTML]{000000} 0.46 & {\cellcolor[HTML]{800026}} \color[HTML]{000000} 1.00 & {\cellcolor[HTML]{FC5B2E}} \color[HTML]{000000} 0.60 & {\cellcolor[HTML]{FC6C33}} \color[HTML]{000000} 0.56 & {\cellcolor[HTML]{FC5B2E}} \color[HTML]{000000} 0.60 & {\cellcolor[HTML]{FD7435}} \color[HTML]{000000} 0.55 & {\cellcolor[HTML]{FD7435}} \color[HTML]{000000} 0.55 & {\cellcolor[HTML]{FD8239}} \color[HTML]{000000} 0.52 & {\cellcolor[HTML]{FD913E}} \color[HTML]{000000} 0.49 & {\cellcolor[HTML]{FEA446}} \color[HTML]{000000} 0.42 \\
\textbf{2023-05} & {\cellcolor[HTML]{FD923E}} \color[HTML]{000000} 0.48 & {\cellcolor[HTML]{FD9C42}} \color[HTML]{000000} 0.45 & {\cellcolor[HTML]{FEA546}} \color[HTML]{000000} 0.42 & {\cellcolor[HTML]{FC5B2E}} \color[HTML]{000000} 0.60 & {\cellcolor[HTML]{800026}} \color[HTML]{000000} 1.00 & {\cellcolor[HTML]{FC6631}} \color[HTML]{000000} 0.58 & {\cellcolor[HTML]{FC6C33}} \color[HTML]{000000} 0.56 & {\cellcolor[HTML]{FD7234}} \color[HTML]{000000} 0.55 & {\cellcolor[HTML]{FD8F3D}} \color[HTML]{000000} 0.49 & {\cellcolor[HTML]{FD923E}} \color[HTML]{000000} 0.48 & {\cellcolor[HTML]{FD7E38}} \color[HTML]{000000} 0.53 & {\cellcolor[HTML]{FD9D43}} \color[HTML]{000000} 0.45 \\
\textbf{2023-06} & {\cellcolor[HTML]{FD953F}} \color[HTML]{000000} 0.47 & {\cellcolor[HTML]{FD9640}} \color[HTML]{000000} 0.47 & {\cellcolor[HTML]{FEAF4B}} \color[HTML]{000000} 0.38 & {\cellcolor[HTML]{FC6C33}} \color[HTML]{000000} 0.56 & {\cellcolor[HTML]{FC6631}} \color[HTML]{000000} 0.58 & {\cellcolor[HTML]{800026}} \color[HTML]{000000} 1.00 & {\cellcolor[HTML]{F74327}} \color[HTML]{000000} 0.65 & {\cellcolor[HTML]{FC6430}} \color[HTML]{000000} 0.58 & {\cellcolor[HTML]{FC6C33}} \color[HTML]{000000} 0.57 & {\cellcolor[HTML]{FC6330}} \color[HTML]{000000} 0.59 & {\cellcolor[HTML]{FD8E3C}} \color[HTML]{000000} 0.50 & {\cellcolor[HTML]{FD9640}} \color[HTML]{000000} 0.47 \\
\textbf{2023-07} & {\cellcolor[HTML]{FEA044}} \color[HTML]{000000} 0.43 & {\cellcolor[HTML]{FD9E43}} \color[HTML]{000000} 0.44 & {\cellcolor[HTML]{FEAC49}} \color[HTML]{000000} 0.40 & {\cellcolor[HTML]{FC5B2E}} \color[HTML]{000000} 0.60 & {\cellcolor[HTML]{FC6C33}} \color[HTML]{000000} 0.56 & {\cellcolor[HTML]{F74327}} \color[HTML]{000000} 0.65 & {\cellcolor[HTML]{800026}} \color[HTML]{000000} 1.00 & {\cellcolor[HTML]{F84628}} \color[HTML]{000000} 0.64 & {\cellcolor[HTML]{F94828}} \color[HTML]{000000} 0.64 & {\cellcolor[HTML]{FC5D2E}} \color[HTML]{000000} 0.60 & {\cellcolor[HTML]{FD7C37}} \color[HTML]{000000} 0.53 & {\cellcolor[HTML]{FD9740}} \color[HTML]{000000} 0.47 \\
\textbf{2023-08} & {\cellcolor[HTML]{FEA044}} \color[HTML]{000000} 0.43 & {\cellcolor[HTML]{FEA546}} \color[HTML]{000000} 0.42 & {\cellcolor[HTML]{FEB34D}} \color[HTML]{000000} 0.37 & {\cellcolor[HTML]{FD7435}} \color[HTML]{000000} 0.55 & {\cellcolor[HTML]{FD7234}} \color[HTML]{000000} 0.55 & {\cellcolor[HTML]{FC6430}} \color[HTML]{000000} 0.58 & {\cellcolor[HTML]{F84628}} \color[HTML]{000000} 0.64 & {\cellcolor[HTML]{800026}} \color[HTML]{000000} 1.00 & {\cellcolor[HTML]{FC612F}} \color[HTML]{000000} 0.59 & {\cellcolor[HTML]{FC572C}} \color[HTML]{000000} 0.61 & {\cellcolor[HTML]{FC6832}} \color[HTML]{000000} 0.57 & {\cellcolor[HTML]{FD7836}} \color[HTML]{000000} 0.54 \\
\textbf{2023-09} & {\cellcolor[HTML]{FD9C42}} \color[HTML]{000000} 0.45 & {\cellcolor[HTML]{FEA747}} \color[HTML]{000000} 0.41 & {\cellcolor[HTML]{FEB852}} \color[HTML]{000000} 0.36 & {\cellcolor[HTML]{FD7435}} \color[HTML]{000000} 0.55 & {\cellcolor[HTML]{FD8F3D}} \color[HTML]{000000} 0.49 & {\cellcolor[HTML]{FC6C33}} \color[HTML]{000000} 0.57 & {\cellcolor[HTML]{F94828}} \color[HTML]{000000} 0.64 & {\cellcolor[HTML]{FC612F}} \color[HTML]{000000} 0.59 & {\cellcolor[HTML]{800026}} \color[HTML]{000000} 1.00 & {\cellcolor[HTML]{FC532B}} \color[HTML]{000000} 0.62 & {\cellcolor[HTML]{FD7E38}} \color[HTML]{000000} 0.53 & {\cellcolor[HTML]{FD9D43}} \color[HTML]{000000} 0.45 \\
\textbf{2023-10} & {\cellcolor[HTML]{FEA546}} \color[HTML]{000000} 0.42 & {\cellcolor[HTML]{FEAF4B}} \color[HTML]{000000} 0.39 & {\cellcolor[HTML]{FEBA55}} \color[HTML]{000000} 0.35 & {\cellcolor[HTML]{FD8239}} \color[HTML]{000000} 0.52 & {\cellcolor[HTML]{FD923E}} \color[HTML]{000000} 0.48 & {\cellcolor[HTML]{FC6330}} \color[HTML]{000000} 0.59 & {\cellcolor[HTML]{FC5D2E}} \color[HTML]{000000} 0.60 & {\cellcolor[HTML]{FC572C}} \color[HTML]{000000} 0.61 & {\cellcolor[HTML]{FC532B}} \color[HTML]{000000} 0.62 & {\cellcolor[HTML]{800026}} \color[HTML]{000000} 1.00 & {\cellcolor[HTML]{FC6330}} \color[HTML]{000000} 0.59 & {\cellcolor[HTML]{FD8439}} \color[HTML]{000000} 0.52 \\
\textbf{2023-11} & {\cellcolor[HTML]{FEA848}} \color[HTML]{000000} 0.41 & {\cellcolor[HTML]{FEA848}} \color[HTML]{000000} 0.41 & {\cellcolor[HTML]{FEBA55}} \color[HTML]{000000} 0.35 & {\cellcolor[HTML]{FD913E}} \color[HTML]{000000} 0.49 & {\cellcolor[HTML]{FD7E38}} \color[HTML]{000000} 0.53 & {\cellcolor[HTML]{FD8E3C}} \color[HTML]{000000} 0.50 & {\cellcolor[HTML]{FD7C37}} \color[HTML]{000000} 0.53 & {\cellcolor[HTML]{FC6832}} \color[HTML]{000000} 0.57 & {\cellcolor[HTML]{FD7E38}} \color[HTML]{000000} 0.53 & {\cellcolor[HTML]{FC6330}} \color[HTML]{000000} 0.59 & {\cellcolor[HTML]{800026}} \color[HTML]{000000} 1.00 & {\cellcolor[HTML]{FD7C37}} \color[HTML]{000000} 0.53 \\
\textbf{2023-12} & {\cellcolor[HTML]{FEAD4A}} \color[HTML]{000000} 0.39 & {\cellcolor[HTML]{FEB54F}} \color[HTML]{000000} 0.37 & {\cellcolor[HTML]{FEC35E}} \color[HTML]{000000} 0.32 & {\cellcolor[HTML]{FEA446}} \color[HTML]{000000} 0.42 & {\cellcolor[HTML]{FD9D43}} \color[HTML]{000000} 0.45 & {\cellcolor[HTML]{FD9640}} \color[HTML]{000000} 0.47 & {\cellcolor[HTML]{FD9740}} \color[HTML]{000000} 0.47 & {\cellcolor[HTML]{FD7836}} \color[HTML]{000000} 0.54 & {\cellcolor[HTML]{FD9D43}} \color[HTML]{000000} 0.45 & {\cellcolor[HTML]{FD8439}} \color[HTML]{000000} 0.52 & {\cellcolor[HTML]{FD7C37}} \color[HTML]{000000} 0.53 & {\cellcolor[HTML]{800026}} \color[HTML]{000000} 1.00 \\
\end{tabular}

%% file: tabulars/left/jaccard_similarity_rfc6890_202301-202312.tex
\begin{tabular}{lrrrrrrrrrrrr}
    \makecell[c]{\textbf{Month}} & \makecell[c]{\rot{\textbf{2023-01}}} & \makecell[c]{\rot{\textbf{2023-02}}} & \makecell[c]{\rot{\textbf{2023-03}}} & \makecell[c]{\rot{\textbf{2023-04}}} & \makecell[c]{\rot{\textbf{2023-05}}} & \makecell[c]{\rot{\textbf{2023-06}}} & \makecell[c]{\rot{\textbf{2023-07}}} & \makecell[c]{\rot{\textbf{2023-08}}} & \makecell[c]{\rot{\textbf{2023-09}}} & \makecell[c]{\rot{\textbf{2023-10}}} & \makecell[c]{\rot{\textbf{2023-11}}} & \makecell[c]{\rot{\textbf{2023-12}}} \\
\textbf{2023-01} & {\cellcolor[HTML]{800026}} \color[HTML]{000000} 1.00 & {\cellcolor[HTML]{FEE38B}} \color[HTML]{000000} 0.19 & {\cellcolor[HTML]{FFE48C}} \color[HTML]{000000} 0.19 & {\cellcolor[HTML]{FEDE82}} \color[HTML]{000000} 0.22 & {\cellcolor[HTML]{FFE895}} \color[HTML]{000000} 0.16 & {\cellcolor[HTML]{FD9F44}} \color[HTML]{000000} 0.44 & {\cellcolor[HTML]{FFE590}} \color[HTML]{000000} 0.17 & {\cellcolor[HTML]{FFEFA5}} \color[HTML]{000000} 0.11 & {\cellcolor[HTML]{FEB24C}} \color[HTML]{000000} 0.38 & {\cellcolor[HTML]{FECA66}} \color[HTML]{000000} 0.30 & {\cellcolor[HTML]{FFF0A8}} \color[HTML]{000000} 0.10 & {\cellcolor[HTML]{FFF5B5}} \color[HTML]{000000} 0.07 \\
\textbf{2023-02} & {\cellcolor[HTML]{FEE38B}} \color[HTML]{000000} 0.19 & {\cellcolor[HTML]{800026}} \color[HTML]{000000} 1.00 & {\cellcolor[HTML]{FFEC9D}} \color[HTML]{000000} 0.14 & {\cellcolor[HTML]{FFE793}} \color[HTML]{000000} 0.17 & {\cellcolor[HTML]{FED976}} \color[HTML]{000000} 0.25 & {\cellcolor[HTML]{FFE895}} \color[HTML]{000000} 0.16 & {\cellcolor[HTML]{FFEFA5}} \color[HTML]{000000} 0.11 & {\cellcolor[HTML]{FFF6B6}} \color[HTML]{000000} 0.06 & {\cellcolor[HTML]{FEDB7B}} \color[HTML]{000000} 0.24 & {\cellcolor[HTML]{FFE48C}} \color[HTML]{000000} 0.19 & {\cellcolor[HTML]{FFEC9D}} \color[HTML]{000000} 0.14 & {\cellcolor[HTML]{FEE38B}} \color[HTML]{000000} 0.19 \\
\textbf{2023-03} & {\cellcolor[HTML]{FFE48C}} \color[HTML]{000000} 0.19 & {\cellcolor[HTML]{FFEC9D}} \color[HTML]{000000} 0.14 & {\cellcolor[HTML]{800026}} \color[HTML]{000000} 1.00 & {\cellcolor[HTML]{FEBF5A}} \color[HTML]{000000} 0.33 & {\cellcolor[HTML]{FFF1A9}} \color[HTML]{000000} 0.10 & {\cellcolor[HTML]{FECE6A}} \color[HTML]{000000} 0.29 & {\cellcolor[HTML]{FFE590}} \color[HTML]{000000} 0.17 & {\cellcolor[HTML]{FEDF83}} \color[HTML]{000000} 0.21 & {\cellcolor[HTML]{FEDE80}} \color[HTML]{000000} 0.22 & {\cellcolor[HTML]{FFEA99}} \color[HTML]{000000} 0.15 & {\cellcolor[HTML]{FFF0A8}} \color[HTML]{000000} 0.10 & {\cellcolor[HTML]{FFE48C}} \color[HTML]{000000} 0.19 \\
\textbf{2023-04} & {\cellcolor[HTML]{FEDE82}} \color[HTML]{000000} 0.22 & {\cellcolor[HTML]{FFE793}} \color[HTML]{000000} 0.17 & {\cellcolor[HTML]{FEBF5A}} \color[HTML]{000000} 0.33 & {\cellcolor[HTML]{800026}} \color[HTML]{000000} 1.00 & {\cellcolor[HTML]{FFEDA0}} \color[HTML]{000000} 0.12 & {\cellcolor[HTML]{FED06C}} \color[HTML]{000000} 0.28 & {\cellcolor[HTML]{FD9F44}} \color[HTML]{000000} 0.44 & {\cellcolor[HTML]{FFEDA0}} \color[HTML]{000000} 0.12 & {\cellcolor[HTML]{FED673}} \color[HTML]{000000} 0.26 & {\cellcolor[HTML]{FEE085}} \color[HTML]{000000} 0.21 & {\cellcolor[HTML]{FFEEA3}} \color[HTML]{000000} 0.12 & {\cellcolor[HTML]{FFE793}} \color[HTML]{000000} 0.17 \\
\textbf{2023-05} & {\cellcolor[HTML]{FFE895}} \color[HTML]{000000} 0.16 & {\cellcolor[HTML]{FED976}} \color[HTML]{000000} 0.25 & {\cellcolor[HTML]{FFF1A9}} \color[HTML]{000000} 0.10 & {\cellcolor[HTML]{FFEDA0}} \color[HTML]{000000} 0.12 & {\cellcolor[HTML]{800026}} \color[HTML]{000000} 1.00 & {\cellcolor[HTML]{FFEC9F}} \color[HTML]{000000} 0.13 & {\cellcolor[HTML]{FFEC9D}} \color[HTML]{000000} 0.13 & {\cellcolor[HTML]{FFF5B3}} \color[HTML]{000000} 0.07 & {\cellcolor[HTML]{FEE084}} \color[HTML]{000000} 0.21 & {\cellcolor[HTML]{FFF5B3}} \color[HTML]{000000} 0.07 & {\cellcolor[HTML]{FEDE80}} \color[HTML]{000000} 0.22 & {\cellcolor[HTML]{FFF1A9}} \color[HTML]{000000} 0.10 \\
\textbf{2023-06} & {\cellcolor[HTML]{FD9F44}} \color[HTML]{000000} 0.44 & {\cellcolor[HTML]{FFE895}} \color[HTML]{000000} 0.16 & {\cellcolor[HTML]{FECE6A}} \color[HTML]{000000} 0.29 & {\cellcolor[HTML]{FED06C}} \color[HTML]{000000} 0.28 & {\cellcolor[HTML]{FFEC9F}} \color[HTML]{000000} 0.13 & {\cellcolor[HTML]{800026}} \color[HTML]{000000} 1.00 & {\cellcolor[HTML]{FEBA55}} \color[HTML]{000000} 0.35 & {\cellcolor[HTML]{FEDE80}} \color[HTML]{000000} 0.22 & {\cellcolor[HTML]{FEDC7C}} \color[HTML]{000000} 0.23 & {\cellcolor[HTML]{FED572}} \color[HTML]{000000} 0.26 & {\cellcolor[HTML]{FFEDA0}} \color[HTML]{000000} 0.12 & {\cellcolor[HTML]{FFE794}} \color[HTML]{000000} 0.16 \\
\textbf{2023-07} & {\cellcolor[HTML]{FFE590}} \color[HTML]{000000} 0.17 & {\cellcolor[HTML]{FFEFA5}} \color[HTML]{000000} 0.11 & {\cellcolor[HTML]{FFE590}} \color[HTML]{000000} 0.17 & {\cellcolor[HTML]{FD9F44}} \color[HTML]{000000} 0.44 & {\cellcolor[HTML]{FFEC9D}} \color[HTML]{000000} 0.13 & {\cellcolor[HTML]{FEBA55}} \color[HTML]{000000} 0.35 & {\cellcolor[HTML]{800026}} \color[HTML]{000000} 1.00 & {\cellcolor[HTML]{FFE793}} \color[HTML]{000000} 0.17 & {\cellcolor[HTML]{FEDE82}} \color[HTML]{000000} 0.22 & {\cellcolor[HTML]{FED673}} \color[HTML]{000000} 0.26 & {\cellcolor[HTML]{FFE590}} \color[HTML]{000000} 0.17 & {\cellcolor[HTML]{FFE590}} \color[HTML]{000000} 0.17 \\
\textbf{2023-08} & {\cellcolor[HTML]{FFEFA5}} \color[HTML]{000000} 0.11 & {\cellcolor[HTML]{FFF6B6}} \color[HTML]{000000} 0.06 & {\cellcolor[HTML]{FEDF83}} \color[HTML]{000000} 0.21 & {\cellcolor[HTML]{FFEDA0}} \color[HTML]{000000} 0.12 & {\cellcolor[HTML]{FFF5B3}} \color[HTML]{000000} 0.07 & {\cellcolor[HTML]{FEDE80}} \color[HTML]{000000} 0.22 & {\cellcolor[HTML]{FFE793}} \color[HTML]{000000} 0.17 & {\cellcolor[HTML]{800026}} \color[HTML]{000000} 1.00 & {\cellcolor[HTML]{FEDA78}} \color[HTML]{000000} 0.24 & {\cellcolor[HTML]{FEE085}} \color[HTML]{000000} 0.21 & {\cellcolor[HTML]{FFE58F}} \color[HTML]{000000} 0.18 & {\cellcolor[HTML]{FFE58F}} \color[HTML]{000000} 0.18 \\
\textbf{2023-09} & {\cellcolor[HTML]{FEB24C}} \color[HTML]{000000} 0.38 & {\cellcolor[HTML]{FEDB7B}} \color[HTML]{000000} 0.24 & {\cellcolor[HTML]{FEDE80}} \color[HTML]{000000} 0.22 & {\cellcolor[HTML]{FED673}} \color[HTML]{000000} 0.26 & {\cellcolor[HTML]{FEE084}} \color[HTML]{000000} 0.21 & {\cellcolor[HTML]{FEDC7C}} \color[HTML]{000000} 0.23 & {\cellcolor[HTML]{FEDE82}} \color[HTML]{000000} 0.22 & {\cellcolor[HTML]{FEDA78}} \color[HTML]{000000} 0.24 & {\cellcolor[HTML]{800026}} \color[HTML]{000000} 1.00 & {\cellcolor[HTML]{FEB04B}} \color[HTML]{000000} 0.38 & {\cellcolor[HTML]{FFEB9C}} \color[HTML]{000000} 0.14 & {\cellcolor[HTML]{FFF1A9}} \color[HTML]{000000} 0.10 \\
\textbf{2023-10} & {\cellcolor[HTML]{FECA66}} \color[HTML]{000000} 0.30 & {\cellcolor[HTML]{FFE48C}} \color[HTML]{000000} 0.19 & {\cellcolor[HTML]{FFEA99}} \color[HTML]{000000} 0.15 & {\cellcolor[HTML]{FEE085}} \color[HTML]{000000} 0.21 & {\cellcolor[HTML]{FFF5B3}} \color[HTML]{000000} 0.07 & {\cellcolor[HTML]{FED572}} \color[HTML]{000000} 0.26 & {\cellcolor[HTML]{FED673}} \color[HTML]{000000} 0.26 & {\cellcolor[HTML]{FEE085}} \color[HTML]{000000} 0.21 & {\cellcolor[HTML]{FEB04B}} \color[HTML]{000000} 0.38 & {\cellcolor[HTML]{800026}} \color[HTML]{000000} 1.00 & {\cellcolor[HTML]{FEDE80}} \color[HTML]{000000} 0.22 & {\cellcolor[HTML]{FFE48D}} \color[HTML]{000000} 0.18 \\
\textbf{2023-11} & {\cellcolor[HTML]{FFF0A8}} \color[HTML]{000000} 0.10 & {\cellcolor[HTML]{FFEC9D}} \color[HTML]{000000} 0.14 & {\cellcolor[HTML]{FFF0A8}} \color[HTML]{000000} 0.10 & {\cellcolor[HTML]{FFEEA3}} \color[HTML]{000000} 0.12 & {\cellcolor[HTML]{FEDE80}} \color[HTML]{000000} 0.22 & {\cellcolor[HTML]{FFEDA0}} \color[HTML]{000000} 0.12 & {\cellcolor[HTML]{FFE590}} \color[HTML]{000000} 0.17 & {\cellcolor[HTML]{FFE58F}} \color[HTML]{000000} 0.18 & {\cellcolor[HTML]{FFEB9C}} \color[HTML]{000000} 0.14 & {\cellcolor[HTML]{FEDE80}} \color[HTML]{000000} 0.22 & {\cellcolor[HTML]{800026}} \color[HTML]{000000} 1.00 & {\cellcolor[HTML]{FFEA9B}} \color[HTML]{000000} 0.14 \\
\textbf{2023-12} & {\cellcolor[HTML]{FFF5B5}} \color[HTML]{000000} 0.07 & {\cellcolor[HTML]{FEE38B}} \color[HTML]{000000} 0.19 & {\cellcolor[HTML]{FFE48C}} \color[HTML]{000000} 0.19 & {\cellcolor[HTML]{FFE793}} \color[HTML]{000000} 0.17 & {\cellcolor[HTML]{FFF1A9}} \color[HTML]{000000} 0.10 & {\cellcolor[HTML]{FFE794}} \color[HTML]{000000} 0.16 & {\cellcolor[HTML]{FFE590}} \color[HTML]{000000} 0.17 & {\cellcolor[HTML]{FFE58F}} \color[HTML]{000000} 0.18 & {\cellcolor[HTML]{FFF1A9}} \color[HTML]{000000} 0.10 & {\cellcolor[HTML]{FFE48D}} \color[HTML]{000000} 0.18 & {\cellcolor[HTML]{FFEA9B}} \color[HTML]{000000} 0.14 & {\cellcolor[HTML]{800026}} \color[HTML]{000000} 1.00 \\
\end{tabular}

%% file: tabulars/left/jaccard_similarity_rfc1112_202301-202312.tex
\begin{tabular}{lrrrrrrrrrrrr}
\makecell[c]{\textbf{Month}} & \makecell[c]{\rot{\textbf{2023-01}}} & \makecell[c]{\rot{\textbf{2023-02}}} & \makecell[c]{\rot{\textbf{2023-03}}} & \makecell[c]{\rot{\textbf{2023-04}}} & \makecell[c]{\rot{\textbf{2023-05}}} & \makecell[c]{\rot{\textbf{2023-06}}} & \makecell[c]{\rot{\textbf{2023-07}}} & \makecell[c]{\rot{\textbf{2023-08}}} & \makecell[c]{\rot{\textbf{2023-09}}} & \makecell[c]{\rot{\textbf{2023-10}}} & \makecell[c]{\rot{\textbf{2023-11}}} & \makecell[c]{\rot{\textbf{2023-12}}} \\
\textbf{2023-01} & {\cellcolor[HTML]{800026}} \color[HTML]{000000} 1.00 & {\cellcolor[HTML]{FEE187}} \color[HTML]{000000} 0.20 & {\cellcolor[HTML]{FFE793}} \color[HTML]{000000} 0.17 & {\cellcolor[HTML]{FEBF5A}} \color[HTML]{000000} 0.33 & {\cellcolor[HTML]{FFE793}} \color[HTML]{000000} 0.17 & {\cellcolor[HTML]{FFEDA0}} \color[HTML]{000000} 0.12 & {\cellcolor[HTML]{FC5B2E}} \color[HTML]{000000} 0.60 & {\cellcolor[HTML]{FFE793}} \color[HTML]{000000} 0.17 & {\cellcolor[HTML]{FFE793}} \color[HTML]{000000} 0.17 & {\cellcolor[HTML]{FEA245}} \color[HTML]{000000} 0.43 & {\cellcolor[HTML]{FD8C3C}} \color[HTML]{000000} 0.50 & {\cellcolor[HTML]{FFEDA0}} \color[HTML]{000000} 0.12 \\
\textbf{2023-02} & {\cellcolor[HTML]{FEE187}} \color[HTML]{000000} 0.20 & {\cellcolor[HTML]{800026}} \color[HTML]{000000} 1.00 & {\cellcolor[HTML]{E2191C}} \color[HTML]{000000} 0.75 & {\cellcolor[HTML]{FC5B2E}} \color[HTML]{000000} 0.60 & {\cellcolor[HTML]{FFE793}} \color[HTML]{000000} 0.17 & {\cellcolor[HTML]{FECE6A}} \color[HTML]{000000} 0.29 & {\cellcolor[HTML]{FFEA9B}} \color[HTML]{000000} 0.14 & {\cellcolor[HTML]{FEAB49}} \color[HTML]{000000} 0.40 & {\cellcolor[HTML]{FFE793}} \color[HTML]{000000} 0.17 & {\cellcolor[HTML]{FFEFA5}} \color[HTML]{000000} 0.11 & {\cellcolor[HTML]{FECE6A}} \color[HTML]{000000} 0.29 & {\cellcolor[HTML]{FD8C3C}} \color[HTML]{000000} 0.50 \\
\textbf{2023-03} & {\cellcolor[HTML]{FFE793}} \color[HTML]{000000} 0.17 & {\cellcolor[HTML]{E2191C}} \color[HTML]{000000} 0.75 & {\cellcolor[HTML]{800026}} \color[HTML]{000000} 1.00 & {\cellcolor[HTML]{FD8C3C}} \color[HTML]{000000} 0.50 & {\cellcolor[HTML]{FED976}} \color[HTML]{000000} 0.25 & {\cellcolor[HTML]{FED976}} \color[HTML]{000000} 0.25 & {\cellcolor[HTML]{FFEDA0}} \color[HTML]{000000} 0.12 & {\cellcolor[HTML]{FC5B2E}} \color[HTML]{000000} 0.60 & {\cellcolor[HTML]{FFEA9B}} \color[HTML]{000000} 0.14 & {\cellcolor[HTML]{FFF1A9}} \color[HTML]{000000} 0.10 & {\cellcolor[HTML]{FEA245}} \color[HTML]{000000} 0.43 & {\cellcolor[HTML]{FEA245}} \color[HTML]{000000} 0.43 \\
\textbf{2023-04} & {\cellcolor[HTML]{FEBF5A}} \color[HTML]{000000} 0.33 & {\cellcolor[HTML]{FC5B2E}} \color[HTML]{000000} 0.60 & {\cellcolor[HTML]{FD8C3C}} \color[HTML]{000000} 0.50 & {\cellcolor[HTML]{800026}} \color[HTML]{000000} 1.00 & {\cellcolor[HTML]{FEBF5A}} \color[HTML]{000000} 0.33 & {\cellcolor[HTML]{FEB24C}} \color[HTML]{000000} 0.38 & {\cellcolor[HTML]{FED976}} \color[HTML]{000000} 0.25 & {\cellcolor[HTML]{FECE6A}} \color[HTML]{000000} 0.29 & {\cellcolor[HTML]{FFEDA0}} \color[HTML]{000000} 0.12 & {\cellcolor[HTML]{FEE187}} \color[HTML]{000000} 0.20 & {\cellcolor[HTML]{FEB24C}} \color[HTML]{000000} 0.38 & {\cellcolor[HTML]{FEB24C}} \color[HTML]{000000} 0.38 \\
\textbf{2023-05} & {\cellcolor[HTML]{FFE793}} \color[HTML]{000000} 0.17 & {\cellcolor[HTML]{FFE793}} \color[HTML]{000000} 0.17 & {\cellcolor[HTML]{FED976}} \color[HTML]{000000} 0.25 & {\cellcolor[HTML]{FEBF5A}} \color[HTML]{000000} 0.33 & {\cellcolor[HTML]{800026}} \color[HTML]{000000} 1.00 & {\cellcolor[HTML]{FEA647}} \color[HTML]{000000} 0.42 & {\cellcolor[HTML]{FFEA9B}} \color[HTML]{000000} 0.14 & {\cellcolor[HTML]{FED976}} \color[HTML]{000000} 0.25 & {\cellcolor[HTML]{FED976}} \color[HTML]{000000} 0.25 & {\cellcolor[HTML]{FD8C3C}} \color[HTML]{000000} 0.50 & {\cellcolor[HTML]{FEC863}} \color[HTML]{000000} 0.31 & {\cellcolor[HTML]{FFEC9D}} \color[HTML]{000000} 0.13 \\
\textbf{2023-06} & {\cellcolor[HTML]{FFEDA0}} \color[HTML]{000000} 0.12 & {\cellcolor[HTML]{FECE6A}} \color[HTML]{000000} 0.29 & {\cellcolor[HTML]{FED976}} \color[HTML]{000000} 0.25 & {\cellcolor[HTML]{FEB24C}} \color[HTML]{000000} 0.38 & {\cellcolor[HTML]{FEA647}} \color[HTML]{000000} 0.42 & {\cellcolor[HTML]{800026}} \color[HTML]{000000} 1.00 & {\cellcolor[HTML]{FFF1A9}} \color[HTML]{000000} 0.10 & {\cellcolor[HTML]{FED976}} \color[HTML]{000000} 0.25 & {\cellcolor[HTML]{FEA245}} \color[HTML]{000000} 0.43 & {\cellcolor[HTML]{FFF3AF}} \color[HTML]{000000} 0.08 & {\cellcolor[HTML]{FEE187}} \color[HTML]{000000} 0.20 & {\cellcolor[HTML]{FEE187}} \color[HTML]{000000} 0.20 \\
\textbf{2023-07} & {\cellcolor[HTML]{FC5B2E}} \color[HTML]{000000} 0.60 & {\cellcolor[HTML]{FFEA9B}} \color[HTML]{000000} 0.14 & {\cellcolor[HTML]{FFEDA0}} \color[HTML]{000000} 0.12 & {\cellcolor[HTML]{FED976}} \color[HTML]{000000} 0.25 & {\cellcolor[HTML]{FFEA9B}} \color[HTML]{000000} 0.14 & {\cellcolor[HTML]{FFF1A9}} \color[HTML]{000000} 0.10 & {\cellcolor[HTML]{800026}} \color[HTML]{000000} 1.00 & {\cellcolor[HTML]{FFEDA0}} \color[HTML]{000000} 0.12 & {\cellcolor[HTML]{FFEDA0}} \color[HTML]{000000} 0.12 & {\cellcolor[HTML]{FEBF5A}} \color[HTML]{000000} 0.33 & {\cellcolor[HTML]{FEB24C}} \color[HTML]{000000} 0.38 & {\cellcolor[HTML]{FFF1A9}} \color[HTML]{000000} 0.10 \\
\textbf{2023-08} & {\cellcolor[HTML]{FFE793}} \color[HTML]{000000} 0.17 & {\cellcolor[HTML]{FEAB49}} \color[HTML]{000000} 0.40 & {\cellcolor[HTML]{FC5B2E}} \color[HTML]{000000} 0.60 & {\cellcolor[HTML]{FECE6A}} \color[HTML]{000000} 0.29 & {\cellcolor[HTML]{FED976}} \color[HTML]{000000} 0.25 & {\cellcolor[HTML]{FED976}} \color[HTML]{000000} 0.25 & {\cellcolor[HTML]{FFEDA0}} \color[HTML]{000000} 0.12 & {\cellcolor[HTML]{800026}} \color[HTML]{000000} 1.00 & {\cellcolor[HTML]{FFEA9B}} \color[HTML]{000000} 0.14 & {\cellcolor[HTML]{FFF1A9}} \color[HTML]{000000} 0.10 & {\cellcolor[HTML]{FEA245}} \color[HTML]{000000} 0.43 & {\cellcolor[HTML]{FEA245}} \color[HTML]{000000} 0.43 \\
\textbf{2023-09} & {\cellcolor[HTML]{FFE793}} \color[HTML]{000000} 0.17 & {\cellcolor[HTML]{FFE793}} \color[HTML]{000000} 0.17 & {\cellcolor[HTML]{FFEA9B}} \color[HTML]{000000} 0.14 & {\cellcolor[HTML]{FFEDA0}} \color[HTML]{000000} 0.12 & {\cellcolor[HTML]{FED976}} \color[HTML]{000000} 0.25 & {\cellcolor[HTML]{FEA245}} \color[HTML]{000000} 0.43 & {\cellcolor[HTML]{FFEDA0}} \color[HTML]{000000} 0.12 & {\cellcolor[HTML]{FFEA9B}} \color[HTML]{000000} 0.14 & {\cellcolor[HTML]{800026}} \color[HTML]{000000} 1.00 & {\cellcolor[HTML]{FFF1A9}} \color[HTML]{000000} 0.10 & {\cellcolor[HTML]{FFEFA5}} \color[HTML]{000000} 0.11 & {\cellcolor[HTML]{FFEFA5}} \color[HTML]{000000} 0.11 \\
\textbf{2023-10} & {\cellcolor[HTML]{FEA245}} \color[HTML]{000000} 0.43 & {\cellcolor[HTML]{FFEFA5}} \color[HTML]{000000} 0.11 & {\cellcolor[HTML]{FFF1A9}} \color[HTML]{000000} 0.10 & {\cellcolor[HTML]{FEE187}} \color[HTML]{000000} 0.20 & {\cellcolor[HTML]{FD8C3C}} \color[HTML]{000000} 0.50 & {\cellcolor[HTML]{FFF3AF}} \color[HTML]{000000} 0.08 & {\cellcolor[HTML]{FEBF5A}} \color[HTML]{000000} 0.33 & {\cellcolor[HTML]{FFF1A9}} \color[HTML]{000000} 0.10 & {\cellcolor[HTML]{FFF1A9}} \color[HTML]{000000} 0.10 & {\cellcolor[HTML]{800026}} \color[HTML]{000000} 1.00 & {\cellcolor[HTML]{FECA66}} \color[HTML]{000000} 0.30 & {\cellcolor[HTML]{FFF3AF}} \color[HTML]{000000} 0.08 \\
\textbf{2023-11} & {\cellcolor[HTML]{FD8C3C}} \color[HTML]{000000} 0.50 & {\cellcolor[HTML]{FECE6A}} \color[HTML]{000000} 0.29 & {\cellcolor[HTML]{FEA245}} \color[HTML]{000000} 0.43 & {\cellcolor[HTML]{FEB24C}} \color[HTML]{000000} 0.38 & {\cellcolor[HTML]{FEC863}} \color[HTML]{000000} 0.31 & {\cellcolor[HTML]{FEE187}} \color[HTML]{000000} 0.20 & {\cellcolor[HTML]{FEB24C}} \color[HTML]{000000} 0.38 & {\cellcolor[HTML]{FEA245}} \color[HTML]{000000} 0.43 & {\cellcolor[HTML]{FFEFA5}} \color[HTML]{000000} 0.11 & {\cellcolor[HTML]{FECA66}} \color[HTML]{000000} 0.30 & {\cellcolor[HTML]{800026}} \color[HTML]{000000} 1.00 & {\cellcolor[HTML]{FEE187}} \color[HTML]{000000} 0.20 \\
\textbf{2023-12} & {\cellcolor[HTML]{FFEDA0}} \color[HTML]{000000} 0.12 & {\cellcolor[HTML]{FD8C3C}} \color[HTML]{000000} 0.50 & {\cellcolor[HTML]{FEA245}} \color[HTML]{000000} 0.43 & {\cellcolor[HTML]{FEB24C}} \color[HTML]{000000} 0.38 & {\cellcolor[HTML]{FFEC9D}} \color[HTML]{000000} 0.13 & {\cellcolor[HTML]{FEE187}} \color[HTML]{000000} 0.20 & {\cellcolor[HTML]{FFF1A9}} \color[HTML]{000000} 0.10 & {\cellcolor[HTML]{FEA245}} \color[HTML]{000000} 0.43 & {\cellcolor[HTML]{FFEFA5}} \color[HTML]{000000} 0.11 & {\cellcolor[HTML]{FFF3AF}} \color[HTML]{000000} 0.08 & {\cellcolor[HTML]{FEE187}} \color[HTML]{000000} 0.20 & {\cellcolor[HTML]{800026}} \color[HTML]{000000} 1.00 \\
\end{tabular}